\documentclass[12pt,preprint]{aastex}

\usepackage{amstex}
\newcommand{\fb}{$f_\mathrm{B}$}
\newcommand\eg{{\it e.g.,}\,}
\newcommand\whz{{W Hz$^{-1}$}}

\begin{document}

\title{Radio-selected Galaxies in Very Rich Clusters at $z \le$ 0.25: 
I.~Multi-wavelength Observations and Data Reduction Techniques}

\author{G. E. Morrison,\altaffilmark{1,2,3,4,5}
F. N. Owen\altaffilmark{1,2} M. J. Ledlow,\altaffilmark{1,6}
W. C. Keel,\altaffilmark{1,7} J. M. Hill,\altaffilmark{8},
W. Voges,\altaffilmark{9}, and T. Herter,\altaffilmark{10}}

\altaffiltext{1}{Visiting Astronomer, Kitt Peak National Observatory
(KPNO), National Optical Astronomy Observatories, operated by the
Association of Universities for Research in Astronomy, Inc.\ under
contract with the National Science Foundation.}

\altaffiltext{2}{National Radio Astronomy Observatory, P. O. Box O, Socorro,
NM 87801; The National Radio Astronomy Observatory is operated by
Associated Universities, Inc., under a cooperative agreement with the
National Science Foundation.; morrison@@ipac.caltech.edu,
fowen@@aoc.nrao.edu}

\altaffiltext{3}{Also Dept. of Physics \& Astronomy,University of New Mexico, Albuquerque, NM
87131}

\altaffiltext{4}{ Vanguard Research, Inc., Scotts Valley, CA 95066}

\altaffiltext{5}{Present address: California Institute of Technology, IPAC, M/S\,100-22,
Pasadena, CA 91125}

\altaffiltext{6}{Gemini Observatory, Southern Operations Center, AURA, Casilla 603,
La Serena Chile; mledlow@@gemini.edu}

\altaffiltext{7}{Dept. of Physics \& Astronomy, University of Alabama,
Tuscaloosa, AL 35487; keel@@bildad.astr.ua.edu}

\altaffiltext{8}{Steward Observatory, University of Arizona, Tucson, AZ 85721;
jhill@@as.arizona.edu}

\altaffiltext{9}{Max-Planck-Institut f\"{u}r extraterrestrische Physik, Postfach
1312, D-85741 Garching bei M\"{u}nchen, Germany; wvoges@@mpe.mpg.de}

\altaffiltext{10}{202 Space Sciences Building, Cornell University, Ithaca, NY 14853-6801;
tlh10@@cornell.edu}

\begin{abstract}

Radio observations were used to detect the `active' galaxy population
within rich clusters of galaxies in a non-biased manner that is not
plagued by dust extinction or the K-correction.  We present wide-field
radio, optical (imaging and spectroscopy), and ROSAT All-Sky Survey
(RASS) X-ray data for a sample of 30 very rich Abell ($R \ge 2$)
cluster with $z \le 0.25$.  The VLA radio data samples the ultra-faint
radio ($L_{1.4} \ge 2\times 10^{22}$ W Hz$^{-1}$) galaxy population
within these extremely rich clusters for galaxies with $M_{\mathrm{R}}
\le -21$. This is the largest sample of low luminosity 20\,cm 
radio galaxies within rich Abell clusters collected to date.

The radio-selected galaxy sample represents the starburst (Star
formation rate $\ge$ 5\,M$_{\sun}$ yr$^{-1}$) and active galactic
nuclei (AGN) populations contained within each cluster. Archival and
newly acquired redshifts were used to verify cluster membership for
most ($\sim 95$$\%$) of the optical identifications. Thus we can
identify all the starbursting galaxies within these clusters,
regardless of the level of dust obscuration that would affect these
galaxies being identified from their optical signature.  Cluster
sample selection, observations, and data reduction techniques for all
wavelengths are discussed.

\end{abstract}


\section{Introduction}

Optical studies of rich clusters to determine the star formation rates
(SFR) of their cluster members have been ongoing since the early
photometric study by \cite{but78a} over 20 years ago. Such studies
have sought to recover the star formation properties of galaxies in
rich clusters via optical photometry, spectroscopy and more recently
from {\it Hubble Space Telescope} (HST) morphological studies. The
discovery by \citeauthor{but78a} (\citeyear{but78a};
\citeyear{but84b}) of the increased fraction of blue galaxies within
the core of selected rich regular clusters indicated significant
evolution in the SFR and/or galaxy type within the central regions of
these massive systems over the last $\sim 5$ Gyr (assuming H$_0$ = 75
km s$^{-1}$, q$_0$ = 0.1) or $z \lesssim 0.4$. The change in SFR of
galaxies in rich clusters may be related to hierarchical
large-scale-structure (LSS) formation and the accretion of field
galaxies.  Where the starburst population occurs within the cluster
environment yields clues to the origin of the perturbing mechanism.
To study this `active' galaxy population (starburst/AGN) requires a
method that will detect this population in a unbiased manner.

Radio observations allow one to detect starbursts and AGN without the
drawbacks of optical observations, such as dust attenuation and the
optical K-correction. In principle, far-IR (FIR) observations would
detect re-emission from the dust, of the UV photons emitted by the OB
stars formed in a massive starburst. Until the launch of SIRTF, FIR
instruments will not have the sensitivity or the resolution to be
effective. However, the well known FIR-radio continuum correlation
based on ongoing star formation allows easier detection of
starbursting systems in distant clusters. Studies of the local radio
luminosity function \citep[\eg][]{con91}, indicate that the
starbursting and spiral galaxy population dominate the 20\,cm radio
luminosity function below 10$^{23}$ W Hz$^{-1}$.

Radio selection of cluster members by their low radio luminosity
yields two kinds of objects - weak AGN and objects with active star
formation. The FIR-radio correlation ties the nonthermal radio
emission to the ongoing star formation \citep{con92}. In starbursts,
radio emission fades faster than blue starlight, so clusters with
large number of radio sources will preferentially be those that have
had a large numbers of recent or ongoing starbursts. This may indicate
the entry of a new group of galaxies into the cluster with star
formation in its members.

This paper is the first in a series studying the ultra-faint
radio-selected galaxy populations associated with rich clusters of
galaxies as a function of redshift. This paper presents the radio,
optical, and X-ray data for a sample of 30 very rich clusters (Abell
richness class $R \ge$2) with $z \le 0.25$ selected from the catalog
of \citet*[][hereafter ACO]{abe89}.  The VLA offers a large field of
view ($\sim$30 arcmin) permitting us to go out to 2.5 Mpc radius from
the cluster core. This allows us to detect infalling starbursting
field galaxies and any supercluster induced effects such as
cluster-cluster mergers. Our detection limit of 2$\times$10$^{22}$ W
Hz$^{-1}$ allows detection of galaxies undergoing moderate to vigorous
star formation with star formation rates (SFR) $\ge$ 5\,M$_{\sun}$
yr$^{-1}$.

In this data paper, we discuss selection of the cluster sample along
with the multi-wavelength observations and the data reduction
techniques for a sample of the richest Abell clusters from 0.02 $\le$
z $\le$ 0.25. These clusters were chosen to be as rich or richer than
our four rich $z = 0.4$ clusters (Abell 370, Cl0024$+$1654,
Cl0939$+$4713, and CL1447$+$2619), which will be presented in future
papers. We also present a radio-selected sample of starburst and AGN
galaxies within these 30 rich clusters. Most of the detected galaxies
had followup spectroscopy to verify cluster membership.

Future papers analyze the radio properties \citep[Paper II;][]{mor00},
the optical properties of the clusters
\citep[Paper III;][]{mor00a}, and spectroscopy of the optical
counterparts \citep[Paper IV;][]{mor00b}.  Deep (15 hr) wide-field VLA
B-array observations of Cl0939+4713 at $z = 0.41$ and subsequent
spectroscopic analysis of the radio-selected galaxy population will
appear in Paper V \citep{mor00c}. HST NICMOS and WFPC2 results of
several of the radio-selected galaxies in Cl0939+4713 were discussed
in \cite{sma99}.

\section{Sample Selection}

A sample was devised to separate the combined effects that epoch and
richness have on galaxy evolution in clusters. The sample consists of
the richest clusters at redshifts ranging from 0.02 $\le z \le$ 0.41
($N_{2.0} \ge 80$ galaxies where $N_{2.0}$ is our ``modified'' ACO
galaxy count. See \cite{mor00a}.). By choosing only the richest
clusters we can determine if all very rich clusters display an excess
number of radio galaxies, or if only the high redshift ones do.

The clusters at $z < 0.25$ were chosen to be a comparison sample for
the four clusters at $z = 0.4$. We selected Abell clusters
predominantly because they have been relatively well studied
\citep{str87, abe89, huc90, zab90, str91} and their statistical biases
are also well-known \citep[e.g.,][]{sco57, sca91}.

One statistical bias found early on by \cite{jus59} is a weak
richness-distance correlation, which may in part explain the
incompleteness of distant ($z >$~0.2) clusters selected by Abell
\citep{luc83}. Also the Abell richness, which is defined to correlate
with galaxy count, is in some cases significantly in error
\citep{mot77,dre80b}. As noted in \cite{sar88} it is generally 
better to use the actual Abell counts instead of the richness class.

Given the larger error in past richness measurements, new galaxy
counts were obtained --whole cluster and core galaxy counts.  The
``modified'' Abell counts, $N_{2.0}$, were re-measured from the CCD
frames, thereby providing a more consistent and quantitative
measurement of the individual cluster's richness \citep[Paper
III;][]{mor00a}.

\subsection{The Redshift Samples}

{\bf The 0.02 $\le z \le$ 0.07 Sample:} The lowest redshift sample
contains the 10 richest nearby clusters\footnote{A3094 is not used
since it was found to be made up of three clusters at different
redshifts \citep{kat96}.}  ($z\,\le\,0.07$). Here the 1.4 GHz,
45\arcsec \, resolution, D-array NRAO VLA sky survey (NVSS) provided
sufficient sensitivity to survey the richest local clusters out to a
radius of 2.5\,Mpc and down to our limiting luminosity of about
2$\times\,10^{22}$ W Hz$^{-1}$.  However, since the local clusters
were not as rich as the clusters in our $z = 0.4$ sample, we must go
to a higher redshifts to find clusters of comparably high richness.

{\bf The 0.10 $\le z \le$ 0.18 sample:} This low-redshift sample of 14
clusters provides the extremely rich clusters that are lacking in the
local sample and are comparable to the $z\,=\,0.40$ sample. This
provides a baseline useful for determining the dependence of redshift
on cluster properties.

{\bf The 0.20 $\le z \le$ 0.25 sample:} Our intermediate redshift
sample contains 12 very rich, X-ray luminous clusters culled from the
ACO catalog. The two clusters that provided the original motivation
for this study, A2125 and A2645 were included in this sample. Some of
these clusters show a significant increase in their blue galaxy
population compared with local-epoch rich clusters. One of these
clusters is A2125, which also shows a substantially larger
radio-selected galaxy population from $10^{22.3-23}$ W Hz$^{-1}$
compared with A2645 \citep{owe99}.

{\bf The $z = $0.4 sample:} This sample will be presented in
a future paper.

\subsection{Cluster Membership requirements}

For an identification to be considered a cluster member, it was
required to have: 1) absolute Cousins R-band magnitude brighter than
$M_{\mathrm{R}} = -21$, 2) radio luminosity greater than
2$\times$10$^{22}$ \whz, 3) radius $\le$ 2.5 Mpc from the X-ray
defined cluster center, and 4) redshift values such that a galaxy's
relative velocity with respect to the cluster center, given by
$\Delta$V~=~$\frac{\Delta z c}{1+z_{cl}}$ indicates that the galaxy is
gravitationally bound to the cluster. Galaxies with $\Delta$V$\leq$
3000 km s$^{-1}$ were classified as cluster members.
 
The absolute R-band magnitude and the 20\,cm radio continuum limits
were chosen to be consistent with our survey limits for our $z = 0.4$
clusters.  The sensitivity limit (hence integration time) for each
cluster was chosen to reach M$_{\mathrm{R}}$ = $-$21.0 and L$_{1.4
GHz}$ = 2$\times$10$^{22}$ \whz.

\section{Observations and Data Reduction}

\subsection{Radio Observations}

Wide-field Very Large Array \citep{tho80} 1.4 GHz radio observations
were obtained for 34 rich clusters spanning 0.02 $\le$ z $\le$ 0.25.
Radio data for the ten rich Abell clusters between 0.02 $\le$ z $\le$
0.07 were from the NVSS \citep{con98}. Fourteen clusters between 0.10
$\le$ z $\le$ 0.18 were observed at 1.4\,GHZ with the VLA in A-array
(1.5\arcsec \, resolution providing $\sigma_{\mathrm{P}} \approx
0.15\arcsec$ for sources at the 5$\sigma$ level, a worst case) and
C-array (15\arcsec \, resolution) for 30 and 15 minutes,
respectively. Only ten were found to have a richness $\ge 2$ after
their richness was re-measured \citep{mor00a} thus the other four were
excluded from the sample\footnote{four excluded clusters (A1437,
A1674, A1918, A1990)}.  An additional ten clusters between 0.20 $\le$
z $\le$ 0.23 were observed with the VLA in A array for 30 minutes. The
radio observations of the $z=0.25$ clusters, A2125 and A2645, are
discussed in detail in \cite{dwa99}. Our radio observations achieved
sufficient depth at large radius to define flux-density-limited
samples.  Thus, all clusters regardless of redshift were observed down
to radio luminosities of at least $L_{1.4\,GHz} \ge 2 \times
10^{22}$\,W Hz$^{-1}$.

The observations were adjusted to the flux density scale of
\cite{baa77} using 3C286 as the primary flux density calibrator.
The AIPS (Astronomical Imaging Processing System) task IMAGR was used
to clean the full primary beam thus achieving a field of view of 30
arcmin. Wide field images incur both bandwidth and time-average
smearing. To minimize the bandwidth smearing, we used the ``4'' mode
in the spectral line mode with 2 polarizations, 2 IFs and
7$\times$3.125 MHz channels \citep{per00}.  We averaged over the 7
channels and 2 IFs together in the gridding process under IMAGR to
make the continuum map. To correct the time-average smearing, which is
a significant problem in the A-array, the averaging time of the VLA
correlator was changed from 10 to 5 seconds.

Self-calibration (phase and amplitude) was performed to increase the
dynamic range and the sensitivity of the radio maps.  This process
included self-editing of the data, where the CLEAN components (from a
suitable model of the sky intensity distribution) from these maps were
subtracted from the (u,v) data set. Having removed the model from the
(u,v) data, the remaining high residuals are known to be unrelated to
the sky intensity distribution and were deleted. The model was then
restored to the clipped (u,v) data.

To correct for non-coplanar baselines, we imaged the (u,v) data in
time ranges that have approximately coplanar baselines. All
significant sources ($\ge 1$ mJy) out to the first sidelobe of the
primary beam were imaged using subfields. This prevented sidelobes
from bright, out-of-field sources from degrading the high sensitivity
image.  We transformed these subimages into the same flat coordinate
system using the AIPS task OHGEO and summed them. The noise in the
resulting images approaches the theoretical value: A-array $\sigma
\sim$\,45$\mu$Jy, and C-array, $\sigma \sim$\,70$\mu$Jy. This process
yields wide field images with the correct sky projection.

Radio source catalogs were generated from the A- and C-array images
using the AIPS task SAD (Search and Destroy). This task finds all
potential sources within a defined region and fits Gaussian components
to these. SAD was used to search for radio sources with peak flux
above 4.5$\sigma$. The images were viewed by eye, to look for sources
missed by SAD. In general, fewer than 3$\%$ of the total number of
sources were missed due to algorithm fitting failure.

\subsection{Optical Observations}

\subsubsection{CCD Imaging and The Digital Sky Survey}

Twenty-two clusters ($z > 0.1$) were observed with the KPNO 0.9 meter
telescope in Cousins $R$-band and 17 were observed in $B$-band. The
CCD detector was a 2048$\times$2048 Textronix chip having a pixel size
on the sky of 0.68 \arcsec, yielding a field of view of 23\arcmin
\, on a side. We dithered and co-added 3$\times$5 minute frames 
in $R$ and 3$\times$10 minute frames in $B$. The stacking of the
individual frames allows the cosmic rays to be rejected. All frames
were bias corrected and flattened using a twilight frame. Photometric
calibration was determined from $\sim$20 Landolt standards each night
\citep{lan92}.

The Gunn-Oke aperture \citep[G-O,][]{gun75} with metric diameter of
26.2 kpc was used in measuring the apparent magnitudes on the $R$
frame. We corrected the magnitudes for Galactic reddening and applied
a K-correction \citep{mor00a}.  Since the photometry was done using a
G-O metric aperture, the very extended galaxies will have their
apparent R-band magnitudes underestimated since the G-O aperture does
not encompass all of the outer halo. Figure 2 in \cite{led95b} shows a
comparison between isophotal magnitudes, which measure out to a fixed
surface brightness value (in this case 24.5\,mag/\sq\arcsec), and G-O
aperture magnitudes. We adjusted the magnitudes of the extended
galaxies to put them all on an approximate isophotal system using the
above calibration by \citeauthor{led95b}.

The 23\arcmin \, field-of-view of the 0.9 meter telescope was too
small for clusters with $z < 0.2$ to cover the 5\,Mpc diameter
region. In those cases, we used the Digital Sky Survey\footnote{The
Digitized Sky Surveys were produced at the Space Telescope Science
Institute under U.S.  Government grant NAG W-2166. The images of these
surveys are based on photographic data obtained using the Oschin
Schmidt Telescope on Palomar Mountain. The plates were processed into
the present compressed digital form with the permission of these
institutions.}  (DSS) to find optical identifications for the radio
sources out to 2.5 Mpc from the cluster center. The photometric
calibration for the DSS was bootstrapped from the $R$-band images.
The lack of accurate photometric calibration and color information are
the primary reasons we undertook the CCD imaging for the $z > 0.1$
clusters.

\subsubsection{Astrometry}

Optical counterparts to the radio sources were found using the
$R$-band images and the DSS E-plate images. The plate solution for the
$R$ image was calculated using the AIPS task XTRAN.  Stars from the
HST Guide Star Catalog CD-ROM (GSC 1.1) between 12.5 to 15 magnitude
were used to set the absolute astrometry for the CCD frame.
Unfortunately, the GSC has an absolute positional uncertainty of about
0.7\arcsec \ in each coordinate \citep{rus90,deu99}. This required us
to set the absolute astrometry to the radio reference frame using the
offsets in right ascension and declination from the optical
identifications. Thus the absolute optical astrometry is about
0.25\arcsec, while the residuals from XTRAN were less than 0.5\arcsec.

The DSS images had their plate solution calculated and applied using
the AIPS task SKYVE. As with the 0.9 meter telescope images, the
absolute astrometry was tied to the radio reference frame using the
offsets from the optical identifications.

\subsubsection{Spectroscopy}

A log of spectroscopic observations of candidate radio galaxy
identifications are shown in Table~\ref{tbl-1}.

The MX multifiber spectrograph on the Steward Observatory 2.3 meter
telescope \citep{hil86} was used to observe up to 32 objects per 1
hour observation. A 400 line mm$^{-1}$ grating was used covering
3700-6950\AA \ at an 8 \AA \ resolution, with a UV-flooded Loral 1200
$\times$ 800 pixel CCD as the detector. Flat-fielding, sky-subtraction
and wavelength calibration were done as described in \cite{hil93}.

Additional spectroscopy was done using the long slit of the RC
spectrograph on the KPNO 4 meter telescope. The 300 lines/mm grating
is blazed at 4000 \AA \ and covered 4300\AA \ - 8600\AA, allowing [N
II] and H$\alpha$ to be observed for 0.10 $\le$ z $\le$ 0.18. The long
slit was usually rotated to accommodate two or more radio galaxies
during the five to fifteen minute exposures. Nightly observations of
standard stars were used to determine a relative flux calibration for
the spectral energy distribution of all the observed
galaxies. Absolute calibrations were not done.

Further spectroscopy was done in the long slits mode of the COSMIC
spectrograph \citep{kel98} and the Double Spectrograph \citep{oke82}
mounted on the Hale 5.1 (P200) meter telescope\footnote{The Hale 5
meter telescope is owned and operated by the California Institute of
Technology}. COSMIC was used with the 300 lines mm$^{-1}$ grating
blazed at 5500 \AA \ giving 8 \AA \ resolution. The wavelength
coverage was 3500-9800 \AA \ which allowed [O\,II] $\lambda$\,3727 and
H$\alpha$ to be observed throughout our redshift range of $0.02 \le z
\le 0.23$. The Double Spectrograph used the 5500 \AA \ dichroic filter
thereby directing the light blueward of 5500 \AA \ to the blue
detector and light redward of the break to the red detector. The blue
detector (1024$^2$ pixel CCD) setup used the 600\,lines mm$^{-1}$
grating blazed at 4000 \AA.  The red detector (1024$^2$ pixel CCD)
setup included the 316\,lines mm$^{-1}$ grating blazed at 7500
\AA. This setup allowed wavelength coverage from 3700-8200 \AA \ with
a break from 5500-5650 \AA. We chose the wavelength of this break so
that both [OII] and H$\alpha$ throughout the $0.02 \leq z \leq 0.23$
redshift range. Due to the poorer sensitivity of the blue camera
compared with the red, we wished to observe as much of the spectrum in
the red camera as possible.

The spectra of the optical identifications were used to determine
cluster membership. The IRAF cross-correlation task, FXCOR
\citep{fit93}, was used to derive redshifts from absorption
features with a K0 III radial velocity standard as the template. For
the MX data, redshift errors were determined using the
\cite{ton79} R-value which represents the cross-correlation strength
of the template spectrum and object spectrum. The velocity error was
calculated using $\Delta$V = 280/(1+R) km s$^{-1}$, which was
determined from multiple MX observations of cluster galaxies by
\cite{hil93}. Spectra with R-values less than 4 were considered
unreliable\footnote{Noted by a symbol in Table~\ref{Z_LOW_OPTICAL_IDS}
as uncertain.}. All spectra were visually inspected to determine if
the calculated redshift agreed with spectroscopic features.

Redshift calibration errors of $\le$ 20 km\,s$^{-1}$ for the spectra
observed with the KPNO 4 meter and the P200 telescopes were determined
from repeated observations of a radial velocity standard on different
nights. This indicates that the errors are primarily related to the
signal-to-noise ratio.  Given the lack of many duplicate spectra to
evaluate the errors, the errors from FXCOR have been assumed.

Additional spectra from the KPNO 2.1-meter telescope using the GoldCam
spectrograph were obtained by Neal Miller. Details are given by
\cite{mil01}.

Spectroscopic classification and analysis of the radio-selected
galaxies will be presented in a future paper \citep{mor00b}.

\subsection{The X-ray Data}

\subsubsection{The X-ray Database}\label{RASS_DATA}

The R\"{o}ntgensatellite, ROSAT \citep{tru83}, conducted an all-sky
survey \citep[RASS--ROSAT All-sky survey;][]{vog99} in the soft X-ray
energy band (0.1 - 2.4 keV) from 1990 August to 1991 February
\citep{vog92b}. The survey was observed in 2$\arcdeg$ wide great
circles that were orthogonal to the solar direction. The resultant
exposure times for these strips varied between $\sim$\,400 \ seconds
at the ecliptic equator and $\sim$\,40,000 \ seconds at the ecliptic
poles. The instrument used for the RASS was a Position Sensitive
Proportional Counter, PSPC-C, \citep{pfe86} as compared to the PSPC-B
used for the pointed observations.  The survey was processed using the
Standard Analysis Software System (SASS) which was developed for RASS
\citep{vog92a} at the Max-Planck-Institute f\"{u}r extraterrestrische
Physik (MPE). Further details on the x-ray database used can be found in 
\citep{vog99}

\subsubsection{X-ray Luminosity}\label{ECF_SECTION}

The X-ray luminosity for each cluster was derived from the photon
counts over the RASS images. The photons were accumulated within a 0.5
Mpc radius circle, $\mathcal{S}~=~S + B$ (source plus background),
centered on the peak thermal X-ray emission from the cluster. This
aperture size was chosen to optimize the signal-to-noise ratio for the
detected cluster's x-ray emission as discussed in \cite{bri93} and to
be comparable with survey work done by others. The background rate was
determined from a source free region of the image.  The center was
measured using the AIPS task JMFIT. This Gaussian fitting routine was
used to measure the position of the peak intensity, integrated
intensity, and FWHM of a source. The fit was made to a smoothed X-ray
image (convolved with a 500 kpc FWHM Gaussian) over a 1.0 Mpc diameter
region at the cluster's redshift.  The counts accumulated within the
aperture were measured using the AIPS task IRING. The recorded average
value was in units of count s$^{-1}$ pixel$^{-1}$ which was then
multiplied by the number of pixels contained in the measured region to
yield a total ROSAT count rate.

The ROSAT count rates were converted into fluxes using the ROSAT PSPC
photon-to-flux conversion factor or ECF \citep{bri93}. We assumed a
thermal model for the spectrum with a temperature,
$\log_{10}(T_{\mathrm{g}}) = 7.3$, and a metal abundance (Z) of
0.3. Neither $T_{\mathrm{g}}$ nor Z were  critical for the ECF in our
energy band, 0.5 - 2.0 keV. Changing the temperature by 2 keV has a
less than 5$\%$ effect on the total luminosity \citep{riz97}.

To get a `total' luminosity for each cluster a correction must be made
for the fact that we were using an aperture that does not contain all
of the x-ray flux from the cluster \citep{bri93}. We corrected the
$L_{\mathrm{X}}$ determined from the counts in the 500 kpc aperture to
the total $L_{\mathrm{X}}$ assuming a core radius of $r_{\mathrm{c}} =
0.2$\,Mpc and using a modified King or $\beta$ model for the surface
brightness distribution. This gave a correction factor of 1.67 to the
aperture luminosity. Table~\ref{X_RAY_RESULTS} shows the x-ray
luminosity for each cluster along with 1$\sigma$ associated Poisson
errors.

\section{Optical Identifications}

\subsection{Optical Identification Criteria: A-array Radio Data}
     
To establish the optical identification criteria for clusters having
VLA A-array data, we must determine the number of optical candidates
within a particular search radius of the individual radio
positions. Given the high accuracy of the A-array positions, the
number of optical identifications found relative to the A-array
position was recorded as a function of search radius. A radio and
optical database was constructed from the AIPS task SAD, which found
all sources on the optical and radio images above a given peak cutoff
value and then used the AIPS task JMFIT to fit Gaussian components to
these sources.

Considering only radio sources with a Gaussian FWHM $\le$ 3\arcsec, a
search was done around each radio source for an optical counterpart,
with the search radius varying from $\le$~1\arcsec~to
$\le$~5\arcsec~in steps of one arcsec. Given the small sizes of the
optical objects ($\le$ 5\arcsec) and radio sources, a separation $\ge$
5\arcsec~was considered to be a random identification. To determine
the random optical association probability for each radio source, we
looked at the optical source densities at each radio position. To have
adequate signal-to-noise to base our coincidence prediction on, we
searched out from each radio position 5\arcsec\, to 50\arcsec\,
looking for optical identifications. By increasing the search radius,
a larger number of optical objects could be selected, thus providing a
sizable data set to base our random optical probability function on.

The coincidence analysis used radio sources found by SAD with a FWHM
$\le$ 3.0\arcsec~and $S_{peak} \ge 4.5\sigma_A$, where $\sigma_A$ is
the rms or ``noise'' of the A array map for that particular
region. The resulting number of radio sources was 221 for A665. Having
limited the radio sources to $\ge 4.5\sigma_A$ detections, we next
limited the magnitude range of acceptable optical identifications.
Identifications were accepted only for optical objects with absolute
magnitude $M_{R} \le -21$ at the cluster redshift. This is one
magnitude fainter than $M_{\star}$ ($M_{\star} = -22$) and contains a
significant fraction of the population of these rich clusters.  This
allowed a larger number of optical identifications to be found thereby
allowing an accurate ($\pm$0.3\arcsec) measurement of the
optical-radio offset.

For radio sources with $\theta \le$ 3\arcsec \ we used the radio and
optical data of A665 to establish the optical identification
criteria. The technique we used here does not depend on the density of
radio sources but on the density of optical sources in a given
field. A665 is the highest richness cluster known and thus gives a
pessimistic answer, {\it i.e.}, an answer on the high side of the
number of random galaxies expected. Thus we may be slightly
overestimating the random galaxies for the other fields.  The data
from the other clusters ($z \ge 0.10$) were checked and compared with
the results from A665.  The predicted number of random identifications
for A665 (using the limit of $M_{R} \le -21$) given a search radius
$\leq$ 1\arcsec~is 0.28 per square arc minute, based on the optical
identifications made in the 5\arcsec-50\arcsec~annulus. Thus, A665
should have 0.05 random objects within a search radius of 1\arcsec \
of all 221 radio sources. For all 20 clusters with $0.1 \le z \le
0.23$ there will be 1.0 random object in our result. Therefore, we
will adopt a search radius $\le$ 1\arcsec~for all radio sources with
angular size $\le 3$\,\arcsec.

A larger search radius ($r \leq$ 3\arcsec) is required for radio
sources with angular sizes from 3\arcsec $\le \theta \le$ 6\arcsec.
There are few sources with sizes between 3\arcsec $< \theta \le$
6\arcsec, which allows us to search out further without greatly
increasing the number of random matches. For the $M_{R} \le -21$ and a
search radius of 3\arcsec, the total number of random identifications
for the 20 clusters is 0.5 objects.

The optical identification for extended radio sources ($\theta$ $\ge$
6\arcsec) was based on known radio galaxy morphologies.  The twin-jet
radio morphology types considered were the Narrow-Angle tail (NAT),
and Wide-Angles tail (WAT). Other morphologies that were not so clearly
delineated by these groups were classified as asymmetrical or complex
radio sources. The most likely optical candidate for the latter cases
usually involve large positional offsets which is further problematic
for very low-surface brightness radio sources. In some cases, the
low-surface brightness sources were not seen in the A-array maps, but
were detected only on the C-array images. For these cases, the
identification was based on the type of radio source we believe them to
be, by comparison with local surveys.

All optical and radio catalogs generated from this study will be made
available through NED\footnote{The NASA/IPAC Extragalactic
Database (NED) is operated by the Jet Propulsion Laboratory,
California Institute of Technology, under contract with the National
Aeronautics and Space Administration.}.

\subsection{Optical Identification Criteria: NVSS Radio Data}

For the richest nearby clusters we used radio data from the
NVSS. These 10 clusters ranged between $0.02 \le z \le 0.07$ and Abell
richness values between $2 \le R \le 3$. The low resolution (beam size
= 45\arcsec) and high sensitivity ($\sigma \approx 0.45$mJy) of the
NVSS is good for detecting extended radio sources but the large
beamsize makes it more difficult to identify optical
counterparts. Based on a procedure similar to that used for the
A-array data set, an acceptable number of random identifications
occurs for a search radius of 15\arcsec. However, for extended radio
structures whose centroids do not agree with the position of the
parent galaxy, contour plots of NVSS sources were overlaid on the DSS
images to secure the optical identification. The reliability of this
method is hard to quantify, but since there were only fewer than 3
such cases per cluster and the optical identifications tend to be
brighter, the chances of there being an incorrect identification is
rather small.

\subsection{Radio Flux Density Criteria}\label{FLUX_CRITERIA}

The flux density requirement for inclusion was $S_{Peak} \ge
4.5\,\sigma_A$ for A-array detections and $S_{Peak} \ge 4.5\,\sigma_C$
for C-array. If these sources proved to have an optical counterpart
then we report radio luminosities derived from the integrated flux
densities. For unresolved and slightly resolved sources, we fit a
Gaussian model using JMFIT. Resolved irregular sources required the
use of the AIPS task TVSTAT. This task determines image statistics
such as mean and rms brightnesses found over irregular selected
regions. A future paper will present the fluxes and Gaussian fits to
the radio sources.

The flux densities were converted into absolute radio luminosities
using the inverse square law,

\begin{equation}
L_{\mathrm{1.4\,GHz}} = 4\pi D^2_{\mathrm{L}} S_{\mathrm{1.4}} \frac{ \ (1+z)^\alpha}{1+z}\,,
\end{equation}
\vspace{2 mm}

\noindent where both $(1+z)^\alpha$ and $\frac{1}{1+z}$ are the {\it
K-correction} terms where the former is the `color' correction and the
latter is the bandwidth correction. The spectral index $\alpha$ is
given by $S_{\nu} \propto \nu^{-\alpha}$, where $\nu$ is 1.4\,GHz and
$\alpha$ is assumed to be 0.8 \citep{con92}. $S_{1.4}$ is the flux
density in Janskys\footnote{1 Jy = $10^{-26}$\,W\,Hz$^{-1}$\,m$^{-2}$}
at 1.4 GHz and $D_{\mathrm{L}}$ is the luminosity distance. 

The sample between 0.02 $\le z \le$ 0.25 was complete to a lower
luminosity limit of $\sim$\,10$^{22.3}$\,W\,Hz$^{-1}$. This requires a
flux limit of 0.2 mJy for our most distant cluster at z\,=\,0.23
(A2111). The histogram of Figure~\ref{fig_1} suggests that we are
complete to this limit for the $0.2 \le z \le 0.23$ clusters.  All
other clusters ($z \lesssim 0.2$) in our sample have radio
observations significantly fainter than this luminosity limit. Note
that A2125 and 2645 are more distant than A2111; however, their radio
observations were deeper, complete to $\sim 10^{22.2}$\,W\,Hz$^{-1}$\,
\citep{dwa99}.

\placefigure{fig_1}

\section{Tables of Optical Identifications}

The cluster samples considered along with optical-radio
identifications are tabulated and presented in this section. Cluster
data obtained from the literature and newly measured cluster
parameters are also presented.

\subsection{Galaxy Clusters Data Tables}

Table~\ref{NVSS_CLS} contains the cluster data for all rich
clusters. All data in this table are taken from the literature as
noted, except the newly determined radio identifications, radio galaxy
fractions, and new redshift data.

Column (1) is the Abell Name \citep{abe58}. Columns (2) and (3) are
the cluster X-ray center \citep{vog99,vog98} in right-ascension and
declination. All are in J2000 epoch coordinates.  Column (4) contains
cluster redshift found in NED. However, clusters with $z \geq 0.1$ had
their cluster redshifts recalculated from the new redshift data if the
NED results were derived from too few existing measurements. Column
(5) and (6) contain the Bahcall richness counts
\citep{bah81,mor00a} and the Abell richness counts \citep{abe89,mor00a},
respectively. Available blue fraction measurements
\citep{but84b,mor00a} are in (7). Column (8) indicates the number of
galaxies with radio emission within 2.5\,Mpc of the cluster center
that have optical identification, while the numbers in parentheses are
the galaxies from this group that have measured redshifts.

Column (9) is the total number of radio galaxies with
$L_{1.4} \le 10^{22.75}$\,W Hz$^{-1}$ normalized by the total number
of galaxies searched, that were within 2.5 Mpc of the cluster center
and with $M_{\mathrm{R}}$ brighter than $-21$. This limit was chosen,
because local radio luminosity functions \citep[\eg][]{con89} suggest
that, statistically, these radio galaxies should be starbursts. Column
(10) is the same as (9) except that the radio luminosities of the
galaxies are $L_{1.4 GHz} \ge 10^{23}$\,W Hz$^{-1}$.

\subsection{Optical Identification Tables}

Table~\ref{Z_LOW_OPTICAL_IDS} contains the radio-selected galaxies
from each cluster.  For the local, rich Abell clusters (168 to 2256)
the absolute magnitudes of the galaxies were calibrated from
\cite{led95b} with the calibrator galaxies noted.

Each cluster listing is titled by the Abell name. Column (1) is the
galaxy number. Columns (2) and (3) are the optical right-ascension and
declination for the epoch J2000. Column (4) is the distance from the
X-ray defined cluster center in Mpc. Column (5) is the galaxy's
redshift from newly obtained spectra or archival data from NED. Column
(6) is the calibrated absolute magnitude derived from the DSS E-plate
image. For Abell clusters 655 to 2111, absolute magnitudes in column
(6) were measured from the $R$-band images. Column (7) is the galaxy's
$B-R$ color indices if available.  Column (8) lists radio luminosity
at 1.4\,GHz in Watts Hz$^{-1}$.

\section{Results}

%-----------------------------------------------------------------------------
\subsection{$R$-band Images}
%-----------------------------------------------------------------------------

The KPNO 0.9\,meter telescope $R$-band images of all the clusters with
$0.10 \le z \le 0.23$ are displayed in Figures~\ref{fig_2a} to
\ref{fig_2s}. The circles and the triangles on the images denote galaxies with
and without redshifts, respectively. Squares indicate questionable
redshifts. The contours represents the X-ray emission from the
intracluster gas.

\placefigure{fig_2a}
\placefigure{fig_2b}
\placefigure{fig_2c}
\placefigure{fig_2d}
\placefigure{fig_2e}
\placefigure{fig_2f}
\placefigure{fig_2g}
\placefigure{fig_2h}
\placefigure{fig_2i}
\placefigure{fig_2j}
\placefigure{fig_2k}
\placefigure{fig_2l}
\placefigure{fig_2m}
\placefigure{fig_2n}
\placefigure{fig_2o}
\placefigure{fig_2p}
\placefigure{fig_2q}
\placefigure{fig_2r}
\placefigure{fig_2s}

%-----------------------------------------------------------------------------
\subsection{Cluster Substructure: Isopleth Maps}
%-----------------------------------------------------------------------------

The dynamical structure of clusters is linked to the initial
conditions present at the time of cluster formation. Dynamical
equilibrium in a cluster environment would require isotropic galaxy
orbits, with a Gaussian distribution of radial velocities. Departure
from a Gaussian would imply the existence of substructure, infall,
cluster-cluster merging, or initial cluster build-up. Since we have
too few redshifts per cluster to do statistical analysis for cluster
substructure we will use a qualitative method. This method will give a
rough measure of a cluster's departure from a relaxed, virialized
state.

Isopleth maps (optical surface number density contour maps) give us a
qualitative view of the substructure within the cluster
environment. However, this method can suffer from foreground and
background contamination and, in the lower density areas, small number
statistics. Contour maps by \cite{gel82} were density maps smoothed on
defined scales (\eg 0.12\,Mpc, 0.24\,Mpc, etc.) to bring out
`significant' subclustering. Constant smoothing scales that are
critically adjusted to show detail were found to emphasize structure
that the observer chooses to reveal \citep{bee92} and therefore too
subjective.

The adaptive kernel method of \cite{sil86} has a variable smoothing
scale which adjusts to the local density of galaxies.  First, an
initial or pilot estimate is made of the global density which has been
pre-smoothed by a kernel whose bandwidth is set by the total number of
points. The next smoothing is done by a kernel of bandwidth set by the
local density obtained from the pilot estimate. Results by
\cite{bee91} suggest that this method is more or less insensitive to
the pilot estimate's kernel-function and to a certain extent, its
bandwidth.

The data from the R-band images was used as input to the Beers FORTRAN
code, employing the adaptively smoothing kernel, to make
Figures~\ref{fig_3a} through \ref{fig_3s}. The contours represent the
surface density of galaxies and the grey scale shows the RASS X-ray
data. The lowest contour level is at 90 galaxies per Mpc$^2$ with an
interval of 30 galaxies per Mpc$^2$, with all clusters having been
background subtracted \citep{mor00a}. In each redshift sample, all
clusters have their X-ray emission on the same scale. This allows them
to be compared directly to see the relative brightness
differences. The optical magnitude limit for the adaptive kernel maps
is $M_{\mathrm{R}} \leq -19$, the same as used in the
\fb~measurements and is fully discussed in \cite{mor00a}.

As is well-known, the optical isopleths follow the x-ray gas
distribution. In some cases, both x-ray and optical distributions
appear to be centrally condensed. A few clusters appear to have smooth
x-ray distributions while the optical distribution on a small scale
seems to show substructure. This lack of substructure in the x-ray
might just be due to the limited resolution (resolution $\sim$
15\arcsec) and sensitivity of the x-ray data. On the other hand, A1278
and A1882, both appear to contain two or more subclumps in the process
of merging, shown both in the x-ray and the optical data. What
relationship this has to the enhanced radio-selected galaxy population
will be covered in a later paper. Finally, one cluster, A2240, appears
to be a superposition of several foreground groups since it is not
seen in the x-ray or optical images. Therefore A2240 will not be
considered further in our study.

\placefigure{fig_3a}
\placefigure{fig_3b}
\placefigure{fig_3c}
\placefigure{fig_3d}
\placefigure{fig_3e}
\placefigure{fig_3f}
\placefigure{fig_3g}
\placefigure{fig_3h}
\placefigure{fig_3i}
\placefigure{fig_3j}
\placefigure{fig_3k}
\placefigure{fig_3l}
\placefigure{fig_3m}
\placefigure{fig_3n}
\placefigure{fig_3o}
\placefigure{fig_3p}
\placefigure{fig_3q}
\placefigure{fig_3r}
\placefigure{fig_3s}

\subsection{Radio Galaxy Fractions}

To separate the richness or galaxy count of a cluster from the number
of radio galaxies it contains, we devise a fractional radio galaxy
measure defined as

\begin{equation}
f_{\mathrm{RG}} = \frac{N(L_{\mathrm{min}} \le L_{1.4} \le L_{\mathrm{max}} )}{N_{\mathrm{2.5}}}\,,
\label{RADIO_GALAXY_FRACTION}
\end{equation}
\vspace{2 mm}

\noindent where $N$ is the number of radio-selected galaxies within
the prescribed range, and $N_{\mathrm{2.5}}$\footnote{The clusters
with $z < 0.1$ did not have CCD data to allow us to re-measure these
cluster's richnesses. Thus we used the Abell counts
($N_{\mathrm{2.0}}$) from \cite{abe89} catalog where we determined
$N_{\mathrm{2.5}}$ = 0.65$\times N_{\mathrm{2.0}}$. Details are in
\cite{mor00a}} is the number of galaxies brighter than
$M_{\mathrm{R}}$~=~$-$21 within 2.5\,Mpc of the cluster center. The
method used to calculate $N_{2.5}$ is the same as used for
$N_{\mathrm{2.0}}$ \citep{mor00a} but with a brighter optical limit
and a search radius that extends an extra 0.5 Mpc beyond the Abell
radius. The $N_{\mathrm{2.5}}$ values are reported in \cite{mor00}.
Also, $L_{\mathrm{min}}$ and $L_{\mathrm{max}}$ define the range of
radio luminosities for the three classes defined in
Table~\ref{RADIO_SELECTED_CLASS}. The different radio galaxy classes
are high luminosity (HLRG), low luminosity (LLRG), and `starburst'
(SBRG) which are discussed in Paper II.

The radio galaxy fractions are reported in
Table~\ref{NVSS_CLS}.

\section{Summary}

We present the largest sample of low luminosity ($L_{1.4 GHz} \ge
2\times 10^{22}$ W Hz$^{-1}$) radio galaxies within extremely rich
Abell clusters obtained to date. The radio observations allow us to
probe the `active' galaxy population of these clusters without the
difficulties experienced in optical observations, such as dust
obscuration and the optical K-correction. The radio-selected galaxy
sample represents the starburst (SFR $\ge$ 5\,M$_{\sun}$ yr$^{-1}$)
and AGN populations contained within each cluster. New redshifts were
used to verify cluster membership for most of the optical
identifications.

A total of 165 radio-selected galaxies have been detected within 2.5
Mpc of the centers of rich Abell clusters with $z \le 0.23$ and
$M_{\mathrm{R}} \le -21$. There are 89 radio-selected galaxies with
luminosities between $2\times 10^{22}$ W Hz$^{-1} \le L_{1.4} \le
10^{23}$ W Hz$^{-1}$.  There are 76 radio-selected galaxies with
$L_{1.4} \ge 10^{23}$ W Hz$^{-1}$. We have obtained 114 new spectra
for clusters with $z \le 0.23$. A large fraction (95$\%$) of the
optical identifications have measured redshifts.

In addition, new optical and X-ray data are presented for each of
these clusters.  We describe the selection of the cluster sample,
observations and the data reduction techniques for all three
wavelengths, including optical spectroscopy.

\acknowledgments

We greatly appreciate the anonymous referee's careful reading and
useful comments.  We would like to thank Jean Eilek, Carol Lonsdale,
and Schuyler Van Dyk. G.E.M would like to thank the night assistances
at the Steward Observatory's Bok telescope, the Palomar Observatory's
Hale telescope, and Palomar Observatory's
day-crew. G.E.M. acknowledges financial support from NRAO predoctoral
fellowship program during which most of this work was
completed. G.E.M. also gratefully acknowledges the support of JPL
(contract $\#$1147-001 and $\#$1166) and also the support of Vanguard
Research, Inc. The author has made extensive use of NASA's
Astrophysics Data Abstract Service (ADS) and NASA/IPAC Extragalactic
Database (NED).
  
%\bibliographystyle{apj}
%\bibliography{apj-jour,refers}

\clearpage
\begin{deluxetable}{c c c c c c c c c}
\tablecaption{Spectroscopy Observing Runs \label{tbl-1}}
\tabletypesize{\scriptsize}
\tablewidth{0pt}
\tablehead{
\colhead{Dates} & \colhead{Observatory} & \colhead{Telescope}& \colhead{Instrument}& \colhead{Grating} & \colhead{Blaze} &
\colhead{Dispersion} & \colhead{Slit Width} & \colhead{Wavelength Range} 
}
\startdata
1996 Feb  &Steward &2.3 meter  & MX      & 400   & 4000 & 2.75  & fiber 2$\arcsec$ & 3700 - 6950 \\
1996 Oct  &Steward &2.3 meter  & MX      & 400   & 4000 & 2.75  & fiber 2$\arcsec$ & 3700 - 6950 \\
1997 Mar  &Steward &2.3 meter  & MX      & 400   & 4000 & 2.75  & fiber 2$\arcsec$ & 3700 - 6950 \\
1996 May  & KPNO   & 4 meter   & RC      & 300   & 4000 & 2.75  & 2$\arcsec$       & 4300 - 8600 \\
1999 Dec  & KPNO   &2.1 meter  & Goldcam & 09    & 4000 & 2.40  & 2$\arcsec$       & 3700 - 8000 \\
1999 Jun  &Palomar & 5.1 meter & DBSP    & Det-B & 4000 & 1.76  & 1.5$\arcsec$     & 3700 - 5500 \\
          &        &           &         & Det-R & 7500 & 2.49  & 1.5$\arcsec$     & 5650 - 8200 \\
1999 Jul  &Palomar & 5.1 meter & DBSP    & Det-B & 4000 & 1.76  & 1.5$\arcsec$     & 3700 - 5500 \\
          &        &           &         & Det-R & 7500 & 2.49  & 1.5$\arcsec$     & 5650 - 8200 \\
1999 Jun  &Palomar & 5.1 meter & COSMIC  & 300   & 5500 & 2.70  & 1.5$\arcsec$     & 3500 - 9800 \\
2000 Feb  &Palomar & 5.1 meter & COSMIC  & 300   & 5500 & 2.70  & 1.5$\arcsec$     & 3500 - 9800 \\
2001 Feb  &Palomar & 5.1 meter & COSMIC  & 300   & 5500 & 2.70  & 1.5$\arcsec$     & 3500 - 9800 \\

\enddata

\tablecomments{Column legend --- (1) Calendar dates of scheduled 
observations; (2) Observatory; (3) Telescope used; (4) Spectrograph
used; (5) Choice of gratings. Note that for Palomar and the DBSP, the
incident light is split into a blue and a red spectrum by a
dichroic. The gratings and detector characteristics differ for the two
systems; (6) Blaze angle of grating, in Angstroms; (7) Resolution of
resulting spectra, in $\mbox{\AA}$ pixel$^{-1}$; (8) Slit width, in
arcseconds; (9) Wavelength range of resulting spectra, in Angstroms.
In the case of the KPNO spectra, the chip size allows a greater
wavelength range but instrument focus degrades at the extremes.}

\end{deluxetable}

\clearpage
\begin{deluxetable}{c c}
\tablecaption{X-ray Results: $0.02 \le z \le 0.25$\label{X_RAY_RESULTS} }
\tabletypesize{\scriptsize}
\tablewidth{0pt}
\tablehead{
\colhead{Name} & \colhead{$L_{\mathrm{X}}$ ($\times10^{43}$ erg s$^{-1}$}) }
\startdata
A0168 & \ 1.29 $\pm$ 0.12 \\
A0754 &10.00 $\pm$ 0.04 \\
A1367 & \ 2.29 $\pm$ 0.08 \\
A1656 & \ 9.33 $\pm$ 0.14 \\
A1795 &19.05 $\pm$ 0.55 \\
A1904 & \ 1.41 $\pm$ 0.19 \\
A2151 & \ 1.59 $\pm$ 0.10 \\
A2199 & \ 6.92 $\pm$ 0.12 \\
A2256 &10.00 $\pm$ 0.21 \\ \hline
A0655 & \ 9.33 $\pm$ 1.38 \\
A0665 &32.36 $\pm$ 3.09 \\
A0795 &16.98 $\pm$ 0.16 \\
A1278 & \ 3.72 $\pm$ 2.19 \\
A1413 &31.62 $\pm$ 0.22 \\
A1689 &64.57 $\pm$ 5.50 \\
A1882 & \ 3.02 $\pm$ 0.93 \\
A1940 & \ 5.75 $\pm$ 0.85 \\
A2218 &20.89 $\pm$ 1.10 \\
A2240 & \ 0.35 $\pm$ 0.24 \\ \hline
A0773 &33.11 $\pm$ 3.72 \\
A0781 & \ 8.32 $\pm$ 2.69 \\
A0983 & \ 6.46 $\pm$ 1.51 \\
A1331 & \ 5.13 $\pm$ 1.41 \\
A1682 &13.80 $\pm$ 2.46 \\
A1704 &18.62 $\pm$ 2.57 \\
A1895 &10.00 $\pm$ 3.31 \\
A1961 & \ 9.77 $\pm$ 2.04 \\
A2111 &17.78 $\pm$ 3.39 \\
A2125 & \ 4.27 $\pm$ 1.18 \\
A2645 &21.88 $\pm$ 4.17 \\  

\enddata

\tablecomments{Column legend --- (1) Name of the ACO cluster; 
(2) X-ray luminosity in units of $10^{43}$ erg s$^{-1}$ with 1$\sigma$
Poissonian errors.}
\end{deluxetable}

\clearpage

\begin{deluxetable}{c c c}
\tablecaption{Radio-selected Galaxy Classes at 1.4\,GHz \label{RADIO_SELECTED_CLASS}}
\tabletypesize{\scriptsize}
\tablewidth{0pt}
\tablehead{
\colhead{Class} & \colhead{$\log (L_{\mathrm{min}})$ W Hz$^{-1}$ } 
& \colhead{$\log (L_{\mathrm{max}})$ W Hz$^{-1}$}
}
\startdata
SBRG & 22.3 & 22.75 \\ 
LLRG & 22.3& 23 \\ 
HLRG & 23& 25 \\
\enddata

\tablecomments{Column legend --- (1) Name of the Radio-selected Galaxy Class; 
(2) The logarithmic lower limit of the radio luminosity in W Hz$^{-1}$
for that class; (3) The logarithmic upper limit of the radio
luminosity in W Hz$^{-1}$ for that class.}
\end{deluxetable}

\clearpage

\begin{deluxetable} {llllrrrlcc}
\tablecaption{Cluster Sample\label{NVSS_CLS}}
\tabletypesize{\scriptsize}
\tablehead{\colhead{Name}&\colhead{R.A.(2000.0)}&\colhead{Dec}&\colhead{Redshift}&\colhead{$N_{0.5}$}&
\colhead{$N_{\mathrm{2.0}}$}&\colhead{$f_{\mathrm{B}}$}&\colhead{Radio}
&\colhead{$f_{\mathrm{SBRG}}$}&\colhead{$f_{\mathrm{HLRG}}$}\\
\colhead{(1)} &\colhead{(2)} &\colhead{(3)} &\colhead{(4)} &\colhead{(5)} &\colhead{(6)} &
\colhead{(7)} &\colhead{(8)} &\colhead{(9)} &\colhead{(10)} }
\startdata
A0168  &01 14 58.3  &$+$00 21 57 &0.045 &\nodata &89 &\nodata &2&    0.02(0.02)& 0.02(0.02) \\
A0754  &09 09 05.5  &$-$09 40 04 &0.053 &29 &92 &\nodata &4 &    0.02(0.02)& 0.05(0.03)\\
A1367  &11 44 43.2  &$+$19 43 42 &0.022 &18 &117 &0.19 &3 &    0.03(0.02)& 0.01(0.01)\\
A1656  &12 59 45.3  &$+$27 55 37 &0.023 &28 &106 &0.03 &9 &    0.07(0.03)& 0.03(0.02)\\
A1795  &13 48 52.7  &$+$26 35 44 &0.062 &27 &115 &\nodata &5(4) &    0.03(0.02)& 0.04(0.02)\\
A1904  &14 22 14.7  &$+$48 30 44 &0.071 &\nodata &83 &\nodata &3 &    0.02(0.02)& 0.04(0.03)\\
A2151  &16 04 39.4  &$+$17 43 25 &0.037 &17 &87 &0.14 &6 &    0.02(0.02)& 0.09(0.04)\\
A2199  &16 28 37.9  &$+$39 32 55 &0.030 &18 &88 &0.04 &7(6) &    0.09(0.04)& 0.04(0.02)\\
A2256  &17 03 48.1  &$+$78 38 23 &0.060 &31 &88 &0.03 &8(7) &    0.04(0.02)& 0.07(0.04)\\
A3094$^*$$^\diamond$&03 11 27.0&$-$26 55 44 &0.068 &\nodata &80 &\nodata &\nodata & \nodata& \nodata\\ \hline
A0655  &08 25 26.8  &$+$47 07 29 &0.128  & 52&191&\nodata&4(3)& 0.02(0.01)& 0.02(0.01)    \\
A0665  &08 30 57.6  &$+$65 53 30 &0.182  & 56& 244&0.11     & 6(5)& 0.02(0.01)& 0.01(0.01)\\
A0795  &09 24 06.4  &$+$14 09 57 &0.138  & 57& 185&\nodata  & 7& 0.01(0.01)& 0.04(0.02)\\
A1278  &11 30 09.1  &$+$20 30 54 &0.129  & 42& 137&0.19 & 6& 0.05(0.03)& 0.01(0.01)\\
A1413  &11 55 18.0  &$+$23 24 17 &0.140  & 55& 177&0.09     & 10(9)& 0.02(0.01)& 0.02(0.01)\\
A1437$^\diamond$  &12 00 24.3  &$+$03 20 00 &0.132  & 16& 69 &\nodata  & \nodata&\nodata&\nodata \\		   
A1674$^\diamond$  &13 03 31.9  &$+$67 29 55 &0.106 &33&    100&\nodata  & \nodata&\nodata&\nodata \\   
A1689  &13 11 30.4  &$-$01 20 19 &0.181  & 70& 158&0.09$^{\dagger}$  &12(10)& 0.05(0.03) &0.09(0.04)\\
A1882  &14 14 39.9  &$-$00 19 57 &0.142  & 35& 247&0.28     & 13(8)& 0.03(0.01)& 0.03(0.01)\\
A1918$^\diamond$  &14 25 21.7  &$+$63 11 28 &0.139  & 12& 30 &\nodata  & \nodata&\nodata&\nodata \\	   
A1940  &14 35 31.4  &$+$55 09 31 &0.140  & 31& 119&\nodata  & 4(3)&0.00(0.00)& 0.06(0.03)\\
A1990$^\diamond$  &14 53 45.2  &$+$28 04 49 &0.127 & 8&  28  &\nodata  & \nodata &\nodata&\nodata \\
A2218  &16 35 52.3  &$+$66 12 40 &0.171  & 63& 209&0.11$^{\dagger}$  &1& 0.00(0.00)& 0.01(0.01)\\
A2240  &16 53 26.6  &$+$66 43 16 &0.139  & 25& 97 &\nodata  & 2 & 0.02(0.02)& 0.02(0.02)\\ \hline
A0773  &09 17 51.8  &$+$51 43 29 &0.220  & 109& 304&0.14 & 8(5)   & 0.02(0.01)& 0.01(0.01)\\
A0781  &09 20 28.1  &$+$30 31 10 &0.215  & 29 & 151&0.25 & 3(0)   & 0.01(0.01) &0.02(0.02)\\
A0983  &10 23 20.1  &$+$59 48 15 &0.207  & 39 & 201&0.07 & 7(5)  &  0.01(0.01)& 0.02(0.01)\\
A1331  &11 38 55.0  &$+$63 34 45 &0.207  &  36& 100&0.08 & 6&  0.03(0.02)& 0.04(0.02)\\
A1430$^\diamond$  &11 59 12.1  &$+$49 47 26 &0.210  &\nodata    &\nodata  &\nodata &\nodata&\nodata&\nodata\\  
A1682  &13 06 48.2  &$+$46 33 22 &0.226  & 55 & 215&0.07 & 10&   0.03(0.01)& 0.03(0.01)\\
A1704  &13 14 26.1  &$+$64 34 40 &0.216  & 22 & 108&0.08 & 5&  0.03(0.02) &0.02(0.02)\\
A1895  &14 13 55.5  &$+$71 18 46 &0.216  & 35 & 204&0.15 & 5 &  0.01(0.01) &0.02(0.01)\\
A1961  &14 44 31.6  &$+$31 13 02 &0.228  & 46 & 286&0.20 & 5&  0.01(0.01)& 0.02(0.01)\\
A2111  &15 39 44.1  &$+$34 24 58 &0.229  & 61 & 271&0.18 & 4(3)&  0.02(0.01)& 0.01(0.01)\\
A2125  &15 41 13.9  &$+$66 14 54 &0.247  & 41 & 256&0.27 &26   &  0.09(0.02)& 0.03(0.01)\\
A2645  &23 41 16.6  &$-$09 01 25 &0.250  & 65 & 223&0.07 & 7(4)    &  0.01(0.01)& 0.03(0.01)\\  

\tablecomments{Column legend --- (1) is the Abell Name; 
(2) and (3) are the cluster X-ray center \citep{vog99,vog98} in
right-ascension and declination. Both are in J2000 epoch coordinates;
(4) cluster redshift found in NED. However, clusters with $z
\geq 0.1$ had their cluster redshifts recalculated from the new
redshift data if the NED results were derived from too few existing
measurements; (5) and (6) contain the Bahcall richness counts
\citep{bah81} and the Abell richness counts \citep{abe89},
respectively; (7) contains available cluster blue fractions; (8)
indicates the number of galaxies with radio emission within 2.5\,Mpc
of the cluster center that have optical identification, while the
numbers in parentheses are the galaxies from this group that have
measured redshifts;  (9) is the total number of radio galaxies with
$L_{1.4} \le 10^{22.75}$\,W Hz$^{-1}$ normalized by the total number
of galaxies searched within 2.5 Mpc of the cluster center and with
$M_{\mathrm{R}}$ brighter than $-21$; (10) is the same as (9) except
that the radio luminosities of the galaxies are $L_{1.4 GHz} \ge
10^{23}$\,W Hz$^{-1}$.}

\tablenotetext{*}{\cite{abe58} optical center used.} 
\tablenotetext{\diamond}{Superposition of poor clusters}
\tablenotetext{\dagger}{Blue fractions from \cite{but84b}}
\enddata
\end{deluxetable}

\clearpage

\begin{deluxetable} {lrrrllrr}
\tablecaption{Optical Identifications\label{Z_LOW_OPTICAL_IDS}}
\tabletypesize{\scriptsize}
\tablehead{\colhead{Name}&\colhead{R.A.(2000.0)}&\colhead{Dec}&\colhead{Mpc}
&\colhead{Redshift}&\colhead{M$_R$ }& \colhead{B-R}&\colhead{log(L$_{1.4}$)}\\
\colhead{(1)} &\colhead{(2)} &\colhead{(3)} &\colhead{(4)} &\colhead{(5)} &\colhead{(6)} &
\colhead{(7)} &\colhead{(8)}}
\startdata
& & & & & & & \\
Local Redshift Sample& & & & & & & \\ 
& & & & & & & \\
A0168& & & & & & & \\
011559$+$005258&01 15 59.61  &$+$00 52 58.0&1.69 &0.0443  &$-$22.2 &\nodata &22.51   \\
011515$+$001248&01 15 15.76  &$+$00 12 48.6&0.49 &0.0452$^{\ddag}$  &$-$22.0$^{\diamond}$  &\nodata&23.23  \\

A0754& & & & & & & \\
090926$-$092247&09 09 26.30  &$-$09 22 47.9&1.02 &0.0590$^{\ddag}$  &$-$22.3  &\nodata&23.71 \\
091017$-$093707&09 10 17.30  &$-$09 37 07.6&1.02 &0.0549$^{\ddag}$  &$-$22.4  &\nodata&23.84 \\
090855$-$094045&09 08 55.30  &$-$09 40 45.7&0.15 &0.0488$^{\ddag}$  &$-$22.1$^{\diamond}$  &\nodata&23.51 \\
090919$-$094154&09 09 19.10  &$-$09 41 54.8&0.22 &0.0542$^{\ddag}$  &$-$22.3  &\nodata&22.69 \\

A1367& & & & & & & \\
114348$+$195810&11 43 48.70  & $+$19 58 10.9 &0.47 &0.0224$^{\ddag}$  &$-$21.4$^{\diamond}$ &\nodata&22.70 \\ 
114505$+$193621&11 45 05.10  & $+$19 36 21.0 &0.22 &0.0217$^{\ddag}$  &$-$22.2   &\nodata&24.69\\
114559$+$202618&11 45 59.90  & $+$20 26 18.8 &1.12 &0.0244$^{\ddag}$  &$-$21.7 &\nodata & 22.35     \\

A1656& & & & & & & \\
125253$+$282216&12 52 53.60  &$+$28 22 16.2&2.46 &0.0236$^{\ddag}$ &$-$21.8 &\nodata&22.32 \\ 
125418$+$270413&12 54 18.80  &$+$27 04 13.4&2.31 &0.0279$^{\ddag}$ &$-$22.7 &\nodata&22.83 \\
130125$+$291848&13 01 25.16  &$+$29 18 48.3&2.24 &0.0235$^{\ddag}$ &$-$22.1 &\nodata&22.63 \\ 
125643$+$271043&12 56 43.50  &$+$27 10 43.1&1.57 &0.0255$^{\ddag}$ &$-$22.2 &\nodata&22.73 \\
125724$+$272951&12 57 24.30  &$+$27 29 51.4&1.05 &0.0246$^{\ddag}$ &$-$22.7 &\nodata&22.90 \\
125805$+$281433&12 58 05.50  &$+$28 14 33.9&0.76 &0.0235$^{\ddag}$ &$-$21.8 &\nodata&22.39 \\
125935$+$275730&12 59 35.40  &$+$27 57 30.2&0.08 &0.0241$^{\ddag}$ &$-$23.9$^{\diamond}$ &\nodata&23.34 \\
130056$+$274725&13 00 56.00  &$+$27 47 25.0&0.46 &0.0266$^{\ddag}$ &$-$22.5 &\nodata&22.33 \\
125923$+$275443&12 59 23.25  &$+$27 54 43.3&0.13 &0.0229$^{\ddag}$ &$-$21.9$^{\diamond}$ &\nodata&23.46  \\

A1795& & & & & & & \\
134849$+$264658&13 48 49.10  &$+$26 46 58.4&0.74 &0.0862  &$-$21.4 &\nodata&22.87 \\
134852$+$263532&13 48 52.50  &$+$26 35 32.8&0.01 &0.0633$^{\ddag}$  &$-$22.8 &\nodata&24.85      \\ 
134816$+$262913&13 48 16.30  &$+$26 29 13.6&0.68 &0.0647$^{\ddag}$  &$-$21.3 &\nodata&22.28      \\
134905$+$262814&13 49 05.00  &$+$26 28 14.9&0.52 &0.0638$^{\ddag}$  &$-$21.6 &\nodata&22.49     \\
%134816$+$260302&13 48 16.90  &$+$26 03 02.6&2.20 &0.1764  &$-$21.2 &\nodata&23.66   \\
134858$+$263336&13 48 58.98  &$+$26 33 36.4&0.17 &0.0608$^{\ddag}$  &$-$21.4 &\nodata&23.43 \\

A1904& & & & & & & \\
142125$+$482933&14 21 25.50  &$+$48 29 33.5&0.61 &0.0709 &$-$22.4 &\nodata&23.39   \\
142256$+$481623&14 22 56.30  &$+$48 16 23.0&1.18 &0.0757$^{\ddag}$ &$-$22.0 &\nodata&22.57  \\
142336$+$482609&14 23 36.60  &$+$48 26 09.9&1.06 &0.0733 &$-$22.6 &\nodata&23.06   \\

A2151& & & & & & & \\
160125$+$180126&16 01 25.90  &$+$18 01 26.7 &2.02 &0.0369$^{\ddag}$&$-$21.5  &\nodata& 22.35      \\
160616$+$181459&16 06 16.00  &$+$18 14 59.0 &1.59 &0.0368$^{\ddag}$ &$-$22.6  &\nodata& 23.76    \\
160508$+$174528&16 05 08.10  &$+$17 45 28.4 &0.29 &0.0333$^{\ddag}$ &$-$22.2  &\nodata& 23.04   \\
160509$+$174344&16 05 09.10  &$+$17 43 44.1 &0.29 &0.0315 &$-$22.5$^{\diamond}$  &\nodata& 24.33  \\
160332$+$171154&16 03 32.00  &$+$17 11 54.2 &1.44 &0.0339$^{\ddag}$ &$-$22.5 &\nodata & 24.10  \\
160426$+$174430&16 04 26.38  &$+$17 44 30.6 &0.13 &0.0411$^{\ddag}$ &$-$22.2$^{\diamond}$  &\nodata&23.51      \\

A2199& & & & & & & \\
162957$+$403744&16 29 57.80  &$+$40 37 44.8 &2.24 &0.0294$^{\ddag}$ &$-$22.7&\nodata &22.34  \\
162549$+$402041&16 25 49.10  &$+$40 20 41.3 &1.94 &0.0288$^{\ddag}$ &$-$21.6 &\nodata&22.43 \\
162550$+$402920&16 25 50.00  &$+$40 29 20.1 &2.19 &0.0291$^{\ddag}$ &$-$22.5 &\nodata&22.75 \\
162838$+$393259&16 28 38.00  &$+$39 32 59.2 &0.00 &0.0310$^{\ddag}$ &$-$23.1$^{\diamond}$ &\nodata&24.82 \\
163032$+$392302&16 30 32.70  &$+$39 23 02.8 &0.82 &0.0306$^{\ddag}$ &$-$21.3 &\nodata&22.51 \\
163349$+$391547&16 33 49.60  &$+$39 15 47.5 &2.11 &0.0317$^{\ddag}$ &$-$22.3 &\nodata&22.30  \\

A2256& & & & & & & \\
%170137$+$790359&17 01 37.10  &$+$79 03 59.2  &1.69 &\nodata &$-$22.3 &\nodata&23.58 \\
170052$+$784121&17 00 52.30  &$+$78 41 21.1  &0.58 &0.0578$^{\ddag}$ &$-$21.5$^{\diamond}$ &\nodata&22.48 \\
170448$+$783830&17 04 48.40  &$+$78 38 30.7  &0.19 &0.0643$^{\ddag}$ &$-$22.3$^{\diamond}$ &\nodata&22.76 \\
165818$+$782933&16 58 18.30  &$+$78 29 33.3  &1.19 &0.0595$^{\ddag}$ &$-$22.3 &\nodata&22.36 \\
170329$+$783754&17 03 29.31  &$+$78 37 54.8  &0.07&0.0587$^{\ddag}$ &$-$22.5$^{\diamond}$  &\nodata&26.99 \\
170302$+$783555&17 03 02.80  &$+$78 35 55.0  &0.21&0.0553$^{\ddag}$ &$-$22.0$^{\diamond}$  &\nodata&26.57 \\
170330$+$783954&17 03 30.01  &$+$78 39 54.0  &0.11&0.0586$^{\ddag}$ &$-$22.0$^{\diamond}$  &\nodata&26.40\\
%170342$+$784530&17 03 42.95  &$+$78 45 30.4  &0.46&0.0577 &$-$21.5$^{\diamond}$  &\nodata&22.83    \\ & & & & & & & \\
& & & & & & & \\
Low Redshift Sample& & & & & & & \\ 
& & & & & & & \\
A0655& & & & & & & \\
%082544$+$470157& 08 25 44.45 &   $+$47 01 57.8  &0.78 &0.36042   &$-$21.5&\nodata&22.42\\
082603$+$471910& 08 26 03.81 &   $+$47 19 10.8  &1.64 &0.1283    &$-$22.1$^{\dagger}$&\nodata&23.99\\
082503$+$470925& 08 25 03.33 &   $+$47 09 25.0  &0.55 &0.1259   &$-$21.5&\nodata&23.10\\
082549$+$471027& 08 25 49.55 &   $+$47 10 27.6  &0.60 &0.1318   &$-$22.3$^{\dagger}$&\nodata&22.55\\

A0665& & & & & & & \\	                  		
083056$+$654912 & 08 30 56.00 &   $+$65 49 12.6  &0.70 &0.1848$^{\star}$    &$-$21.7&2.07   &22.33 \\
083048$+$655436 & 08 30 48.33 &   $+$65 54 36.9  &0.24 &0.1846    &$-$21.2&1.44   &22.57  \\
083034$+$655151 & 08 30 34.29 &   $+$65 51 51.5  &0.47 &0.1828$^{\star}$    &$-$22.0&2.19   &23.07  \\
%082957$+$660207 & 08 29 57.57 &   $+$66 02 07.0  &1.73 &0.2177&$-$21.1&1.40   &22.44  \\
%083041$+$655810 & 08 30 41.70 &   $+$65 58 10.0  &0.81 &0.2010    &$-$22.9&2.25   &23.33   \\
083055$+$654947 & 08 30 55.94 &   $+$65 49 47.2  &0.61 &0.1868$^{\star}$    &$-$21.7&2.15   &22.82  \\
	   
A0795& & & & & & & \\           		
092431$+$140945 & 09 24 31.01 &   $+$14 09 45.8  &0.79 &0.1358    &$-$22.0&\nodata&23.08  \\
092415$+$140741 & 09 24 15.40 &   $+$14 07 41.4  &0.41 &0.1402    &$-$22.5&\nodata&22.87    \\
092412$+$140712 & 09 24 12.11 &   $+$14 07 12.1  &0.41 &0.1353    &$-$21.8&\nodata&22.37    \\
092405$+$141020 & 09 24 05.31 &   $+$14 10 20.7  &0.06 &0.1354$^{\star}$    &$-$22.8&\nodata&24.62   \\
092435$+$141416 & 09 24 35.59 &   $+$14 14 16.4  &1.09 &0.1397    &$-$22.1&\nodata&23.68  \\
092428$+$141408 & 09 24 28.90 &   $+$14 14 08.6  &0.91 &0.1386    &$-$22.5&\nodata&23.27   \\
092352$+$141657 & 09 23 52.76 &   $+$14 16 57.8  &1.02 &0.1418    &$-$22.2&\nodata&24.55   \\

A1278& & & & & & & \\		         		
113034$+$203617 & 11 30 34.11 &   $+$20 36 17.6  &0.99 &0.1364    &$-$22.1&1.89   &22.61  \\
113032$+$202129 & 11 30 32.58 &   $+$20 21 29.0  &1.36 &0.1304    &$-$21.9&1.70   &23.14  \\
113003$+$202122 & 11 30 03.67 &   $+$20 21 22.3  &1.20 &0.1296    &$-$22.2&1.97   &22.60    \\
112954$+$202510 & 11 29 54.55 &   $+$20 25 10.1  &0.83 &0.1375    &$-$22.7&1.94   &22.92   \\
112943$+$202620 & 11 29 43.29 &   $+$20 26 20.4  &0.94 &0.1306    &$-$22.3&1.67   &22.31  \\
113020$+$203207 & 11 30 20.45 &   $+$20 32 07.0  &0.36 &0.1360   &$-$22.5&1.35   &22.50   \\
	
A1413& & & & & & & \\	         		
115526$+$233731 & 11 55 26.91 &   $+$23 37 31.6  &1.79 &0.1439    &$-$22.2$^{\dagger}$&\nodata&22.96   \\
115515$+$232754 & 11 55 15.12 &   $+$23 27 54.0  &0.49 &0.1533$^{\star}$    &$-$22.1&1.40   &22.51  \\
115508$+$232622 & 11 55 08.98 &   $+$23 26 22.9  &0.39 &0.1440$^{\star}$    &$-$22.0&2.01   &23.95  \\
115458$+$232520 & 11 54 58.45 &   $+$23 25 20.3  &0.61 &0.1318$^{\star}$    &$-$22.7&1.88   &22.85   \\
115520$+$231745 & 11 55 20.98 &   $+$23 17 45.6  &0.87 &0.1438    &$-$21.9&1.96   &22.83   \\
115517$+$232417 & 11 55 17.99 &   $+$23 24 17.9  &0.00 &0.1378$^{\star}?$    &$-$23.2&2.08   &23.21   \\
115526$+$232049 & 11 55 26.24 &   $+$23 20 49.1  &0.53 &0.1368   &$-$21.8&1.12   &22.42    \\
115517$+$232214 & 11 55 17.07 &   $+$23 22 14.2  &0.27 &0.1435   &$-$21.3&1.96   &22.84    \\
115516$+$230724 & 11 55 16.38 &   $+$23 07 24.0  &2.25 &0.1409   &$-$21.4$^{\dagger}$&\nodata&22.76  \\
%115403$+$232022 & 11 54 03.98 &   $+$23 20 22.0  &2.32 &0.05001ha   &$-$22.9$^{\dagger}$&\nodata&22.91   \\
	
A1689& & & & & & & \\         		
131102$-$013147 & 13 11 02.85 &   $-$01 31 47.7  &2.19 &0.1870    &$-$22.1&\nodata&24.08  \\
%131055$-$012725 & 13 10 55.58 &   $-$01 27 25.2  &1.83 &4.6La   &$-$21.0&\nodata&23.43  \\
131208$-$012149 & 13 12 08.71 &   $-$01 21 49.4  &1.58 &0.2954$^{\triangleleft}$&$-$21.0&\nodata&23.15  \\
131145$-$012337 & 13 11 45.40 &   $-$01 23 37.2  &0.82 &0.1868    &$-$21.9&\nodata&22.61  \\
131143$-$011920 & 13 11 43.40 &   $-$01 19 20.4  &0.55 &0.1828$^{\star}$    &$-$22.0&\nodata&24.11  \\
131137$-$011809 & 13 11 37.97 &   $-$01 18 09.0  &0.47 &0.1803$^{\star}$    &$-$21.8&\nodata&22.41  \\
131129$-$012117 & 13 11 29.01 &   $-$01 21 17.4  &0.17 &0.1943$^{\star}$    &$-$21.3&\nodata&23.28  \\
131131$-$011933 & 13 11 31.41 &   $-$01 19 33.6  &0.13 &0.1877$^{\star}$    &$-$22.2&\nodata&24.11  \\
131132$-$011959 & 13 11 32.68 &   $-$01 19 59.2  &0.11 &0.2018$^{\star}$    &$-$22.5&\nodata&23.54  \\
131130$-$012029 & 13 11 30.07 &   $-$01 20 29.4  &0.03 &0.1778$^{\star}$    &$-$22.7&\nodata&23.86  \\
131130$-$012044 & 13 11 30.58 &   $-$01 20 44.5  &0.07 &0.1918$^{\star}$    &$-$22.0&\nodata&22.40  \\
131125$-$012037 & 13 11 25.36 &   $-$01 20 37.7  &0.21 &0.1922$^{\star}$    &$-$21.8&\nodata&22.27  \\

A1882& & & & & & & \\ 	         		
141419$-$001953 & 14 14 19.89 &   $-$00 19 53.0  &0.67 &0.1440$^{\star}$   &$-$21.6&1.59   &22.41   \\
141457$-$002059 & 14 14 57.73 &   $-$00 20 59.5  &0.62 &0.1371$^{\star}$   &$-$22.9&2.13   &23.30   \\
141508$-$002936 & 14 15 08.43 &   $-$00 29 36.0  &1.62 &0.1395$^{\star}$    &$-$23.2&2.06   &22.60    \\
141506$-$002910 & 14 15 06.47 &   $-$00 29 10.7  &1.53 &0.1366    &$-$22.3&2.09   &23.37   \\
141504$-$002926 & 14 15 04.17 &   $-$00 29 26.0  &1.52 &0.1377    &$-$22.2&2.07   &23.69   \\
141441$-$002753 & 14 14 41.88 &   $-$00 27 53.8  &1.07 &0.1434$^{\star}$    &$-$22.2&1.95   &22.56    \\
141429$-$002230 & 14 14 29.13 &   $-$00 22 30.2  &0.50 &0.1344$^{\star}$   &$-$22.9&2.00   &23.64   \\
141416$-$002105 & 14 14 16.25 &   $-$00 21 05.5  &0.81 &0.1392$^{\star}$    &$-$22.0&1.98   &22.91    \\
141432$-$002450 & 14 14 32.26 &   $-$00 24 50.4  &0.71 &0.1378$^{\star}$    &$-$22.2&1.96   &22.32   \\
141431$-$003043 & 14 14 31.07 &   $-$00 30 43.9  &1.48 &0.1384$^{\star}$    &$-$21.9&1.80   &23.02   \\
%141441$-$001704 & 14 14 41.83 &   $-$00 17 04.4  &0.39 &0.0376   &$-$22.1&0.96   &22.61    \\
%141427$-$001143 & 14 14 27.00 &   $-$00 11 43.2  &1.19 &0.0494   &$-$23.2&1.03   &23.04   \\
%141424$-$001128 & 14 14 24.29 &   $-$00 11 28.2  &1.26 &0.0543   &$-$23.0&1.23   &22.80   \\

A1940& & & & & & & \\ 		         		
143504$+$551148 & 14 35 04.12 &   $+$55 11 48.3  &0.60 &0.1398$^{\star}$   &$-$22.7&\nodata&23.25  \\
143528$+$550752 & 14 35 28.42 &   $+$55 07 52.3  &0.23 &0.1394$^{\star}$    &$-$22.8&\nodata&25.32    \\
143502$+$551250 & 14 35 02.08 &   $+$55 12 50.5  &0.71 &0.1391   &$-$22.8&\nodata&23.01    \\
143501$+$545457 & 14 35 01.88 &   $+$54 54 57.9  &2.02 &\nodata    &$-$21.0$^{\dagger}$&\nodata&23.29     \\

A2218& & & & & & & \\ 		         		
163547$+$661444 & 16 35 47.28 &   $+$66 14 44.6  &0.33 &0.1689    &$-$22.3&\nodata&23.65   \\

A2240& & & & & & & \\ 		         		
165614$+$663934 & 16 56 14.00 &   $+$66 39 34.6  &2.25 &0.1445    &$-$22.5$^{\dagger}$&\nodata&22.70   \\
165445$+$665925 & 16 54 45.30 &   $+$66 59 25.4  &2.37 &0.1358   &$-$22.9$^{\dagger}$&\nodata&22.87  \\ 
& & & & & & & \\
Intermediate Redshift Sample& & & & & & & \\ 
& & & & & & & \\
A0773& & & & & & & \\ 
%091837$+$515040 &09 18 37.91 &$+$51 50 40.4 &1.91 &0.4347    &$-$21.5&2.44   &24.04\\
091811$+$514447 &09 18 11.40 &$+$51 44 47.5 &0.62 &0.2181$^{\triangleleft}$&$-$21.2&1.79   &22.40   \\
091805$+$514324 &09 18 05.56 &$+$51 43 24.1 &0.40 &0.2101    &$-$21.9&1.57   &22.59   \\
091751$+$514951 &09 17 51.08 &$+$51 49 51.3 &1.20 &0.2428$^{\triangleleft}$&$-$21.4&1.75   &22.63   \\
091801$+$514415 &09 18 01.38 &$+$51 44 15.2 &0.32 &0.2283    &$-$21.0&2.28   &23.40   \\
091755$+$514259 &09 17 55.35 &$+$51 42 59.0 &0.14 &0.2221    &$-$22.7&2.21   &22.34   \\
091655$+$513917 &09 16 55.83 &$+$51 39 17.1 &1.82 &0.2242    &$-$22.7&2.19   &23.24   \\
091745$+$514308 &09 17 45.17 &$+$51 43 08.0 &0.21 &0.2215    &$-$21.8&2.22   &22.60   \\

A0781& & & & & & & \\ 
092122$+$303403 &09 21 22.96 &$+$30 34 03.4 &2.26 &0.1401$^{\triangleleft}$  &$-$21.2$^{\dagger}$ &\nodata&22.42     \\
%092122$+$302910 &09 21 22.10 &$+$30 29 10.1 &2.19 &0.4230&$-$21.0$^{\dagger}$ &\nodata&23.29  \\ 
%092108$+$302926 &09 21 08.33 &$+$30 29 26.4 &1.64 &0.2945 &$-$21.3&1.90    &23.51 \\ 

A0983& & & & & & & \\ 
102424$+$595143 &10 24 24.07 &$+$59 51 43.1 &1.58 &0.2048    &$-$22.5&2.19   &24.40    \\
102335$+$595142 &10 23 35.52 &$+$59 51 42.1 &0.72 &0.2088    &$-$21.5&1.84   &22.53 \\
102313$+$600041 &10 23 13.06 &$+$60 00 41.9 &2.26 &0.4145$^{\triangleleft}$&$-$21.7$^{\dagger}$&\nodata&22.73 \\
102307$+$594809 &10 23 07.98 &$+$59 48 09.6 &0.28 &0.1988    &$-$22.3&2.18   &24.46      \\
%102346$+$593915 &10 23 46.84 &$+$59 39 15.1 &1.74 &0.1642    &$-$21.2&1.30   &22.40  \\
102308$+$594834 &10 23 08.40 &$+$59 48 34.9 &0.27 &0.1991    &$-$22.4&2.13   &23.00    \\
102313$+$594804 &10 23 13.58 &$+$59 48 04.3 &0.15 &0.1986    &$-$22.7&2.24   &22.86      \\ 

A1331& & & & & & & \\ 
114029$+$633628 &11 40 29.67 &$+$63 36 28.2 &1.93 &0.2094    &$-$22.2 &1.51  &22.71 \\
113848$+$633522 &11 38 48.10 &$+$63 35 22.2 &0.18 &0.2072$^{\star}$    &$-$22.6 &2.24  &24.19 \\
113858$+$633527 &11 38 58.28 &$+$63 35 27.1 &0.14 &0.2092$^{\star}$    &$-$22.1 &2.12  &24.30 \\
113901$+$633720 &11 39 01.77 &$+$63 37 20.4 &0.49 &0.2032    &$-$21.7 &1.29  &22.45  \\
113850$+$634034 &11 38 50.29 &$+$63 40 34.8 &1.06 &0.2120    &$-$21.8 &2.13  &22.86  \\
113808$+$633524 &11 38 08.06 &$+$63 35 24.3 &0.95 &0.1954    &$-$21.6 &1.60  &24.19 \\

A1682& & & & & & & \\ 
130645$+$463330 &13 06 45.68 &$+$46 33 30.8 &0.09 &0.2190$^{\star}$    &$-$23.3 &2.24  &25.30 \\
130649$+$463333 &13 06 49.94 &$+$46 33 33.5 &0.07 &0.2304$^{\star}$    &$-$22.0 &2.26  &23.31 \\
130555$+$464130 &13 05 55.54 &$+$46 41 30.2 &2.34 &0.2475    &$-$22.3 &2.17  &24.47 \\
130636$+$463736 &13 06 36.85 &$+$46 37 36.1 &0.90 &0.2300    &$-$22.2 &1.64  &22.92 \\
130620$+$463821 &13 06 20.87 &$+$46 38 21.5 &1.32 &0.2562    &$-$21.4 &2.06  &22.39 \\
130713$+$462904 &13 07 13.19 &$+$46 29 04.4 &1.17 &0.2249    &$-$22.3 &2.06  &22.39 \\
130738$+$463152 &13 07 38.52 &$+$46 31 52.4 &1.69 &0.2237    &$-$22.2 &2.26  &22.85 \\
130700$+$463151 &13 07 00.20 &$+$46 31 51.9 &0.49 &0.2248    &$-$22.7 &2.33  &22.58 \\
130643$+$462437 &13 06 43.91 &$+$46 24 37.1 &1.69 &0.2454    &$-$21.6 &1.56  &22.56 \\
130643$+$462812 &13 06 43.40 &$+$46 28 12.2 &1.01 &0.2309    &$-$23.1 &2.29  &24.15 \\

A1704& & & & & & & \\ 
131428$+$643351&13 14 28.35&$+$64 33 51.1 &0.16 &0.2206    &$-$21.4 &1.25 &22.44 \\
131410$+$643324&13 14 10.42&$+$64 33 24.2 &0.39 &0.2090    &$-$21.8 &1.34 &22.98 \\
131424$+$643431&13 14 24.55&$+$64 34 31.6 &0.04 &0.2196$^{\star}$    &$-$22.4 &1.81 &22.30 \\
131357$+$642555&13 13 57.66&$+$64 25 55.6 &1.73 &0.2212    &$-$22.5 &2.16 &23.30 \\
131608$+$643326&13 16 08.09&$+$64 33 26.2 &2.05 &0.2046    &$-$21.4$^\dagger$ &\nodata &22.87   \\

A1895& & & & & & & \\ 
141514$+$711319&14 15 14.49 &$+$71 13 19.2 &1.56 &0.2129    &$-$22.8 &2.20  &23.84 \\
141441$+$711346&14 14 41.61 &$+$71 13 46.6 &1.16 &0.2170    &$-$22.2 &2.03  &22.57 \\
141415$+$711730&14 14 15.76 &$+$71 17 30.3 &0.38 &0.2137    &$-$22.7 &2.29  &24.39 \\
141355$+$711721&14 13 55.24 &$+$71 17 21.1 &0.26 &0.2129    &$-$22.0 &2.19  &23.79 \\
141317$+$711352&14 13 17.08 &$+$71 13 52.5 &1.08 &0.2151    &$-$22.3 &2.17  &22.94 \\
               
A1961& & & & & & & \\ 
144515$+$311934&14 45 15.00 &$+$31 19 34.0 &2.20 &0.2242    &$-$21.9 &2.28  &22.78\\ 
144431$+$311335&14 44 31.84 &$+$31 13 35.8 &0.11 &0.2333    &$-$23.1 &2.35  &23.58\\
144433$+$312139&14 44 33.48 &$+$31 21 39.6 &1.68 &0.2339    &$-$21.4 &2.30  &23.03\\
144412$+$310329&14 44 12.40 &$+$31 03 29.5 &2.02 &0.2275    &$-$23.0 &2.28  &23.88\\
144423$+$311552&14 44 23.91 &$+$31 15 52.5 &0.64 &0.2287    &$-$22.2 &2.36  &22.65\\
      
A2111& & & & & & & \\         
153954$+$341707&15 39 54.94 &$+$34 17 07.9 &1.59 &0.2312    &$-$21.5 &2.61  &22.53\\
153955$+$342012&15 39 55.12 &$+$34 20 12.9 &1.03 &\nodata   &$-$21.6 &2.37  &24.14\\
153949$+$342640&15 39 49.38 &$+$34 26 40.5 &0.39 &0.2290$^{\star}$    &$-$21.8 &1.66  &22.74\\
153933$+$342224&15 39 33.08 &$+$34 22 24.6 &0.67 &0.2296    &$-$21.8 &2.05  &22.45\\ 

\tablecomments{Column legend --- (1) is the galaxy
number; (2) and (3) are the optical right-ascension and declination
for the epoch J2000; (4) is the distance from the X-ray defined
cluster center in Mpc; (5) is the galaxy's redshift; (6) is the
absolute magnitude; (7) is the B$-$R color index of the galaxy; (8)
lists radio luminosity at 1.4\,GHz in \whz.}

\tablenotetext{\diamond}{Data from \cite{led95b} was used to calibrate the
magnitude zero point on DSS image.}
\tablenotetext{\dagger}{The CCD image was used to calibrate the
magnitude zero point on the DSS image.}
\tablenotetext{\triangleleft}{Measured redshift had a \cite{ton79} R $\le$ 4, and 
is therefore unreliable.}
\tablenotetext{\ddag}{Redshift from NED.}
\tablenotetext{\star}{Another measurement of this redshift is reported in NED.}
\enddata
\end{deluxetable}
\clearpage

\clearpage

\figcaption[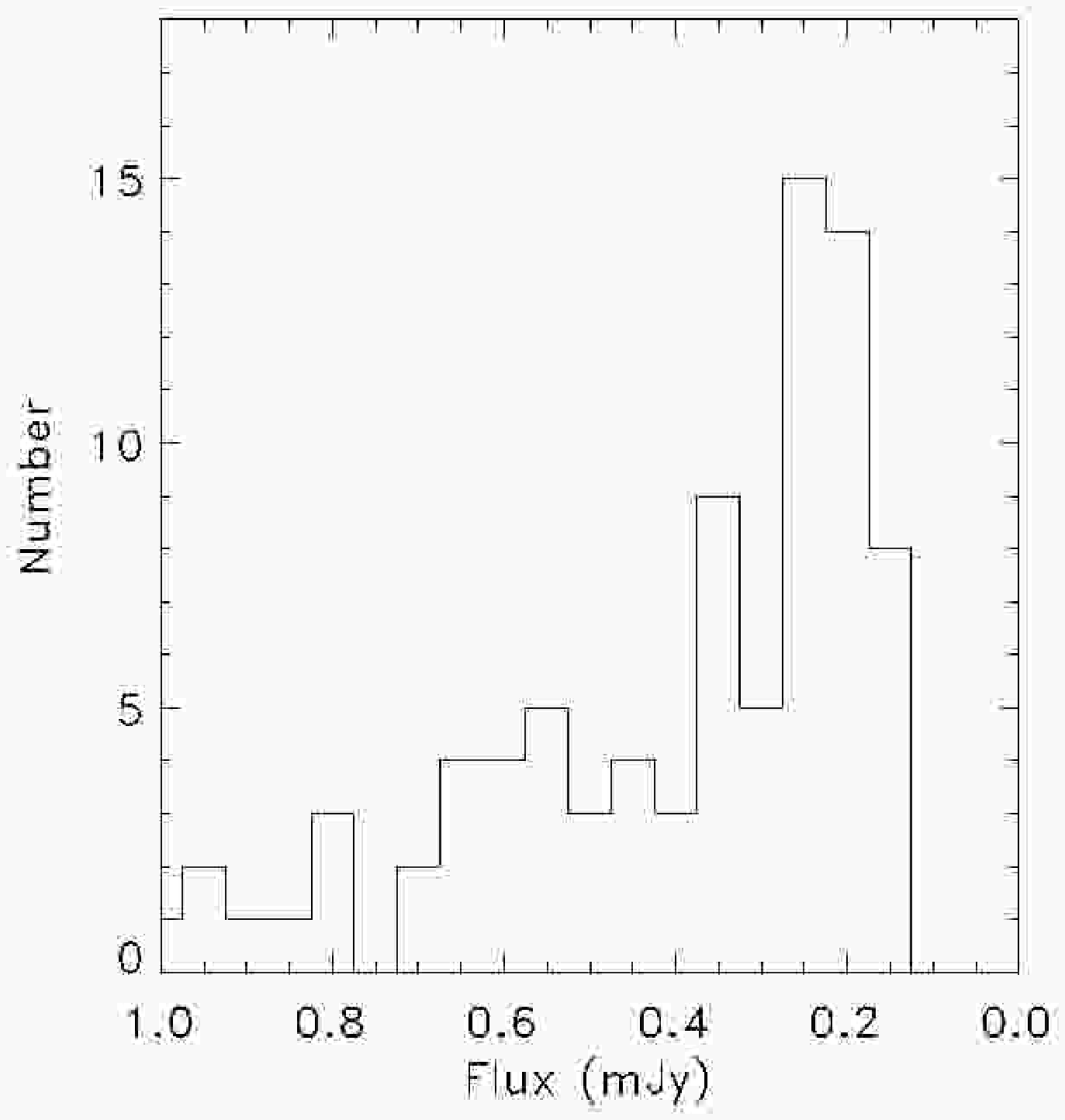]{Histogram of radio fluxes for $z \sim 0.2$
clusters showing that the completeness limit is
0.2\,mJy. \label{fig_1}}

\figcaption[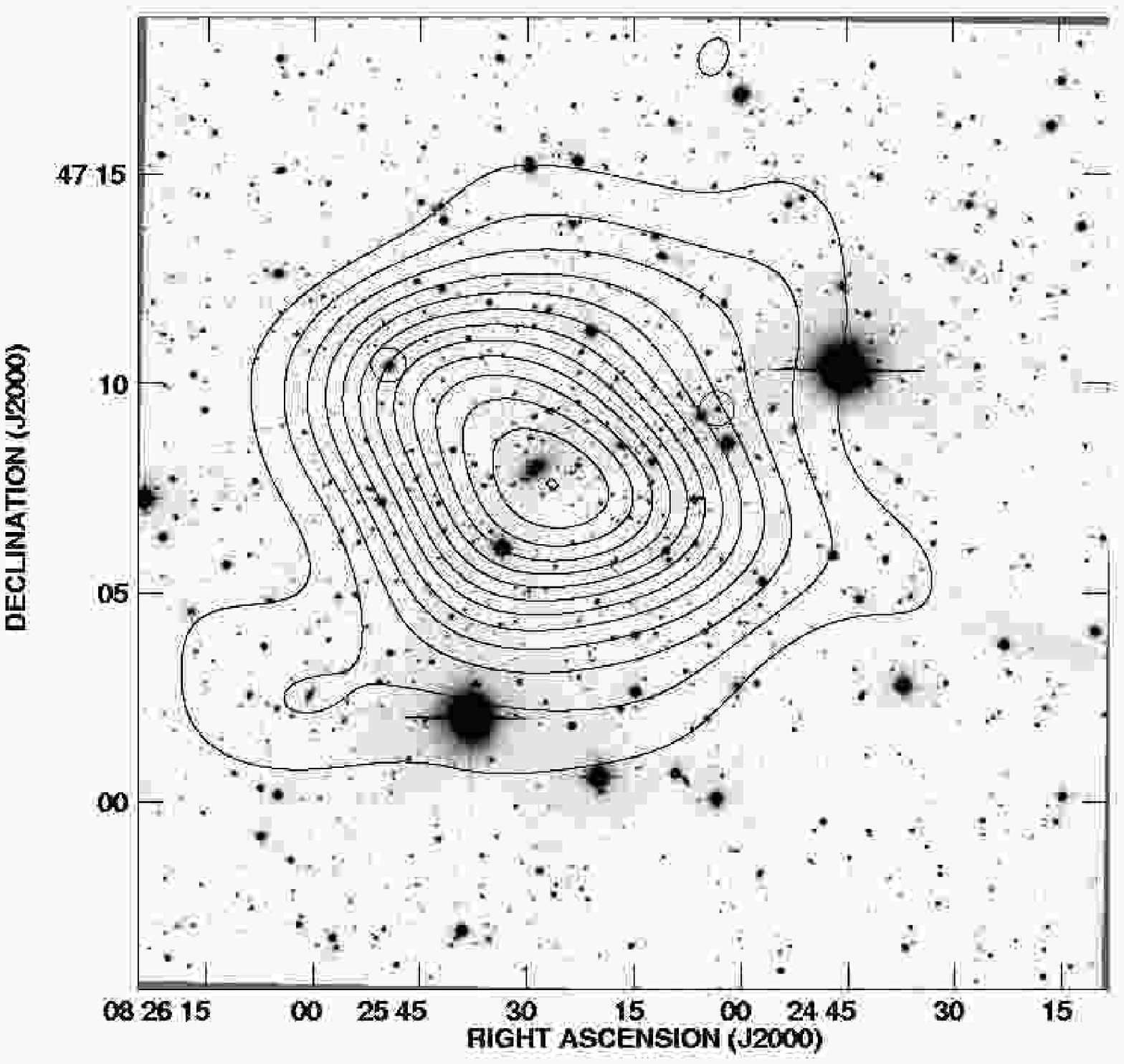]{R-Band image (greyscale) of Abell 655 overlaid with X-ray (ROSAT PSPC-C, 
0.5-2.0 keV) image (contours).  Circles and triangles on the image
denote galaxies detected in the radio (L$_{\mathrm{1.4\,GHz}}$ =
10$^{22.3}$\,\whz) with and without redshifts,
respectively. \label{fig_2a}}

\figcaption[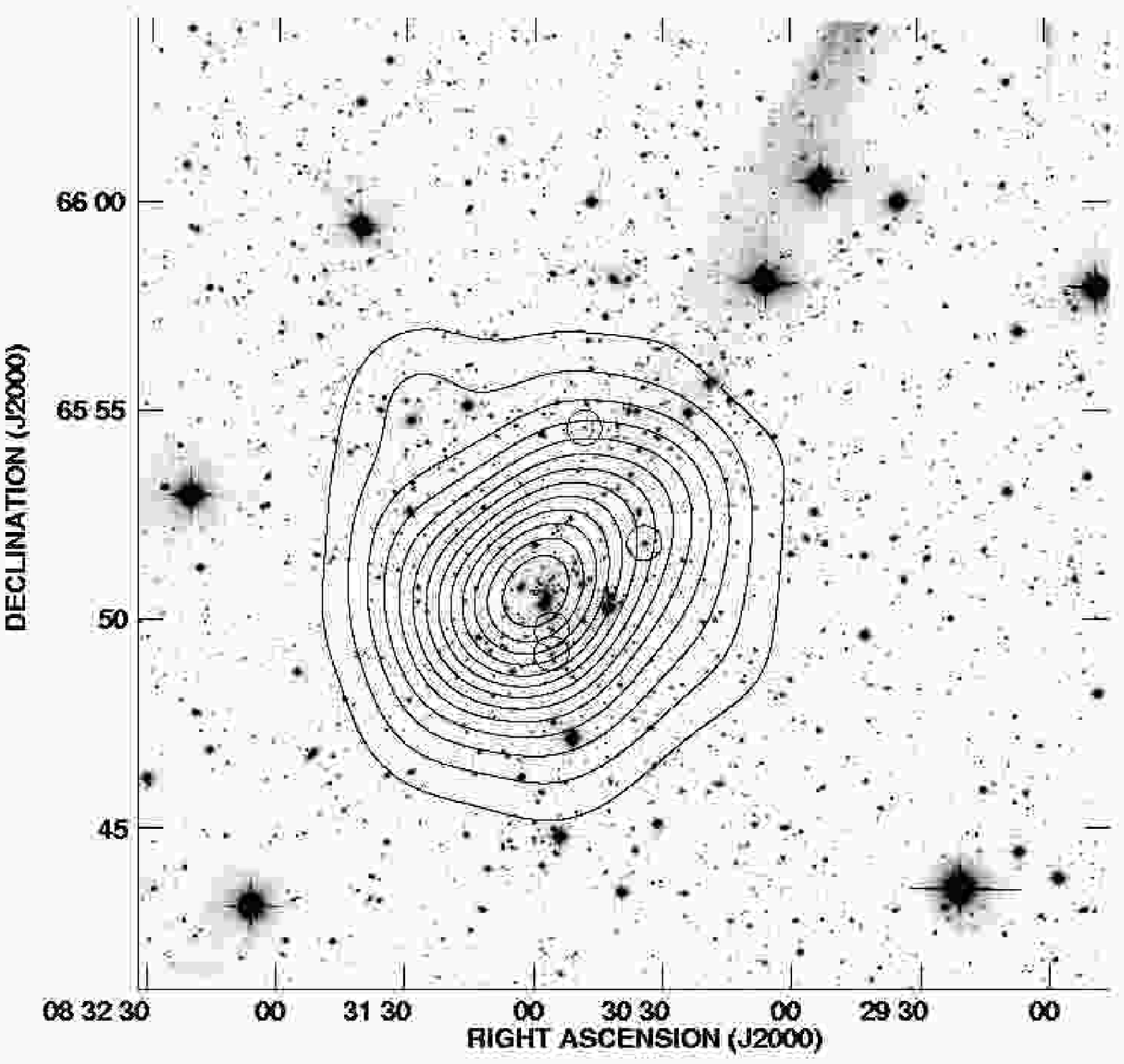]{Abell 665 See Figure \ref{fig_2a} for details. \label{fig_2b}}

\figcaption[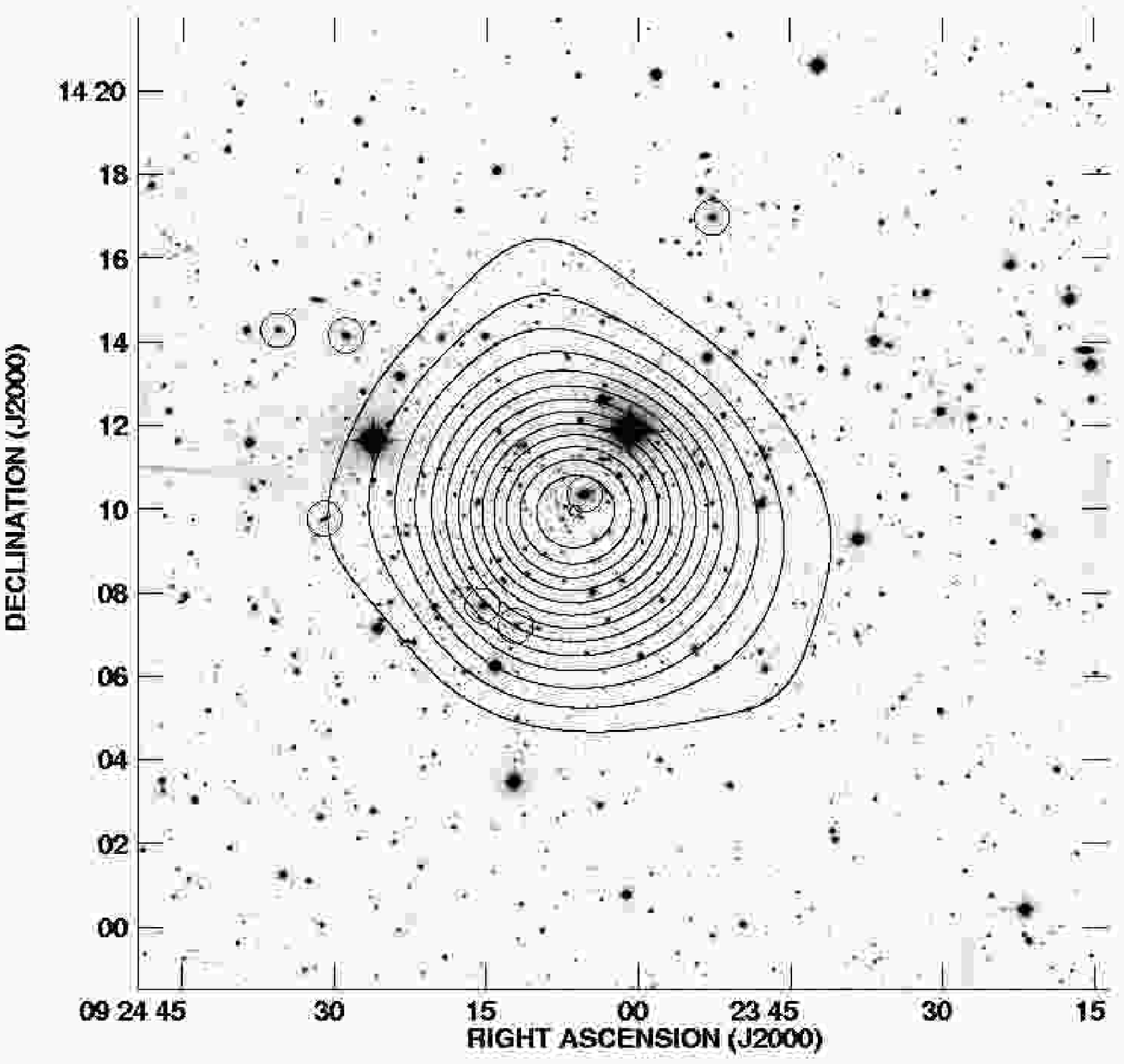]{Abell 795 See Figure \ref{fig_2a} for details. \label{fig_2c}}

\figcaption[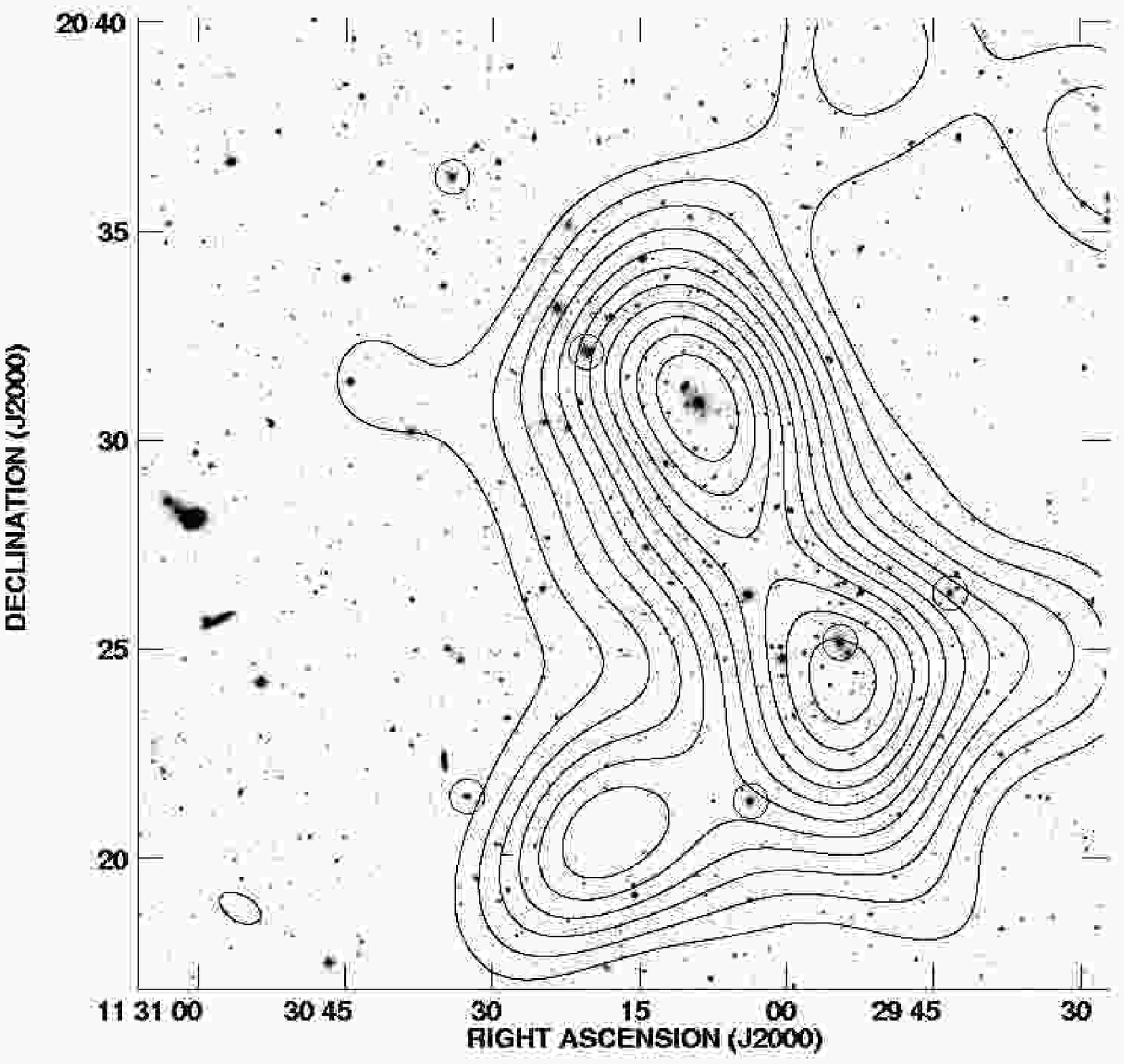]{Abell 1278 See Figure \ref{fig_2a} for details. \label{fig_2d}}

\figcaption[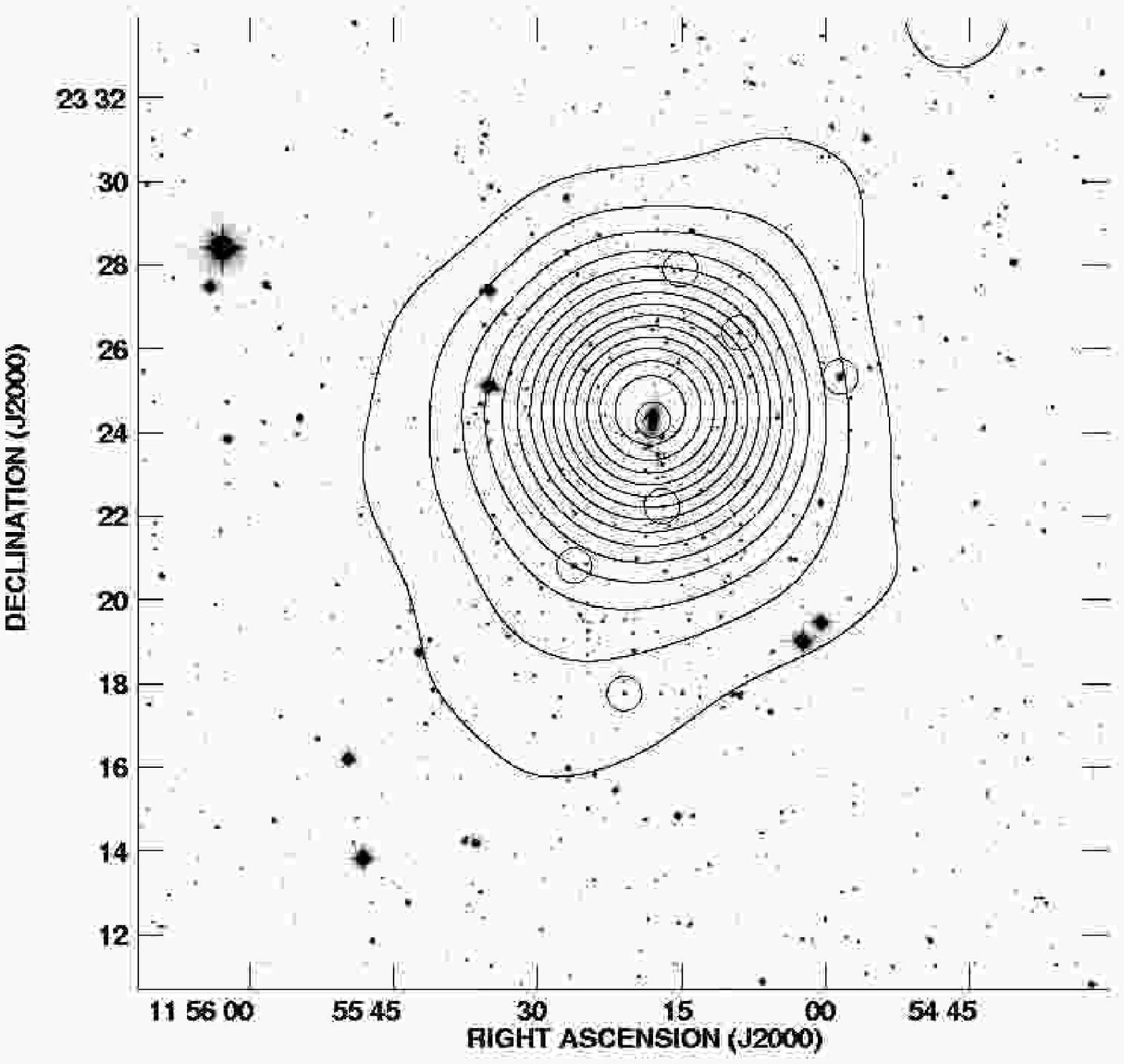]{Abell 1413 See Figure \ref{fig_2a} for details. \label{fig_2e}}

\figcaption[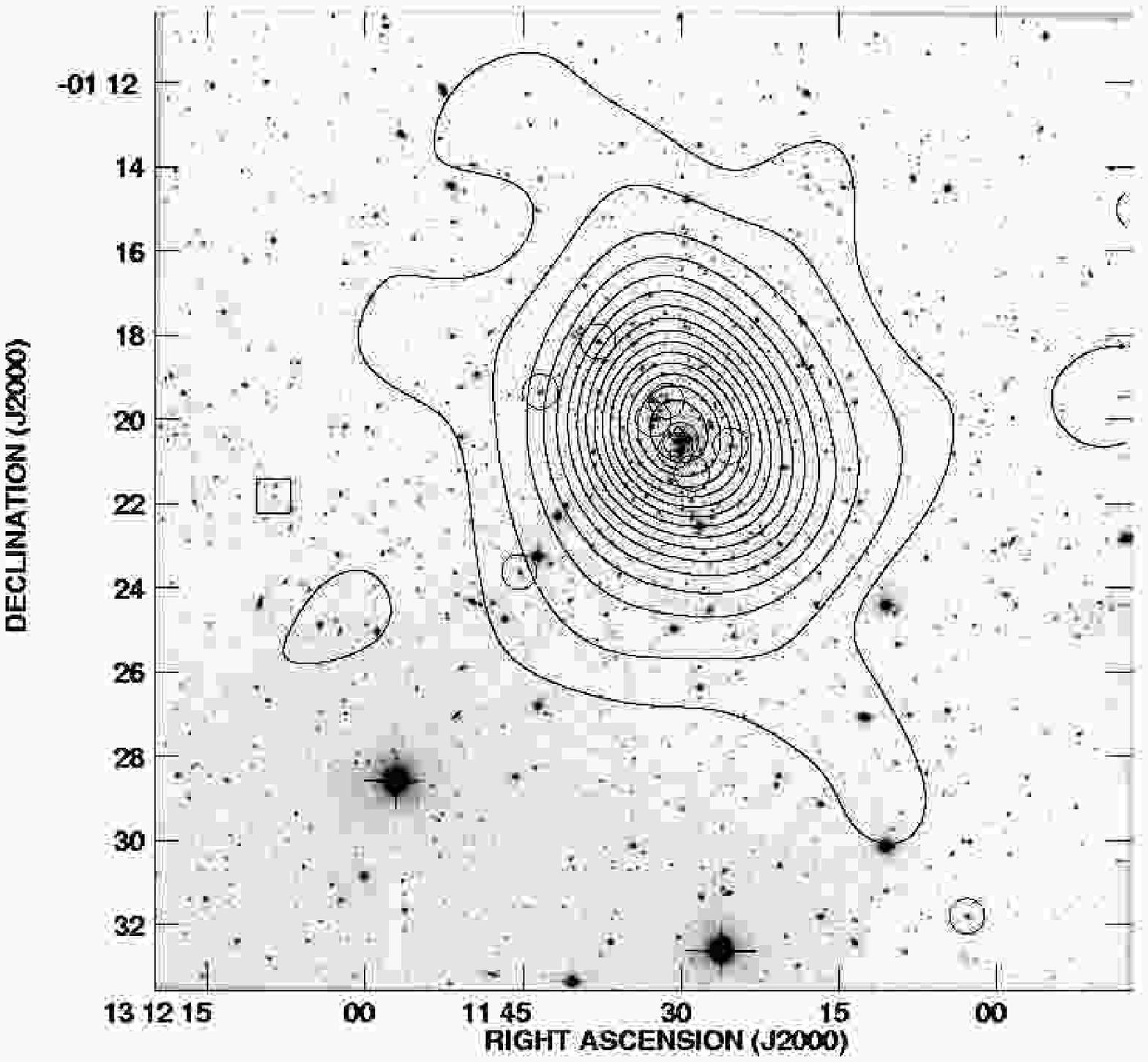]{Abell 1689 See Figure \ref{fig_2a} for details. \label{fig_2f}}

\figcaption[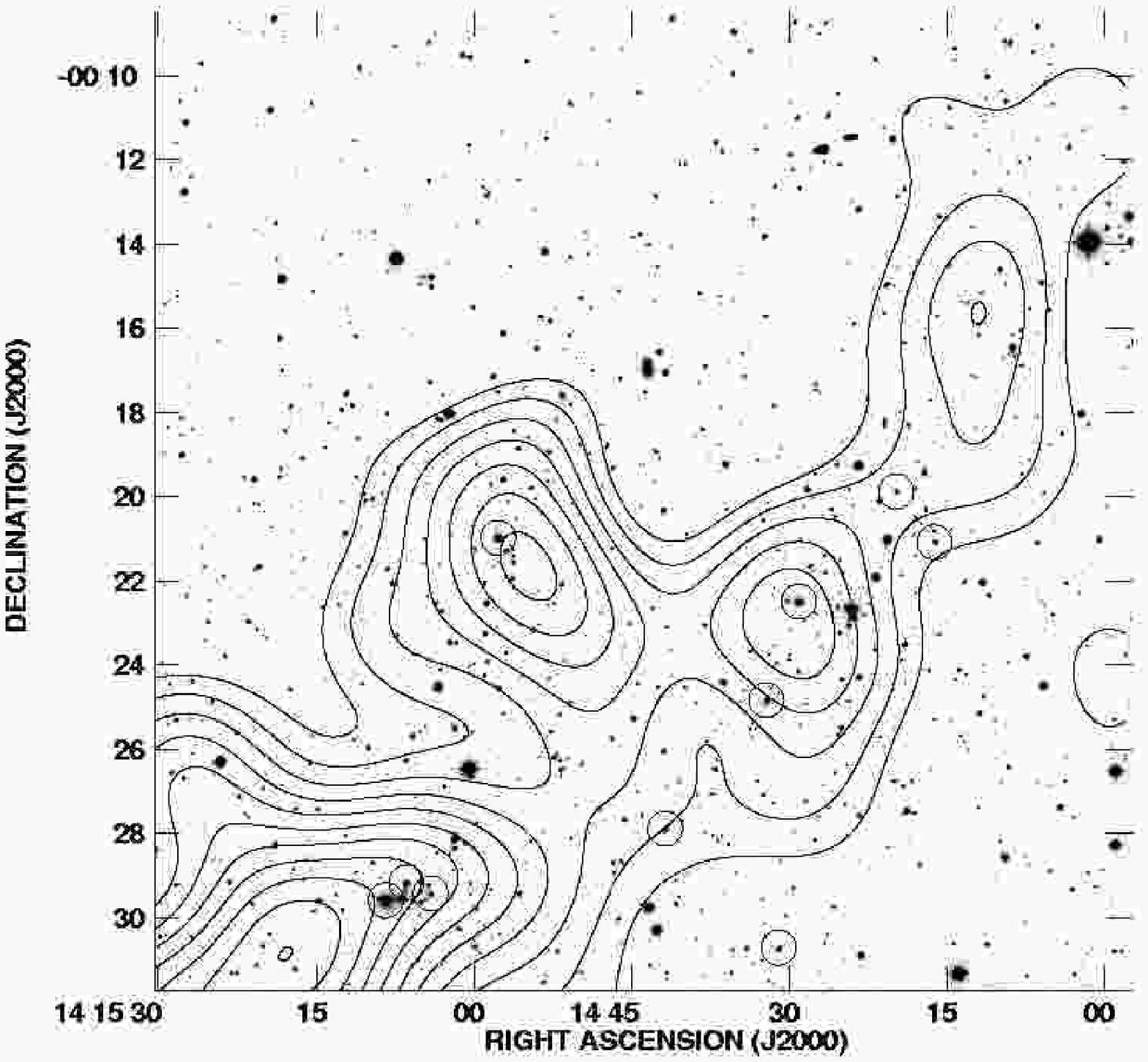]{Abell 1882 See Figure \ref{fig_2a} for details. \label{fig_2g}}

\figcaption[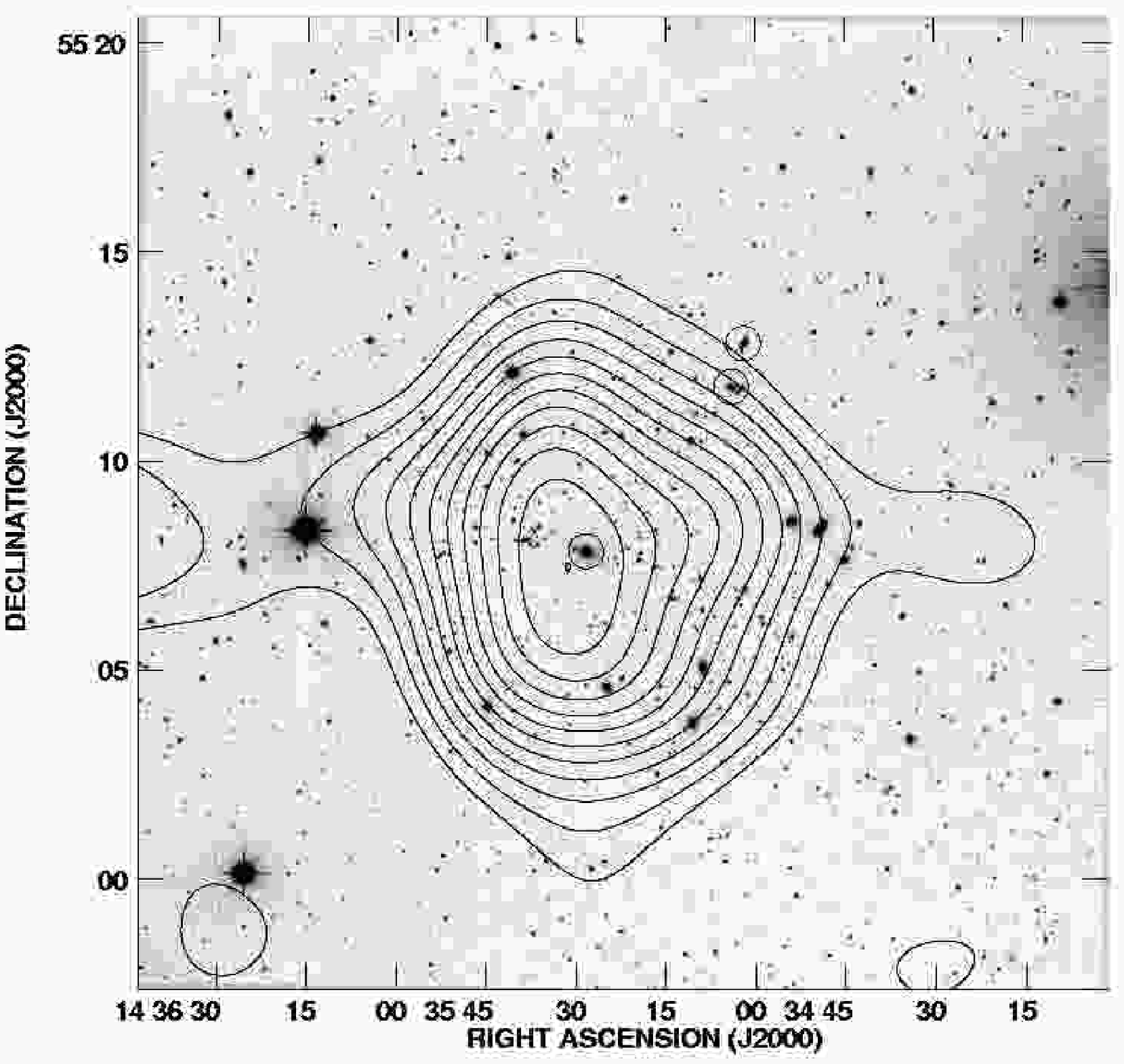]{Abell 1940 See Figure \ref{fig_2a} for details. \label{fig_2h}}

\figcaption[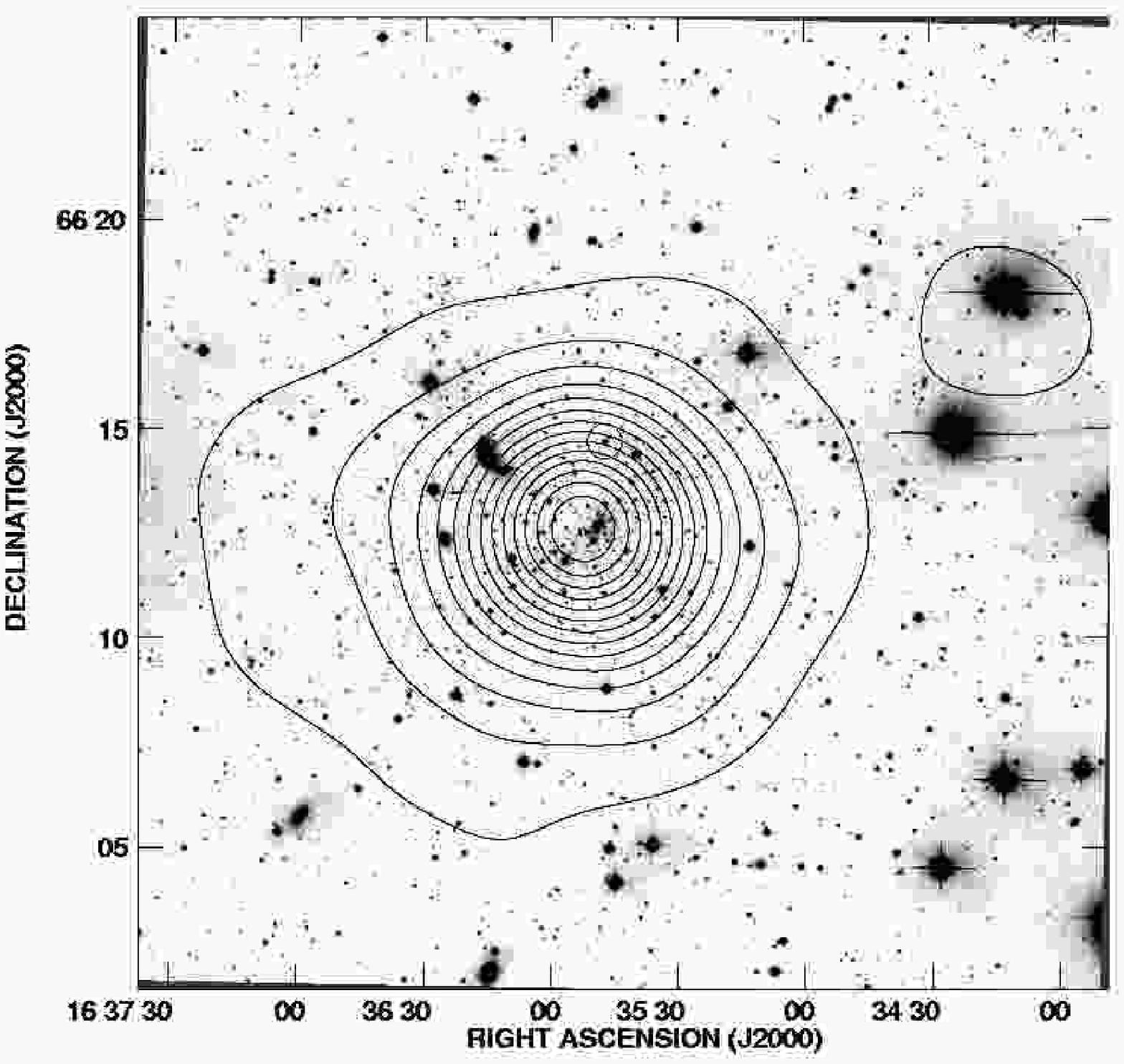]{Abell 2218 See Figure \ref{fig_2a} for details. \label{fig_2i}}

\figcaption[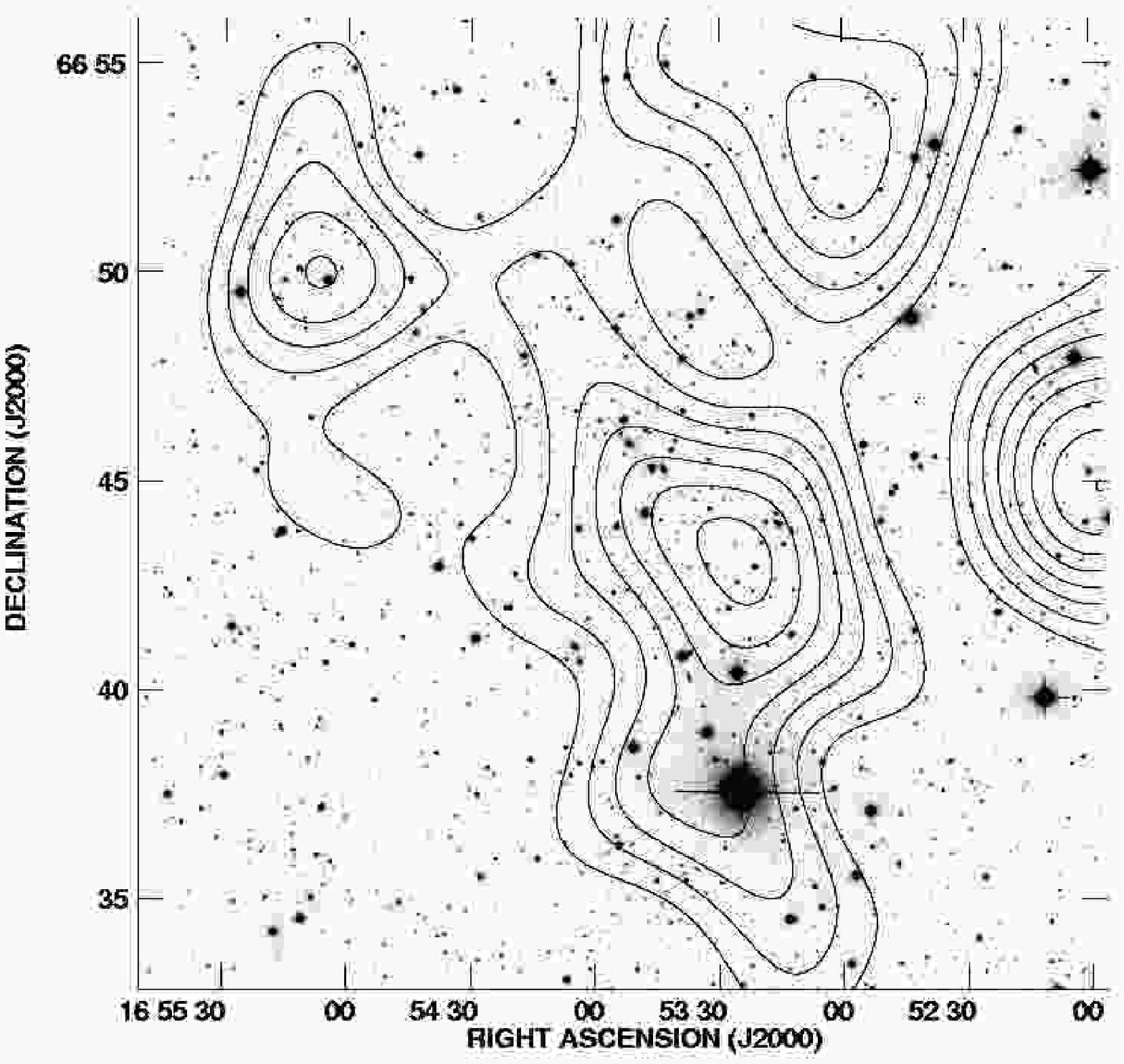]{Abell 2240 See Figure \ref{fig_2a} for details. \label{fig_2j}}

\figcaption[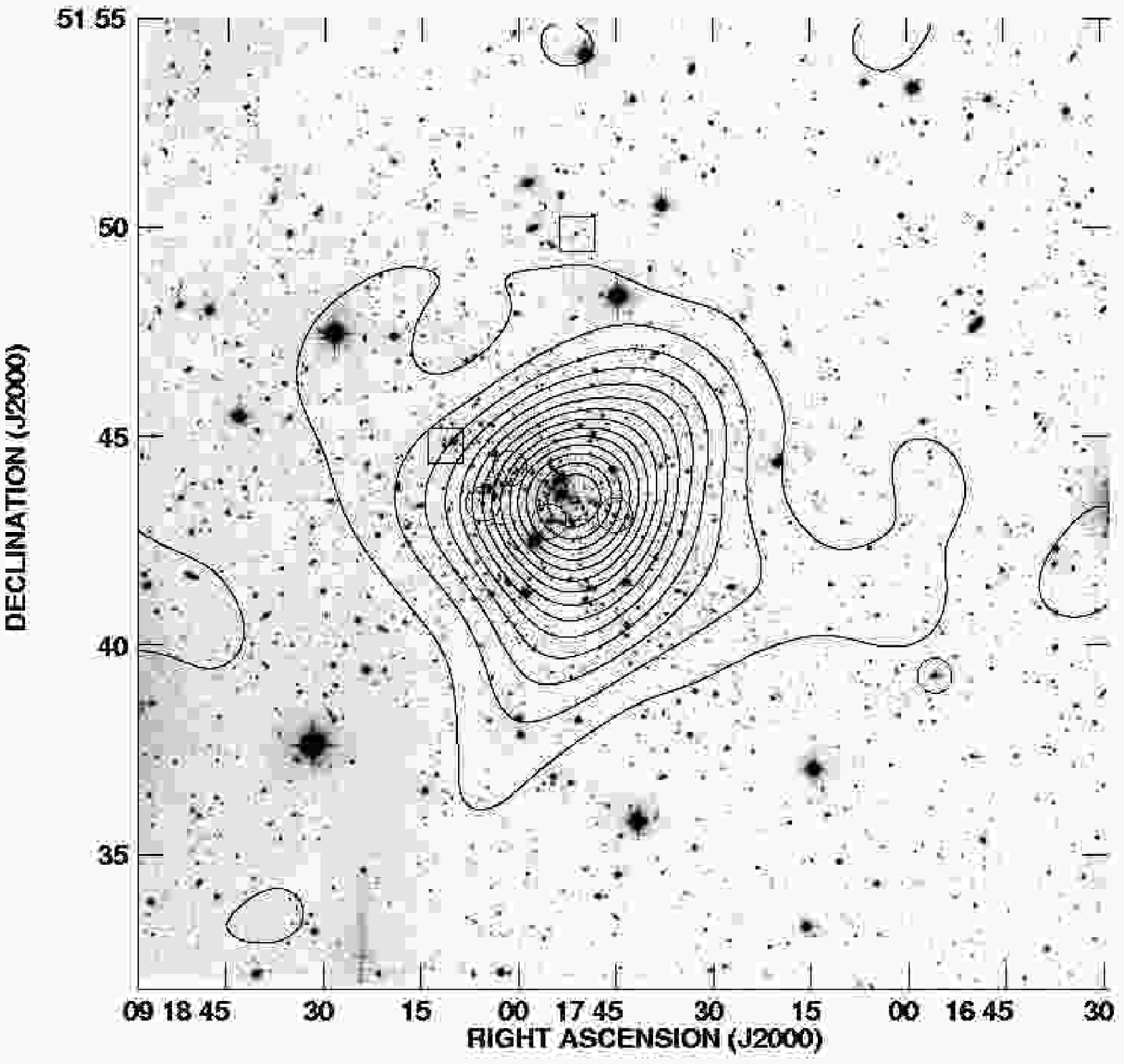]{Abell 773 See Figure \ref{fig_2a} for details. \label{fig_2k}}

\figcaption[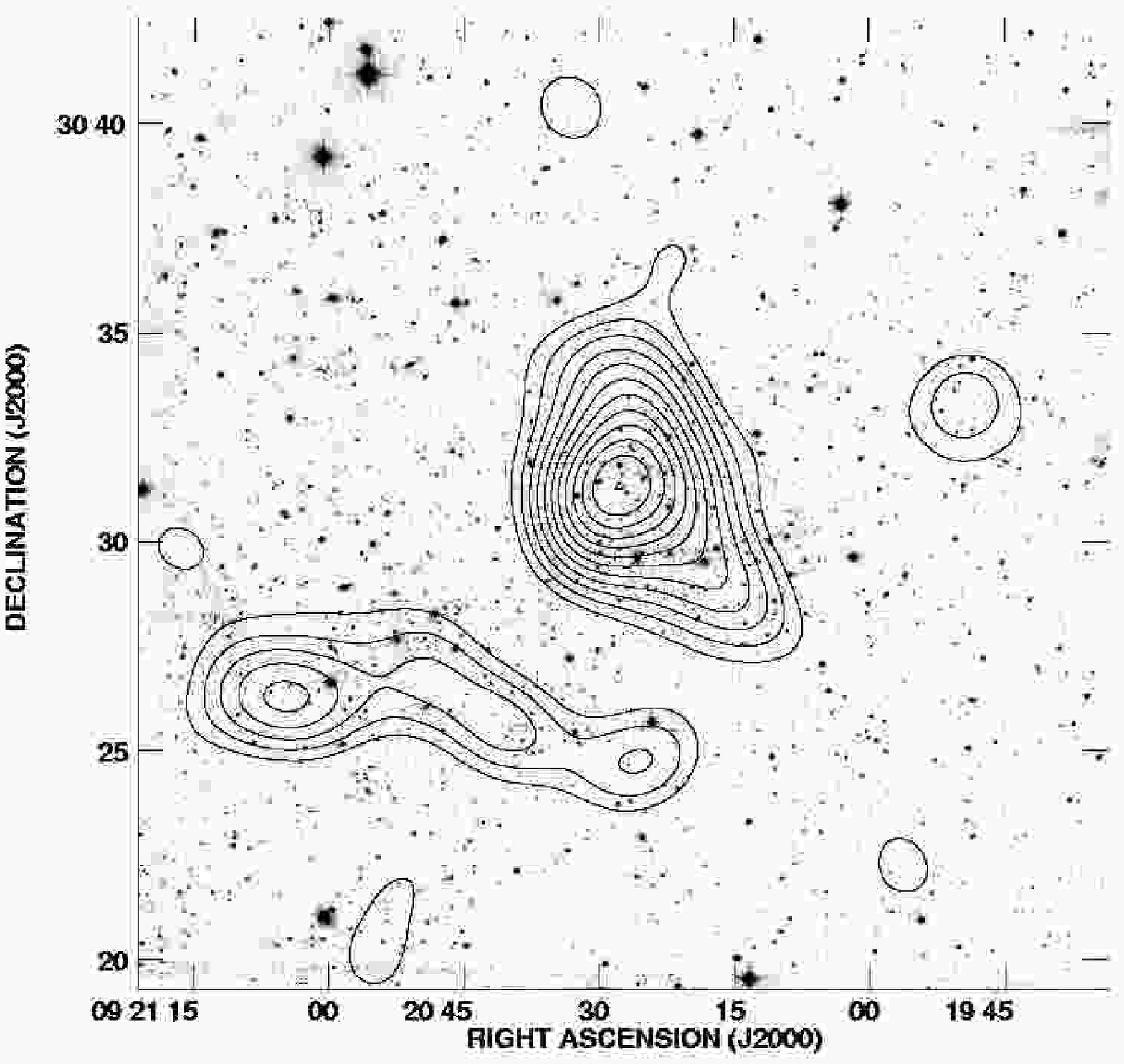]{Abell 781 See Figure \ref{fig_2a} for details. \label{fig_2l}}

\figcaption[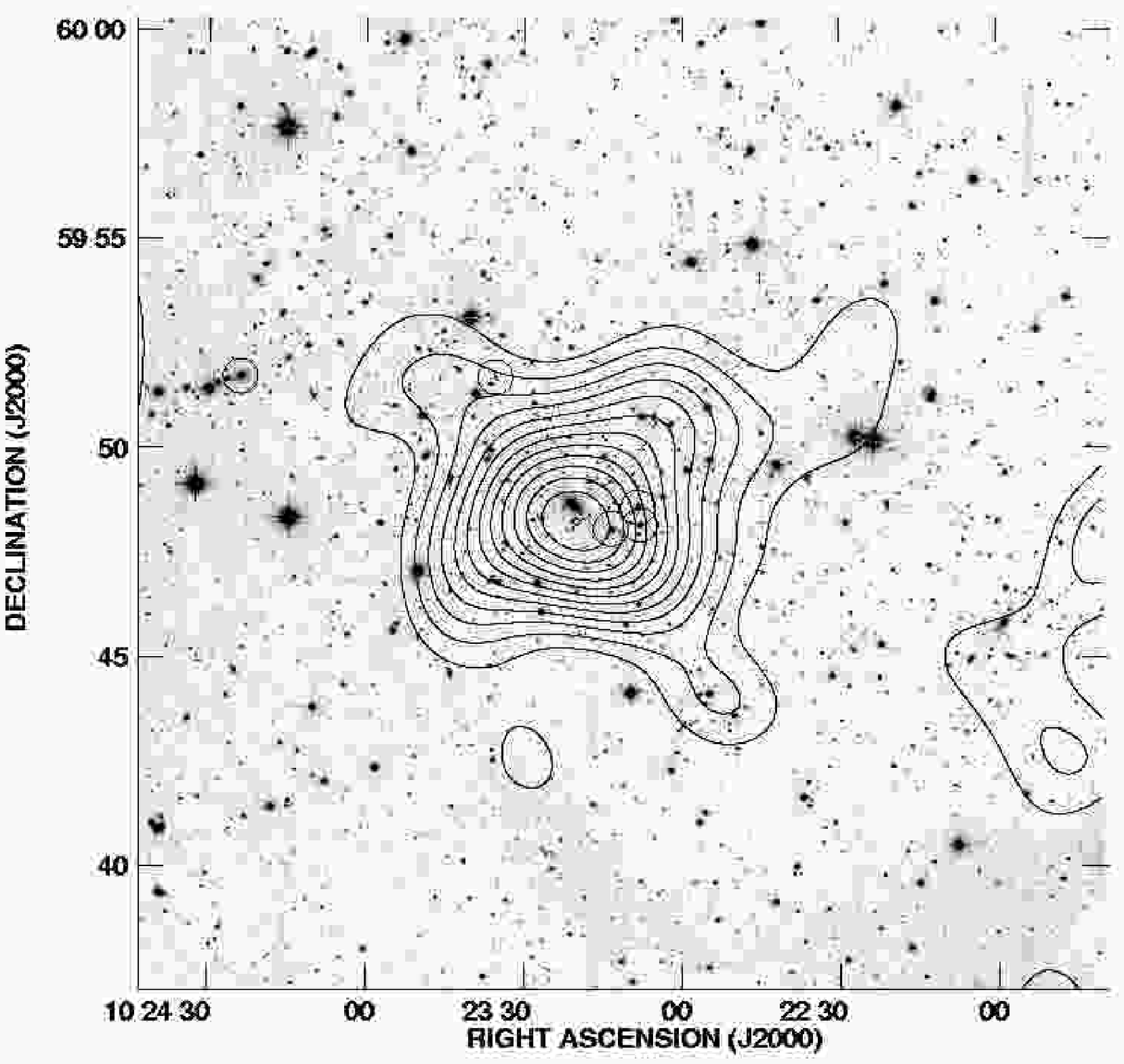]{Abell 983 See Figure \ref{fig_2a} for details. \label{fig_2m}}

\figcaption[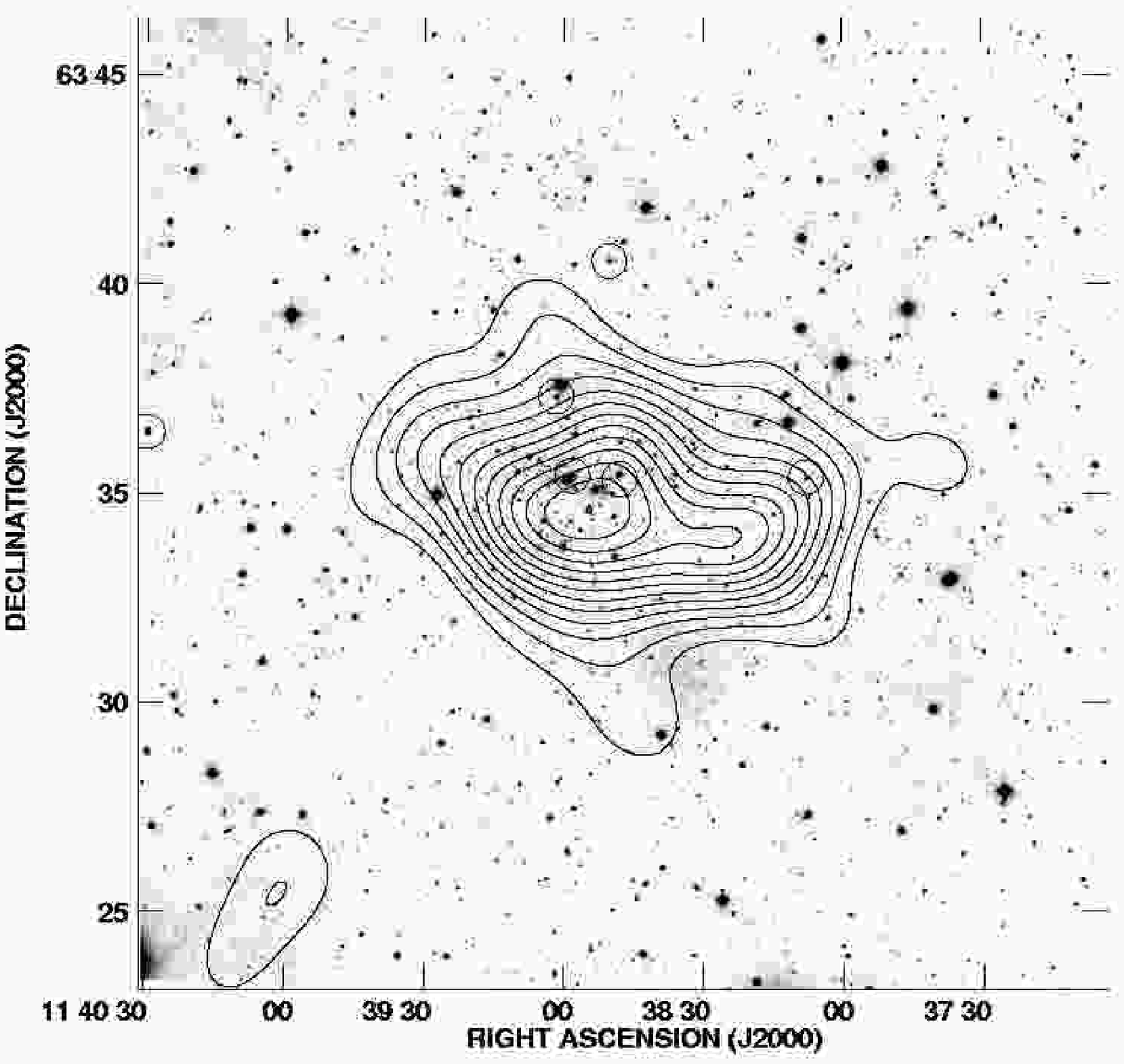]{Abell 1331 See Figure \ref{fig_2a} for details. \label{fig_2n}}

\figcaption[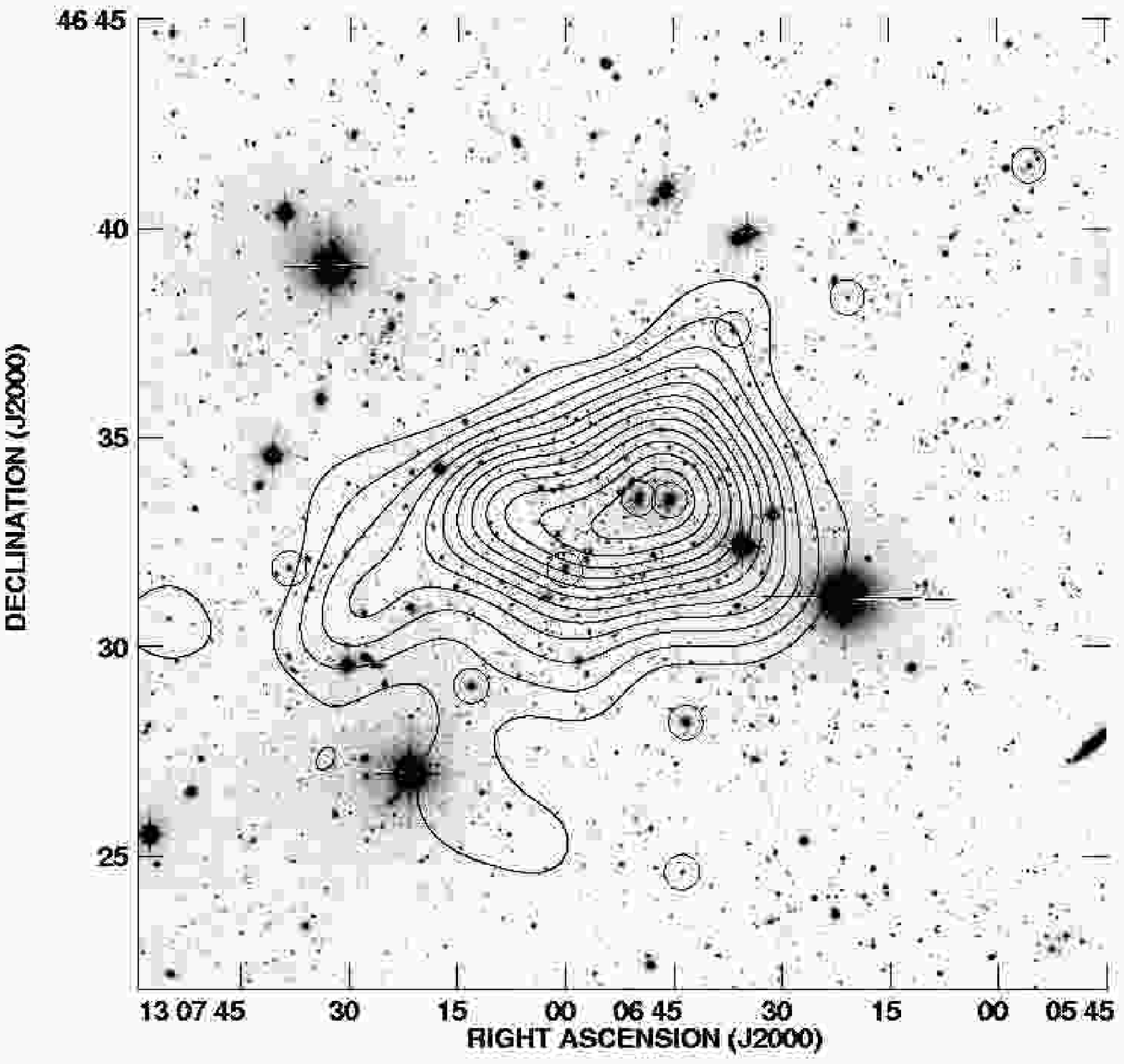]{Abell 1682 See Figure \ref{fig_2a} for details. \label{fig_2o}}

\figcaption[R2p.ps]{Abell 1704 See Figure \ref{fig_2a} for details. \label{fig_2p}}

\figcaption[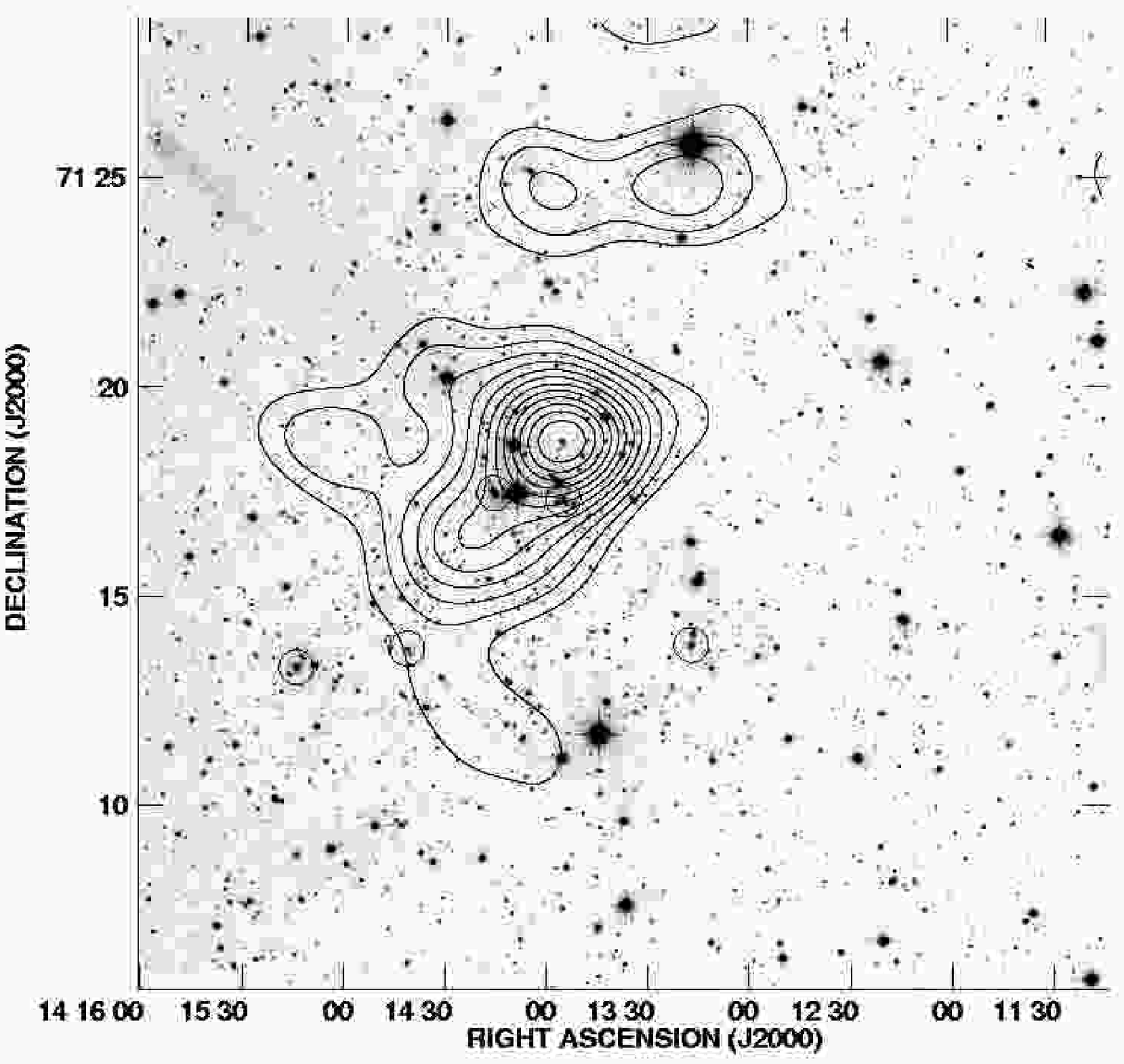]{Abell 1895 See Figure \ref{fig_2a} for details. \label{fig_2q}}

\figcaption[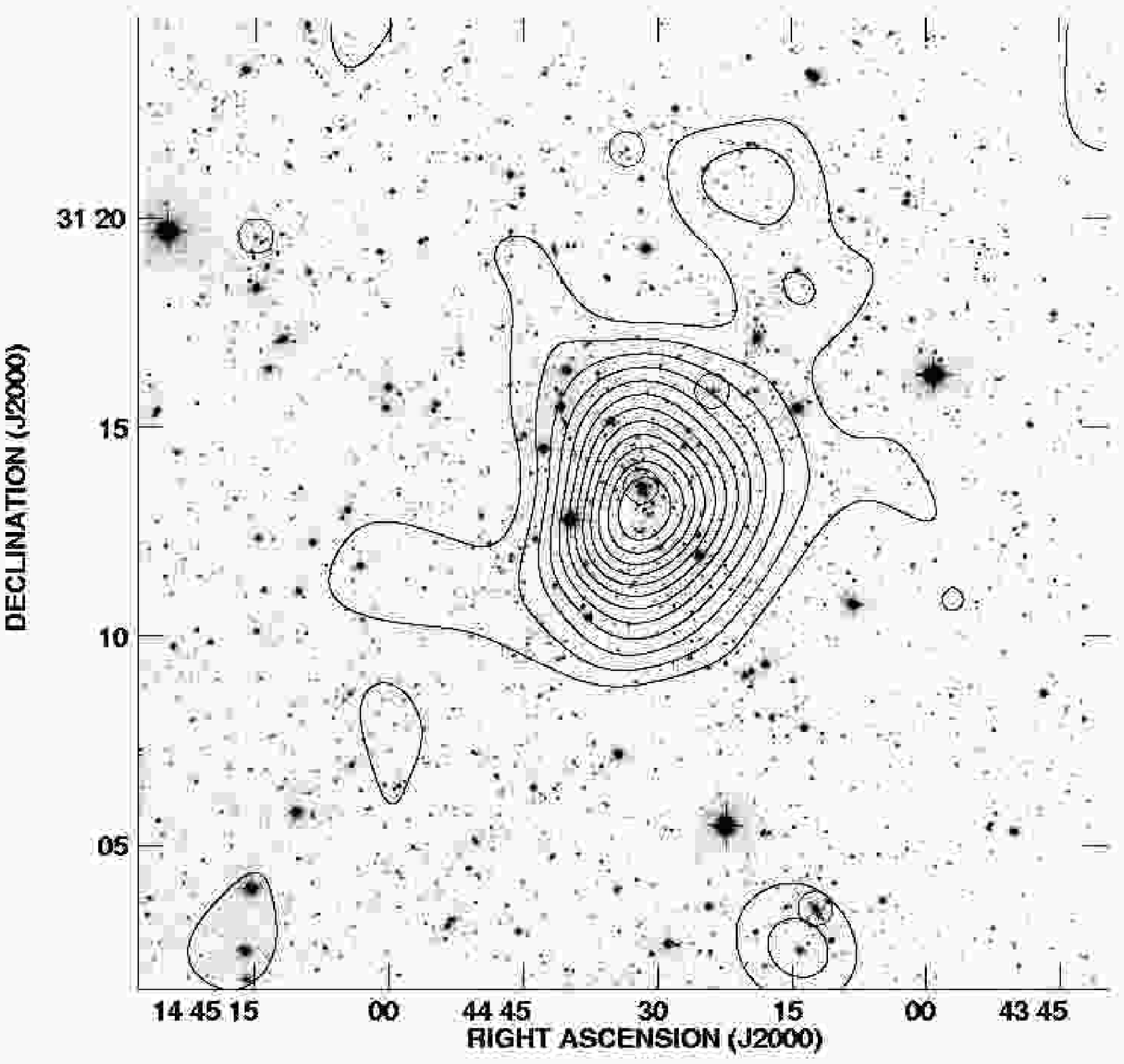]{Abell 1961 See Figure \ref{fig_2a} for details. \label{fig_2r}}

\figcaption[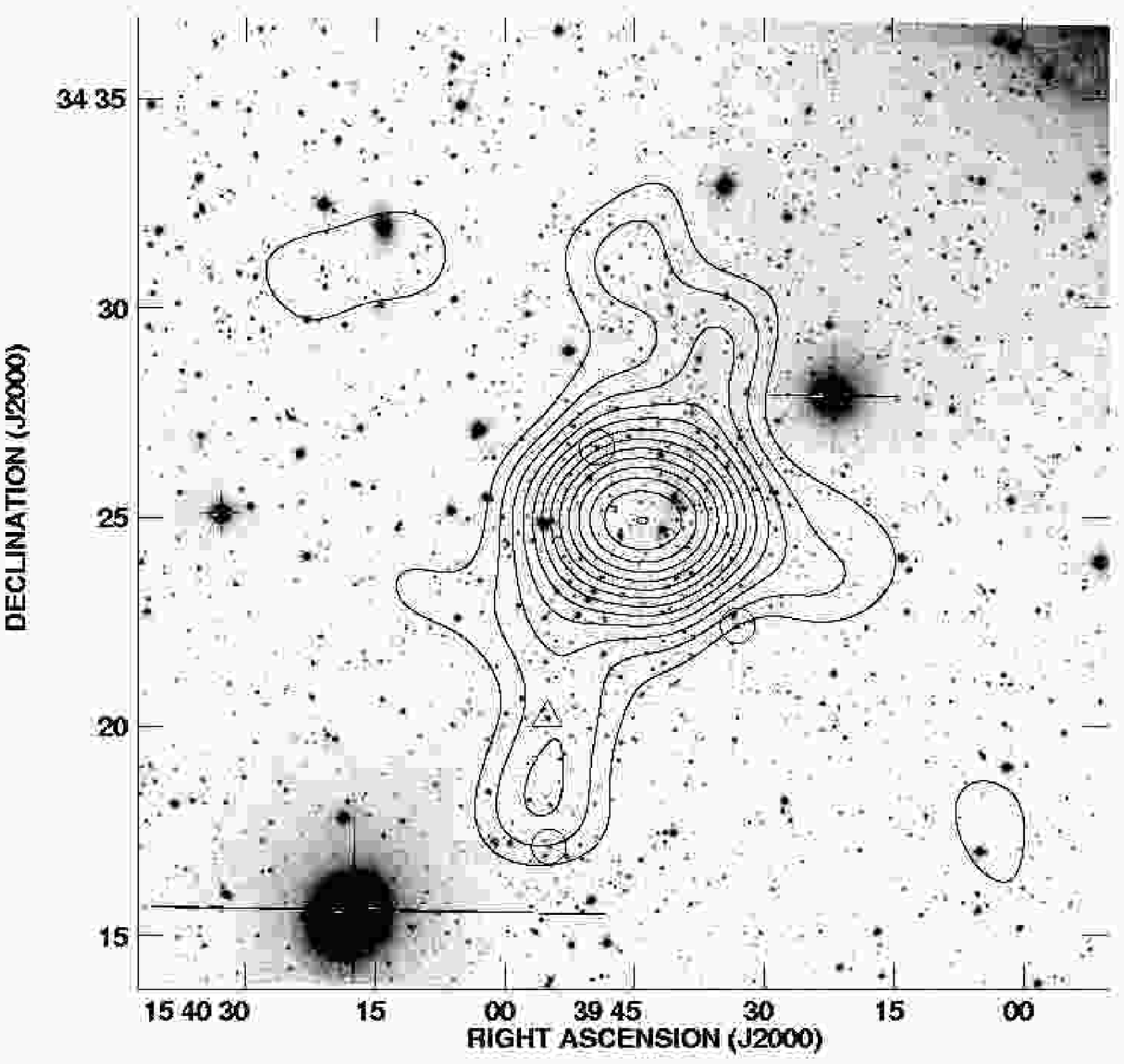]{Abell 2111 See Figure \ref{fig_2a} for details. \label{fig_2s}}

\figcaption[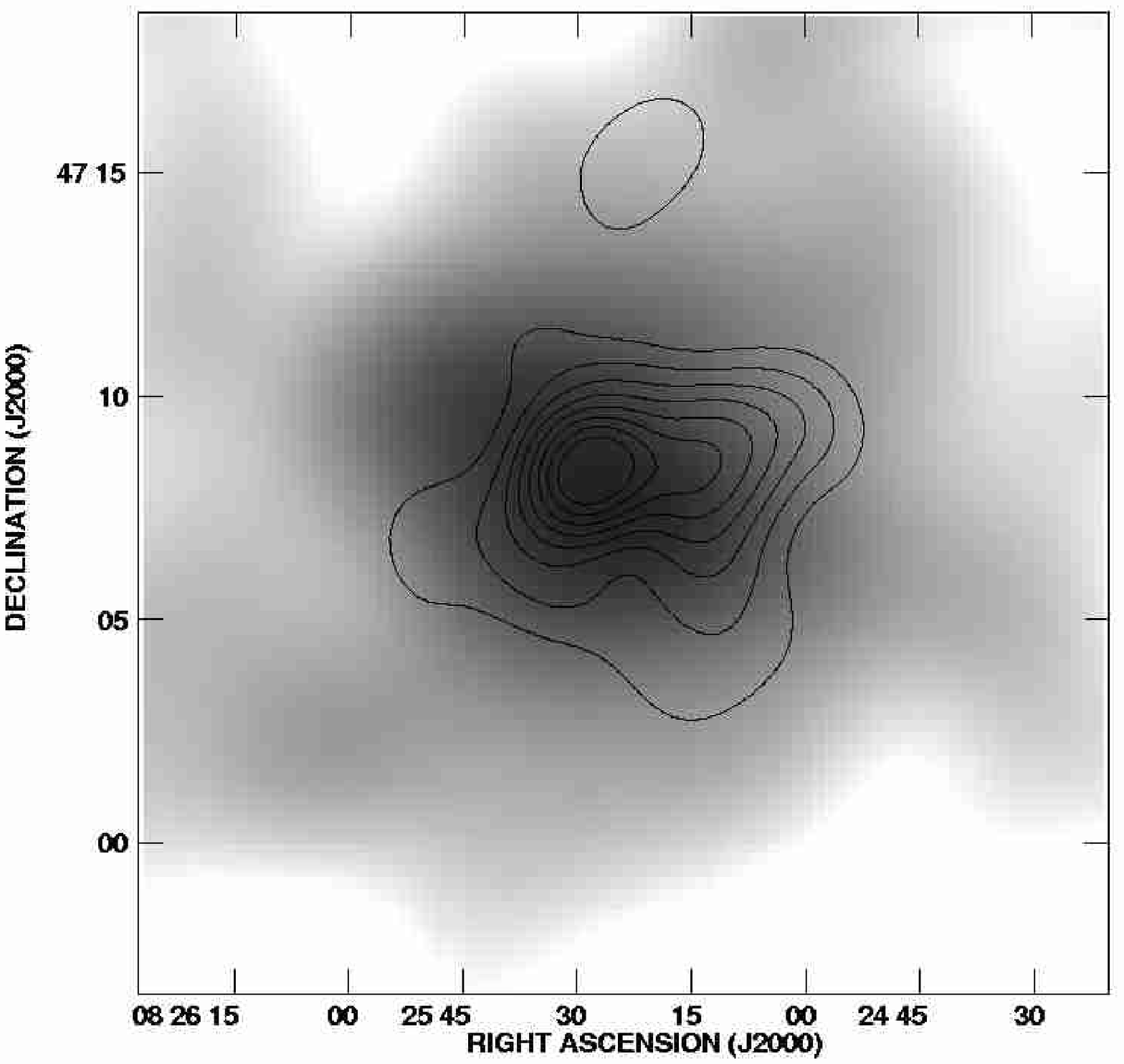]{X-ray (ROSAT PSPC-C, 
0.5-2.0 keV) image (greyscale) of Abell 655 overlaid with isopleth
(constant optical galaxy density) contours. \label{fig_3a}}

\figcaption[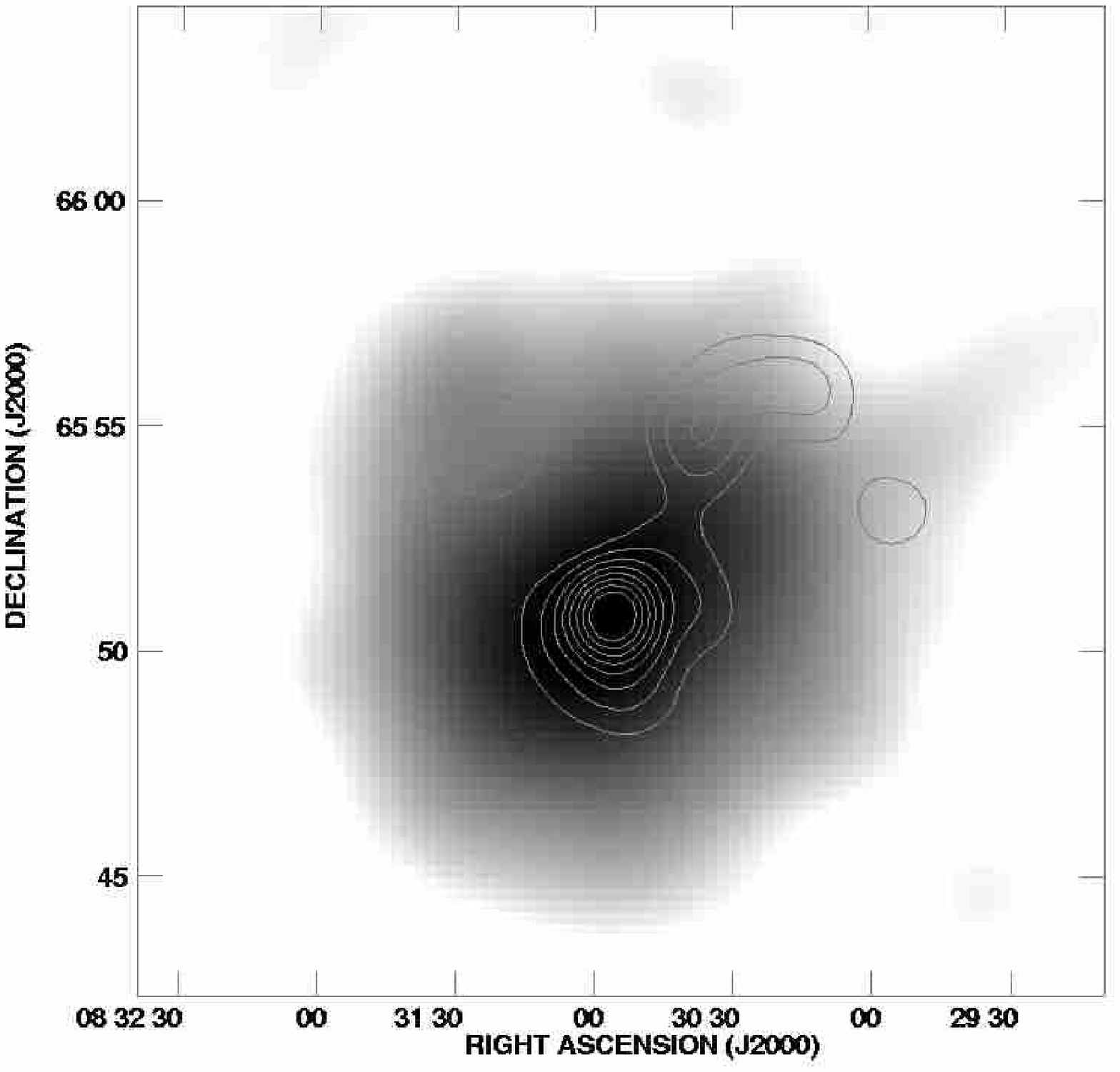]{Abell  665  See Figure \ref{fig_3a} for details. \label{fig_3b}}
\figcaption[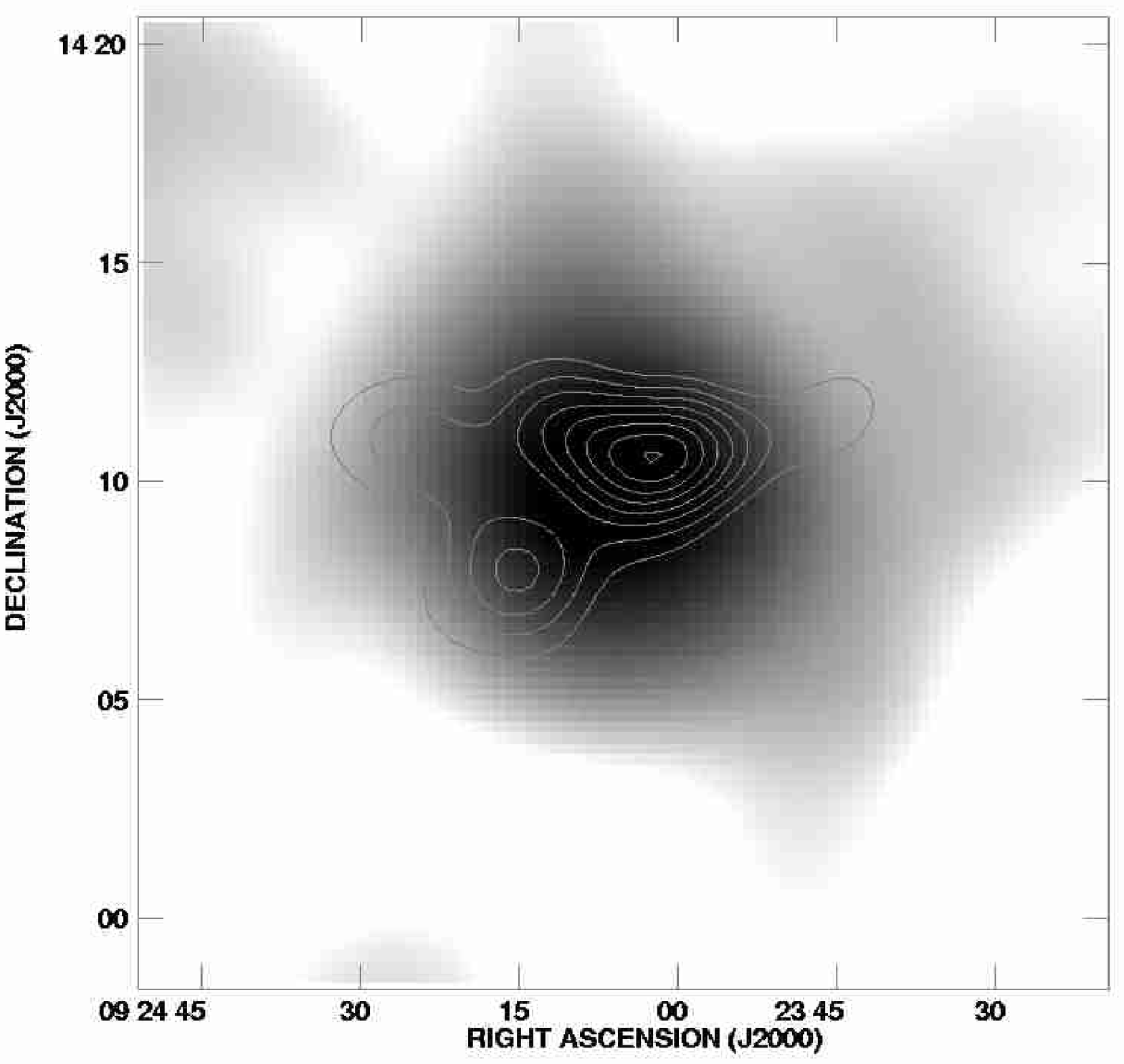]{Abell  795  See Figure \ref{fig_3a} for details. \label{fig_3c}}
\figcaption[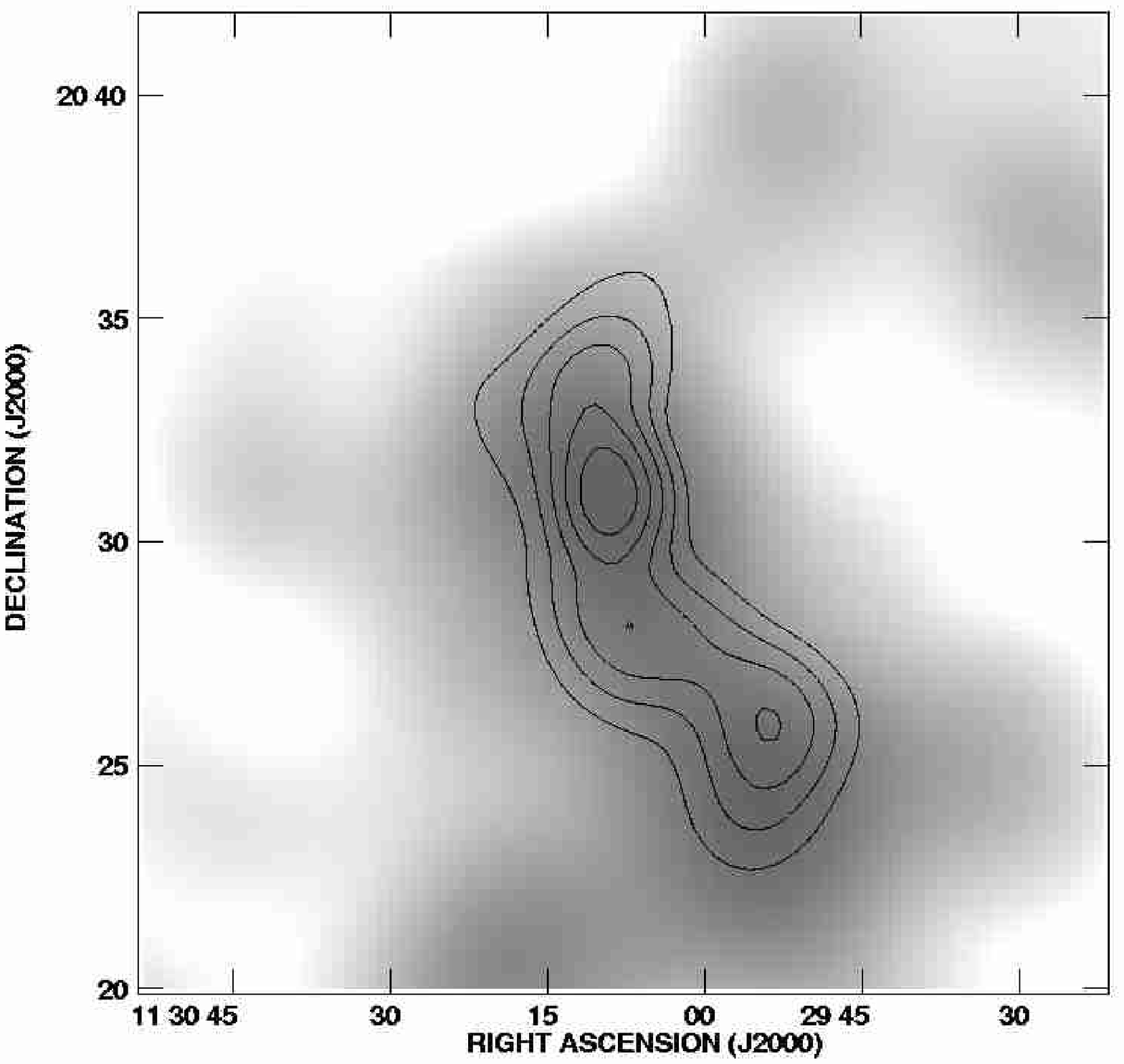]{Abell 1278  See Figure \ref{fig_3a} for details. \label{fig_3d}}
\figcaption[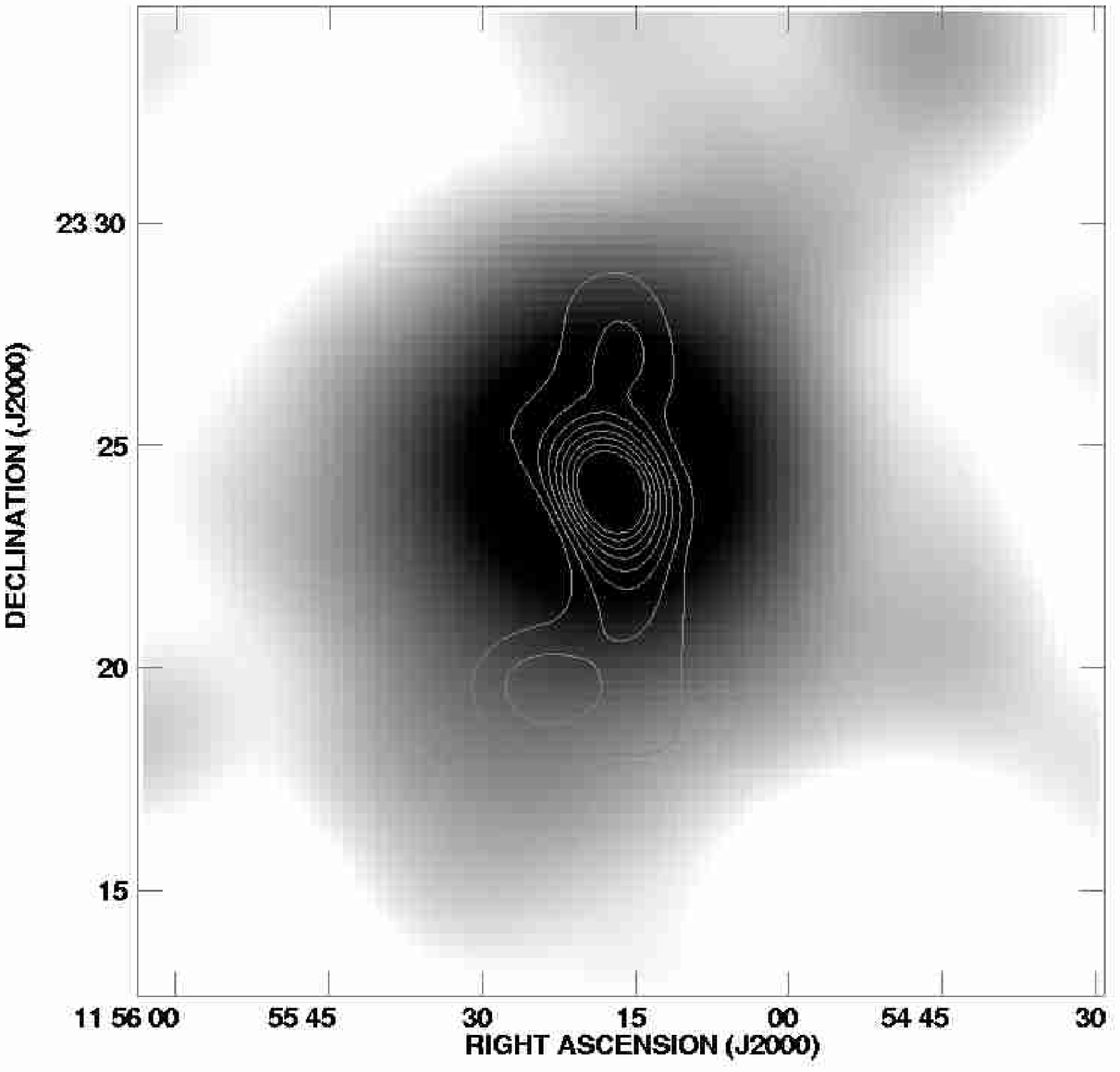]{Abell 1413  See Figure \ref{fig_3a} for details. \label{fig_3e}}
\figcaption[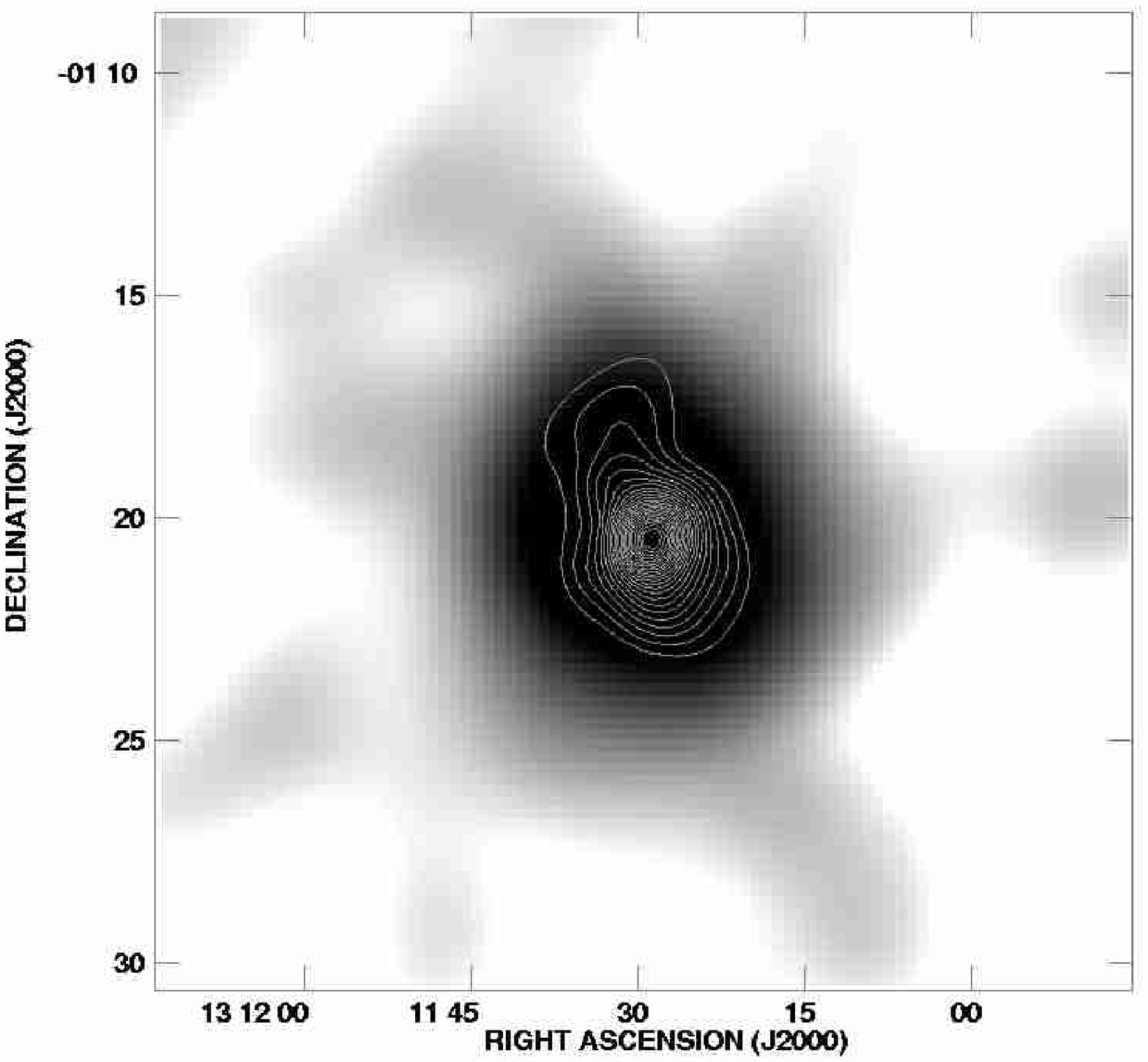]{Abell 1689  See Figure \ref{fig_3a} for details. \label{fig_3f}}
\figcaption[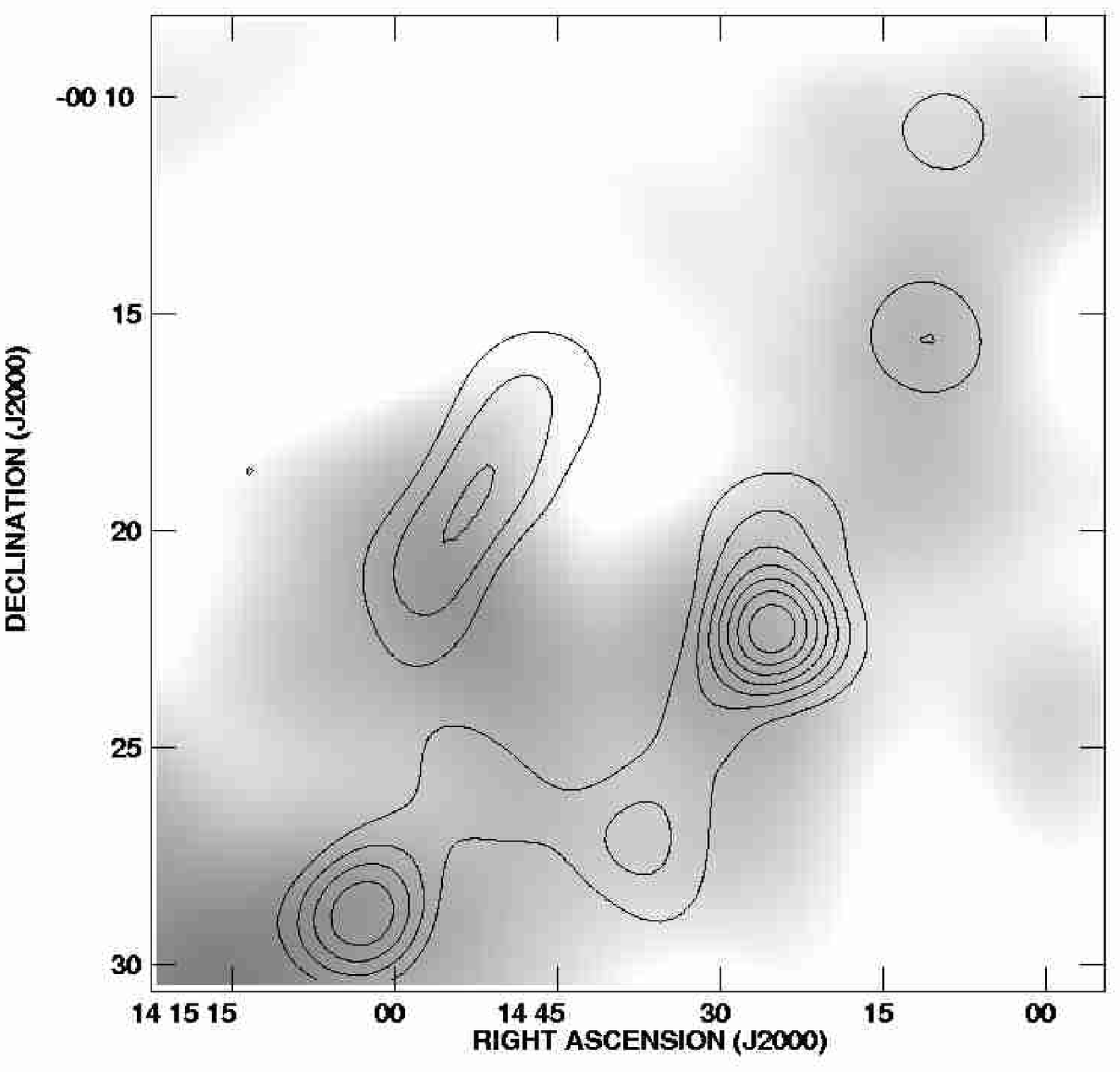]{Abell 1882  See Figure \ref{fig_3a} for details. \label{fig_3g}}
\figcaption[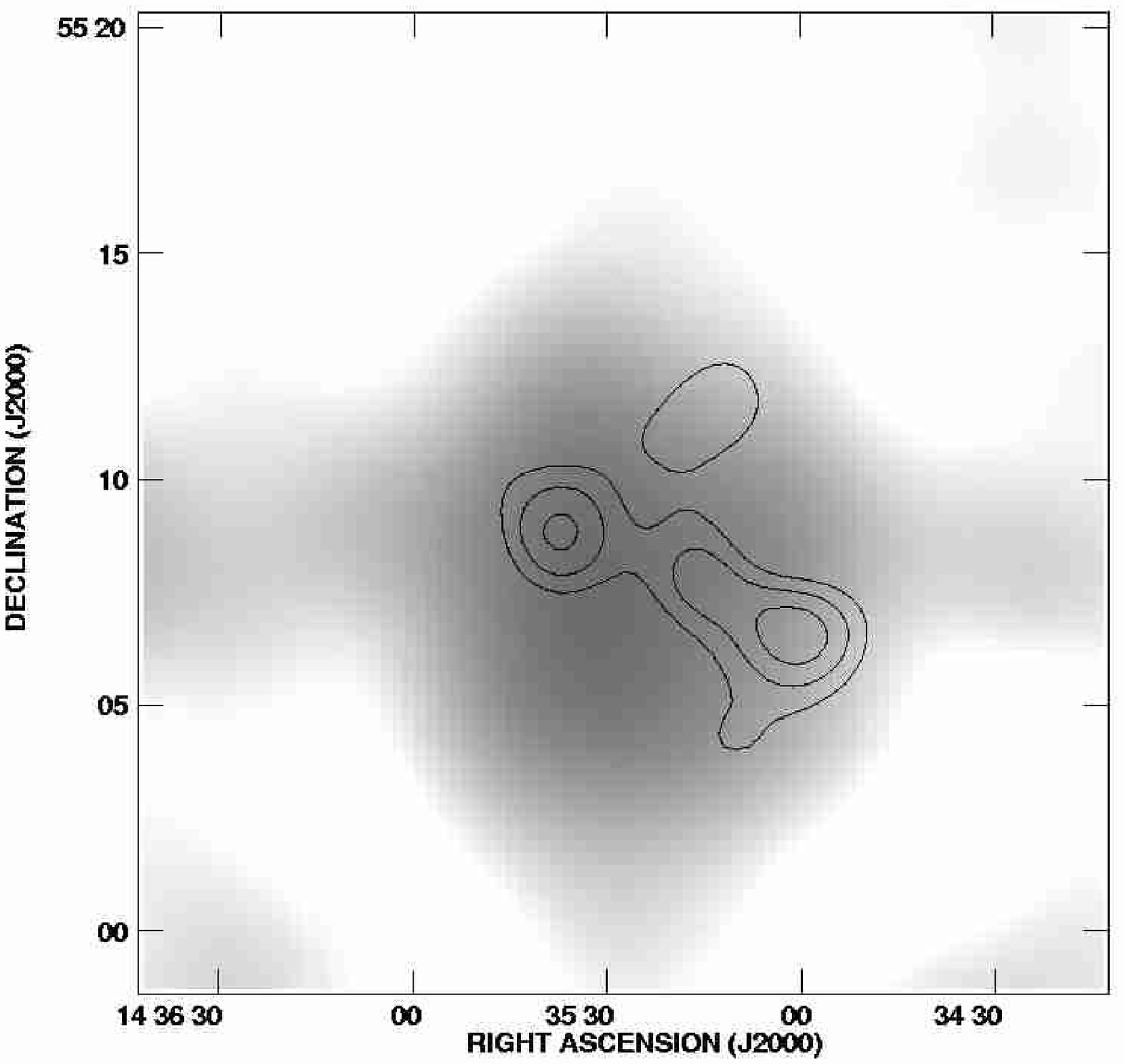]{Abell 1940  See Figure \ref{fig_3a} for details. \label{fig_3h}}
\figcaption[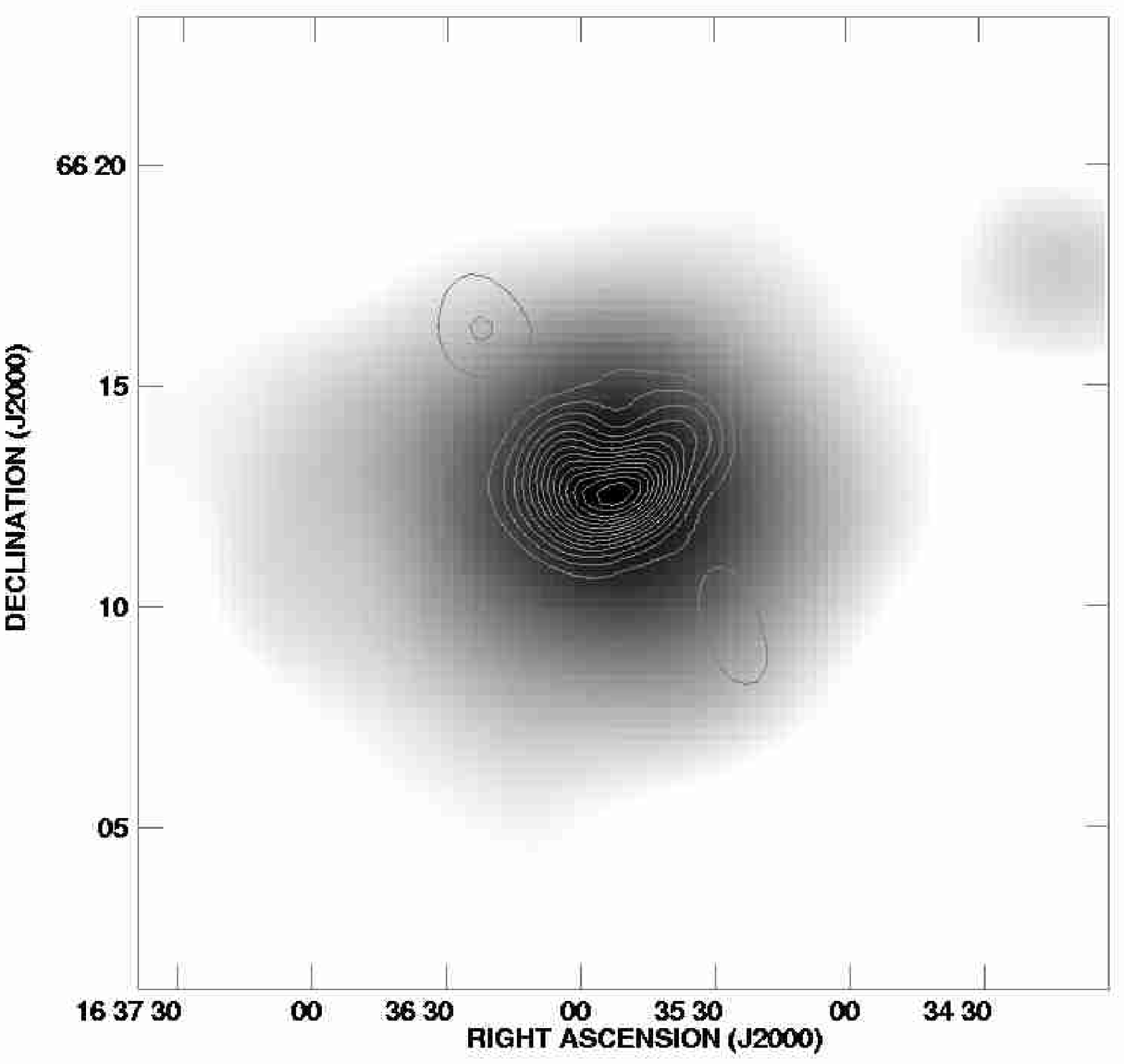]{Abell 2218  See Figure \ref{fig_3a} for details. \label{fig_3i}}
\figcaption[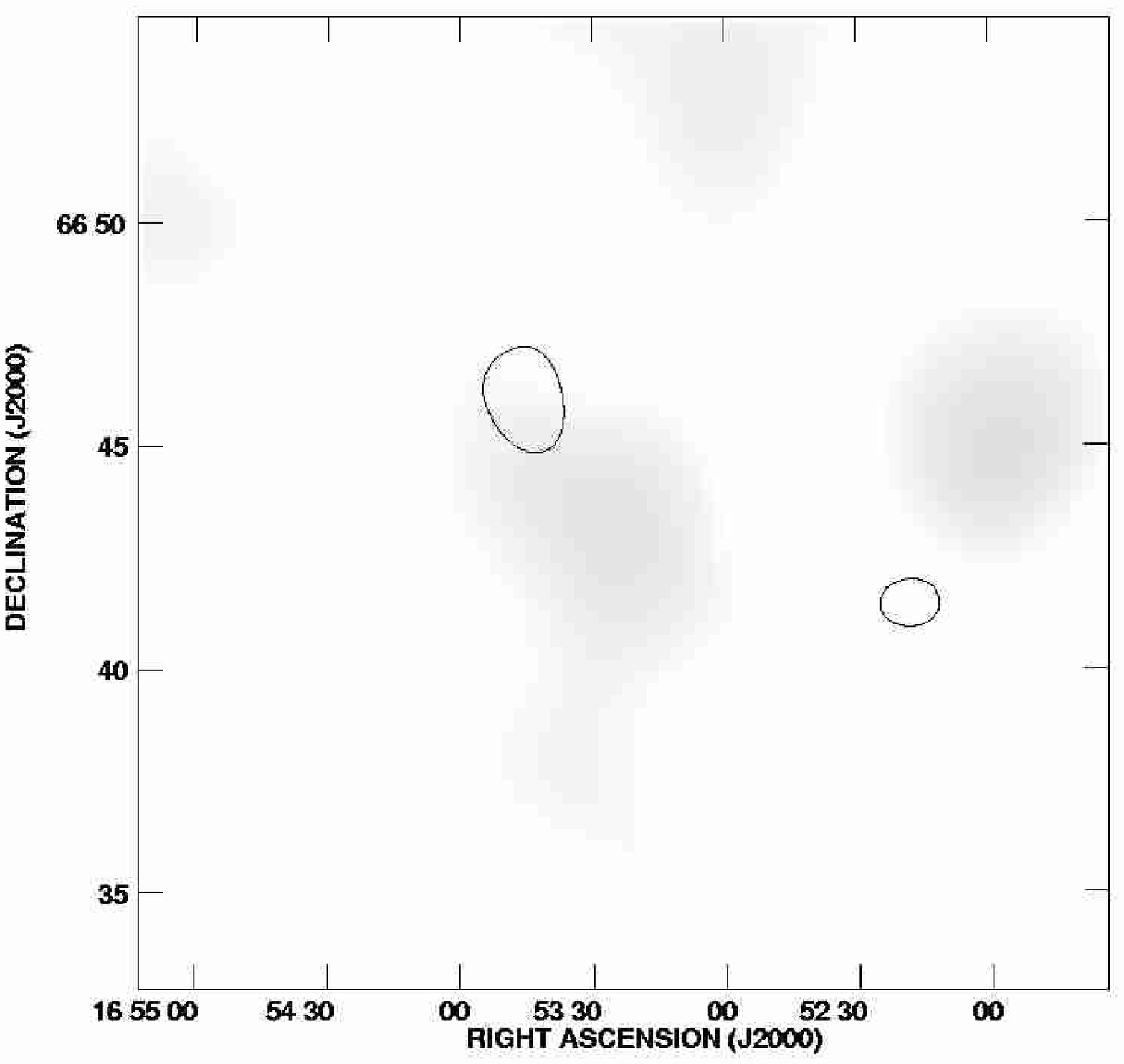]{Abell 2240  See Figure \ref{fig_3a} for details. \label{fig_3j}}
\figcaption[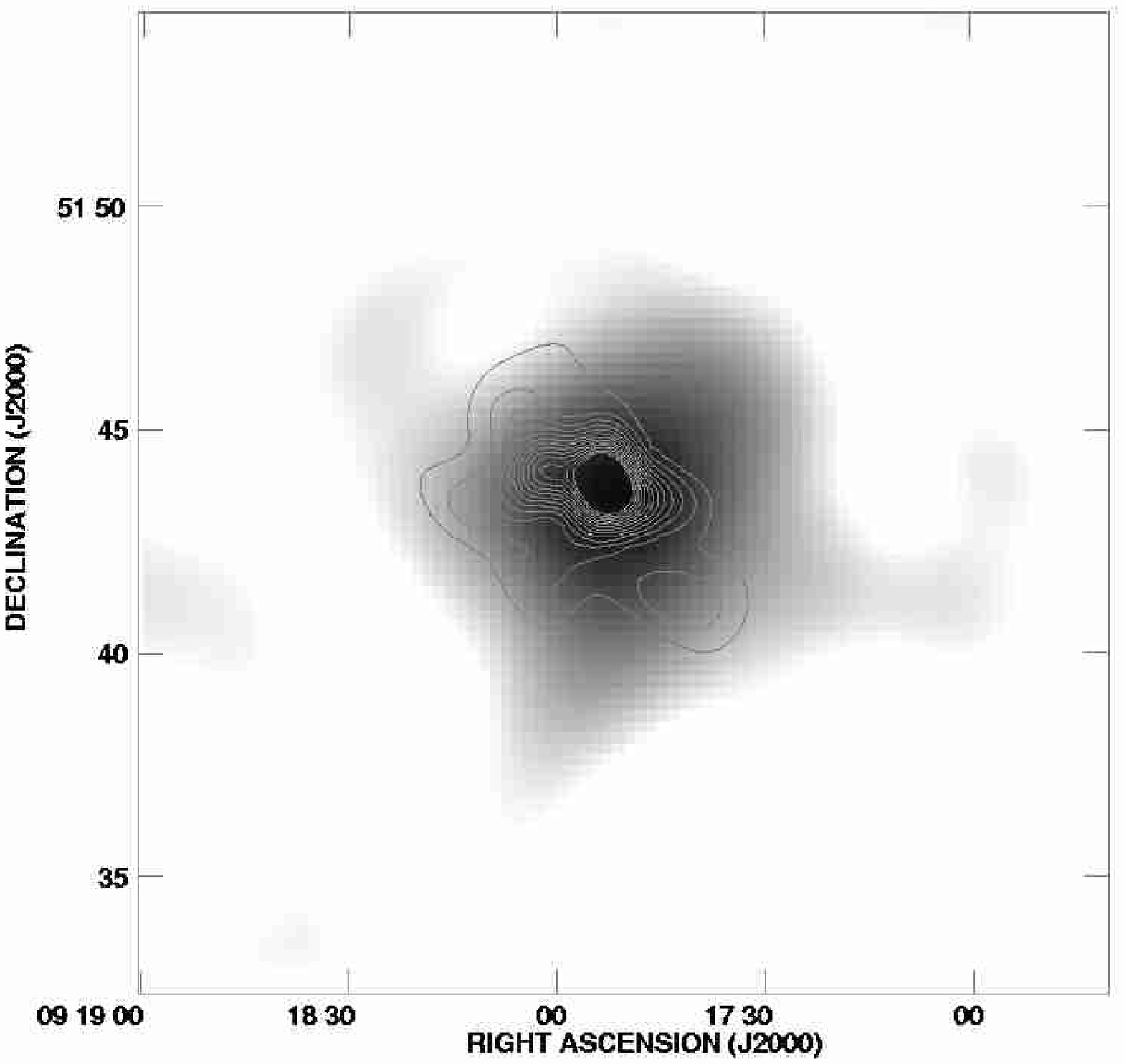]{Abell  773  See Figure \ref{fig_3a} for details. \label{fig_3k}}
\figcaption[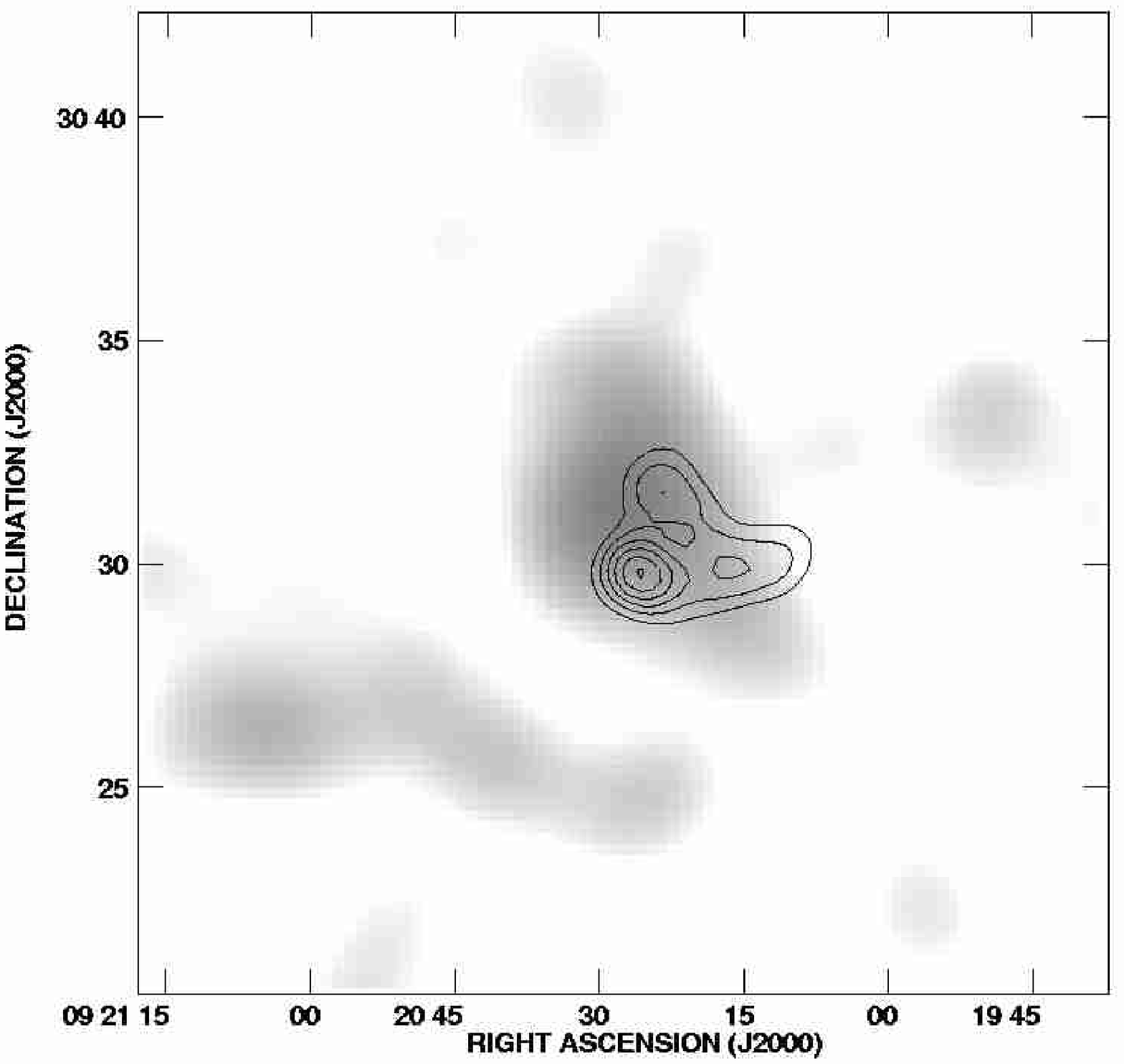]{Abell  781  See Figure \ref{fig_3a} for details. \label{fig_3l}}
\figcaption[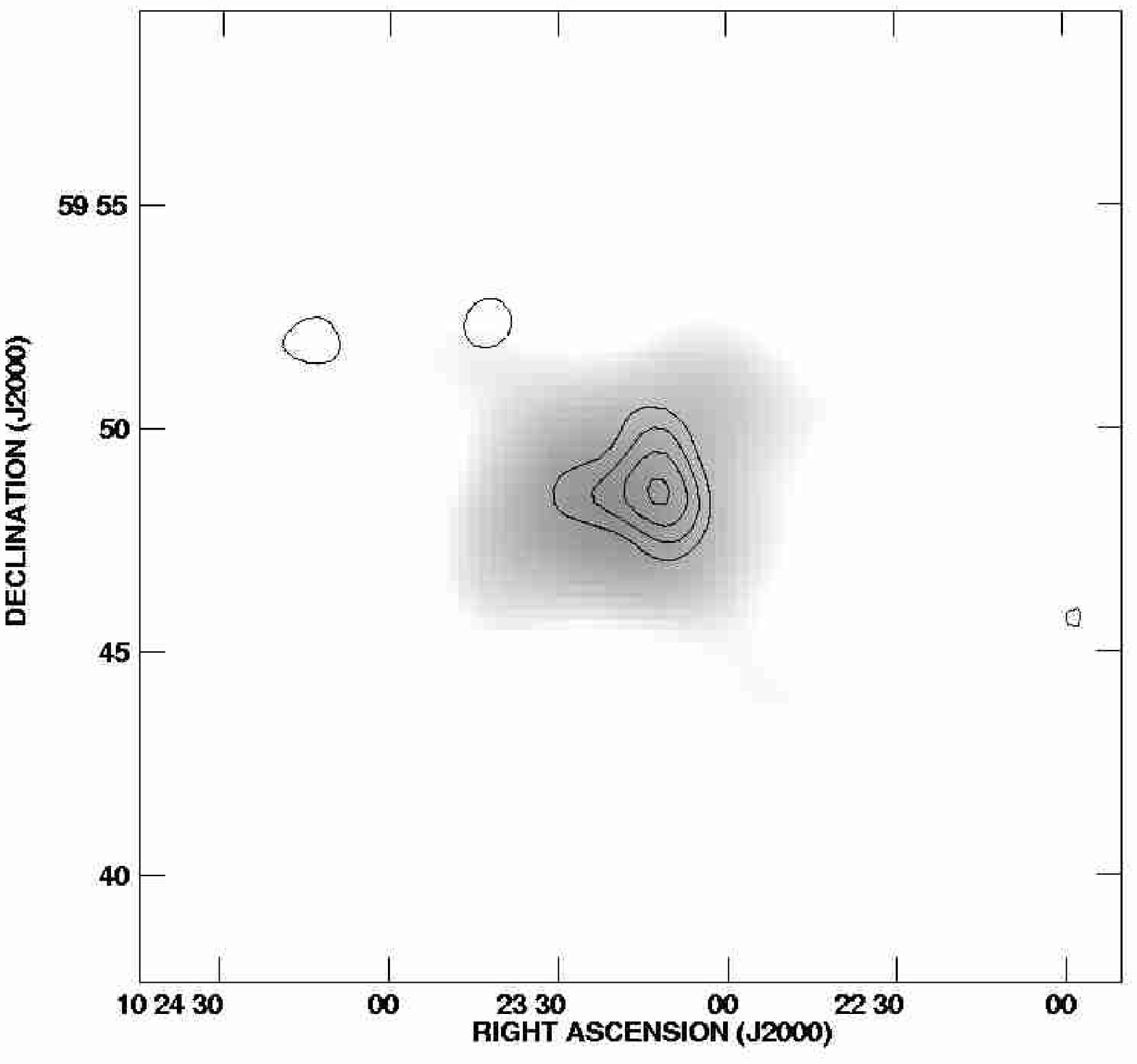]{Abell  983  See Figure \ref{fig_3a} for details. \label{fig_3m}}
\figcaption[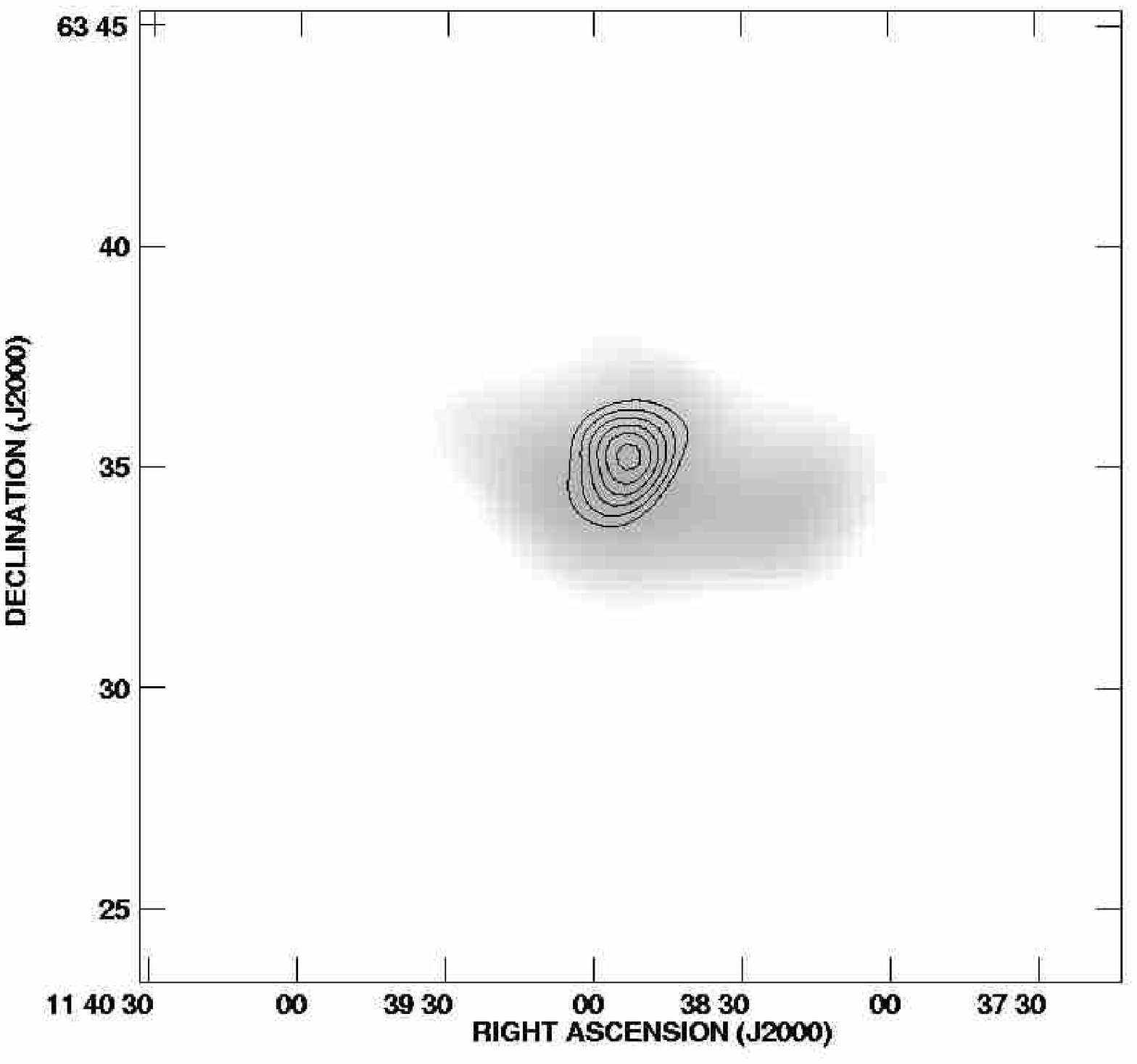]{Abell 1331  See Figure \ref{fig_3a} for details. \label{fig_3n}}
\figcaption[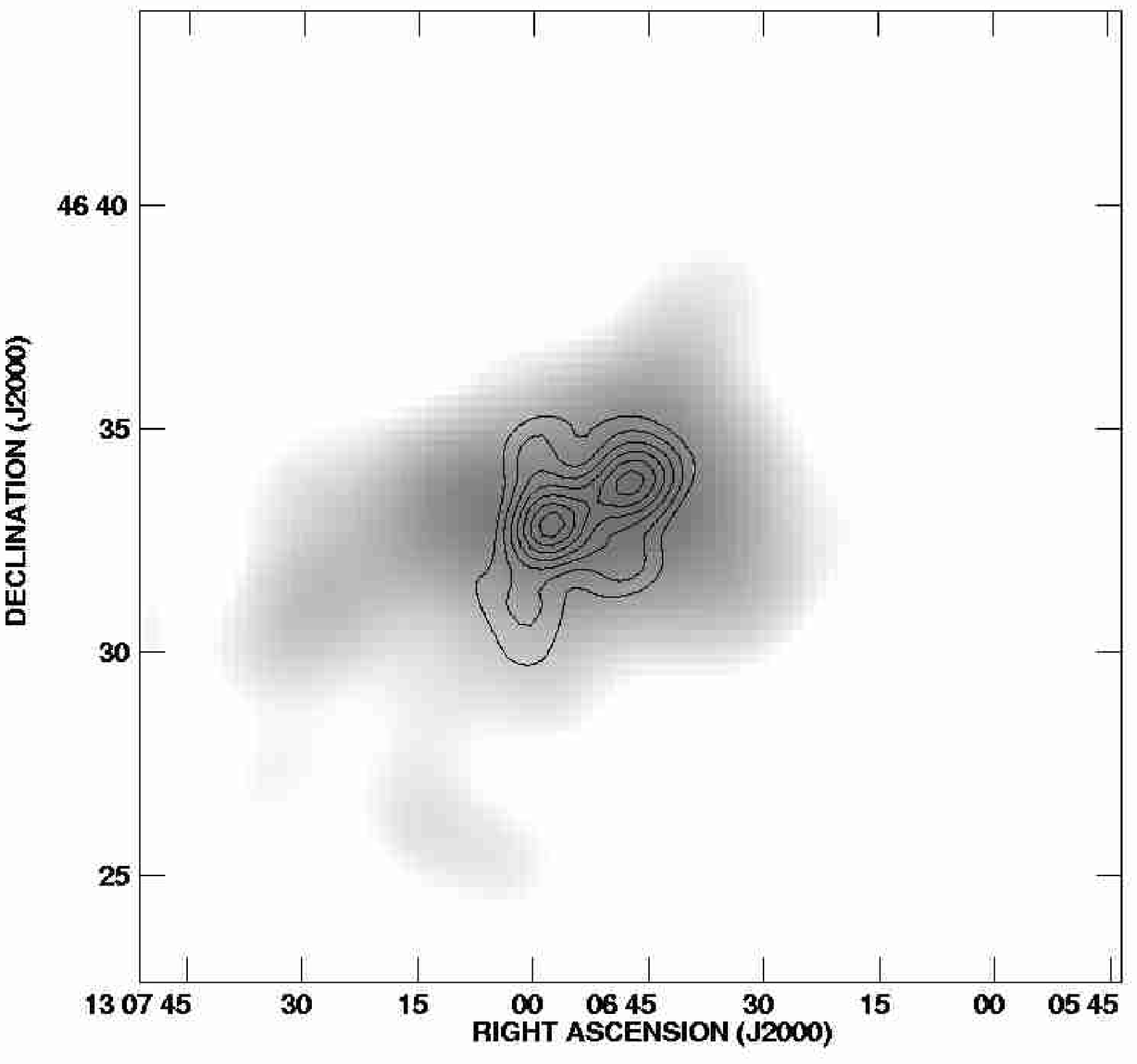]{Abell 1682  See Figure \ref{fig_3a} for details. \label{fig_3o}}
\figcaption[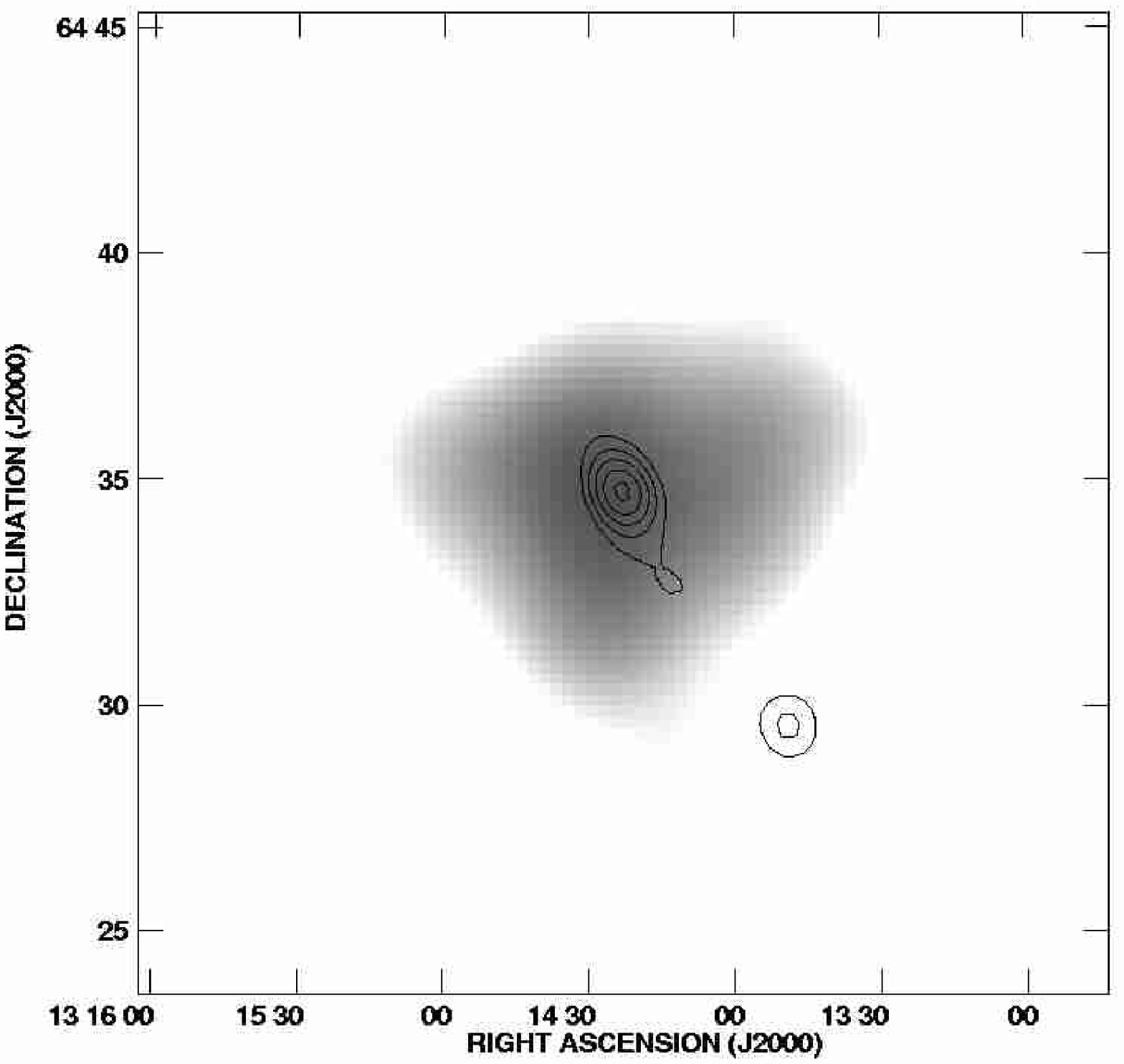]{Abell 1704  See Figure \ref{fig_3a} for details. \label{fig_3p}}
\figcaption[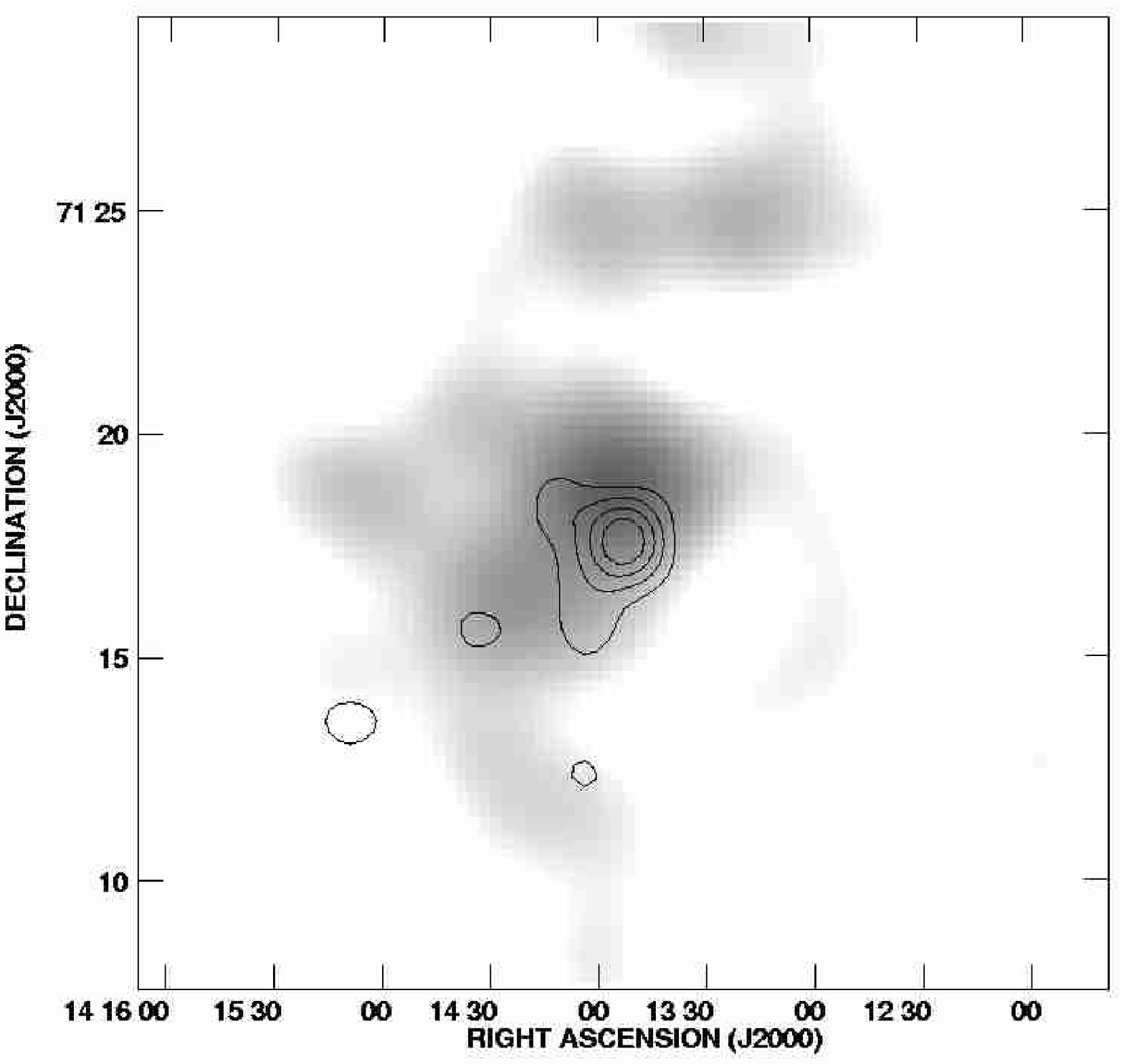]{Abell 1895  See Figure \ref{fig_3a} for details. \label{fig_3q}}
\figcaption[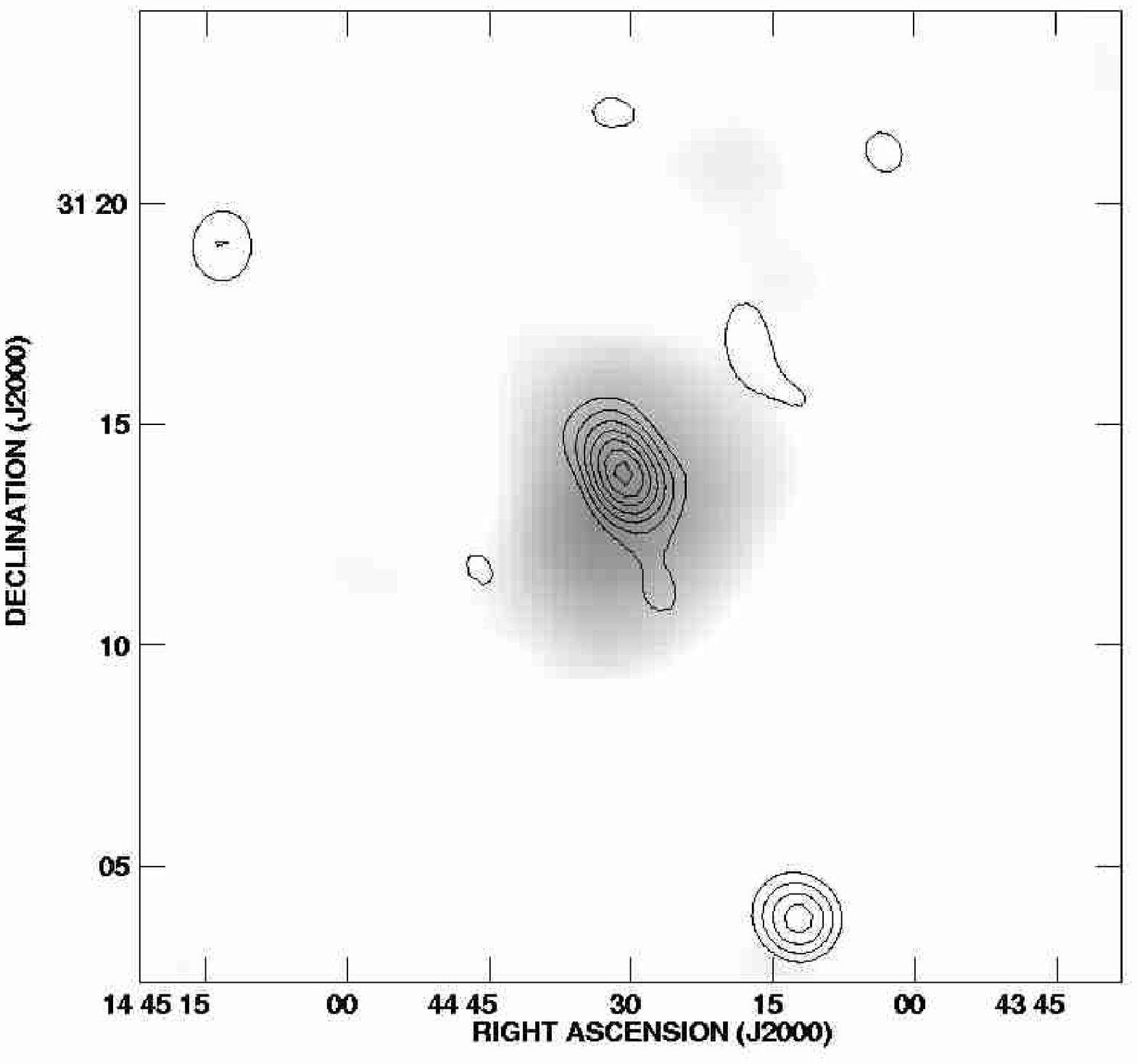]{Abell 1961  See Figure \ref{fig_3a} for details. \label{fig_3r}}
\figcaption[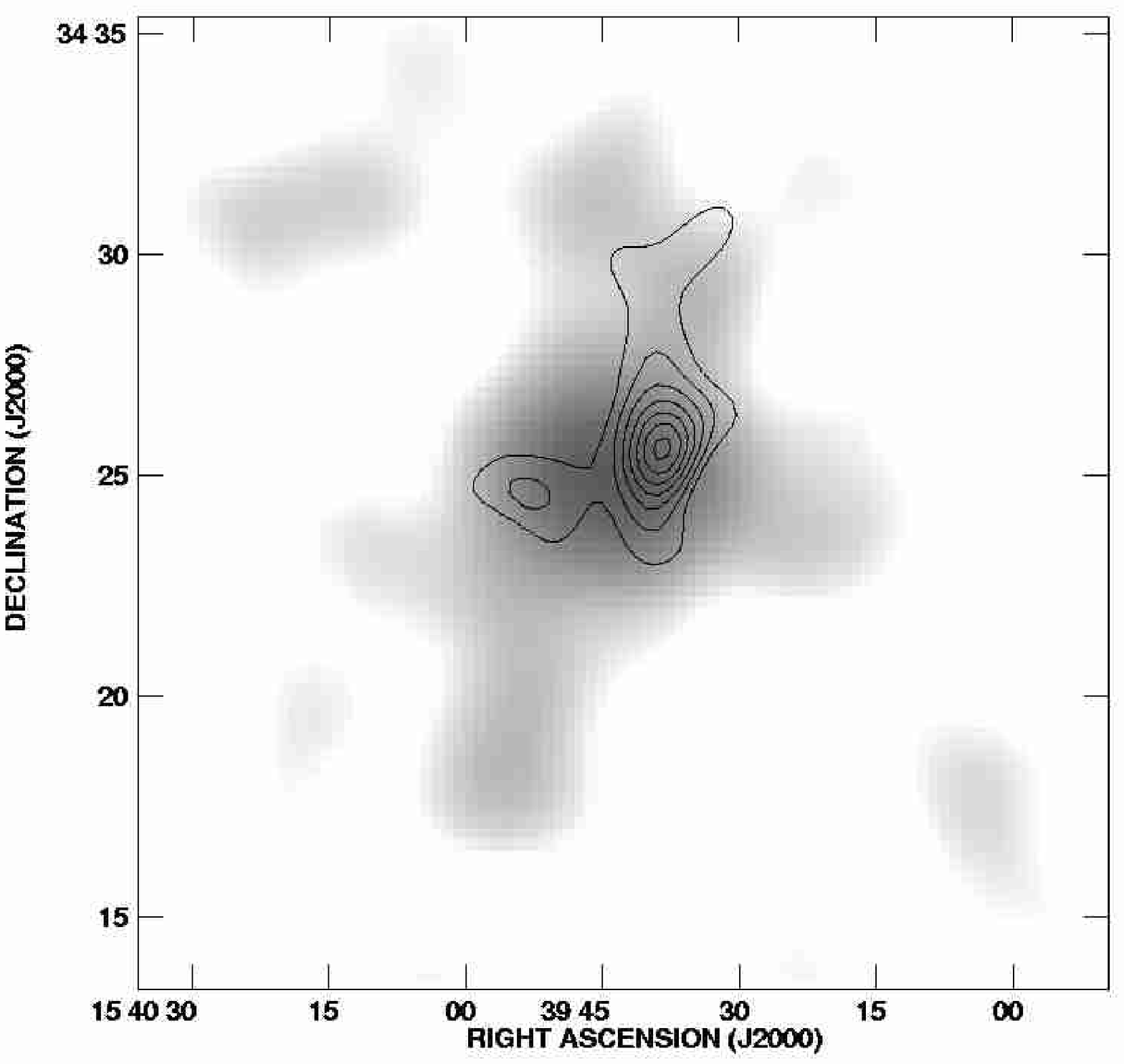]{Abell 2111  See Figure \ref{fig_3a} for details. \label{fig_3s}}

\clearpage

\begin{figure}
\figurenum{1}
\epsscale{1.0}
\plotone{f1.ps}
\end{figure}
\clearpage

\begin{figure}
\figurenum{2a}
\epsscale{1.0}
\plotone{f2a.ps}
\end{figure}

\begin{figure}
\figurenum{2b}
\epsscale{1.0}
\plotone{f2b.ps}
\end{figure}

\begin{figure}
\figurenum{2c}
\epsscale{1.0}
\plotone{f2c.ps}
\end{figure}

\begin{figure}
\figurenum{2d}
\epsscale{1.0}
\plotone{f2d.ps}
\end{figure}
\clearpage

\begin{figure}
\figurenum{2e}
\epsscale{1.0}
\plotone{f2e.ps}
\end{figure}

\begin{figure}
\figurenum{2f}
\epsscale{1.0}
\plotone{f2f.ps}
\end{figure}

\begin{figure}
\figurenum{2g}
\epsscale{1.0}
\plotone{f2g.ps}
\end{figure}

\begin{figure}
\figurenum{2h}
\epsscale{1.0}
\plotone{f2h.ps}
\end{figure}
\clearpage

\begin{figure}
\figurenum{2i}
\epsscale{1.0}
\plotone{f2i.ps}
\end{figure}

\begin{figure}
\figurenum{2j}
\epsscale{1.0}
\plotone{f2j.ps}
\end{figure}

\begin{figure}
\figurenum{2k}
\epsscale{1.0}
\plotone{f2k.ps}
\end{figure}

\begin{figure}
\figurenum{2l}
\epsscale{1.0}
\plotone{f2l.ps}
\end{figure}
\clearpage

\begin{figure}
\figurenum{2m}
\epsscale{1.0}
\plotone{f2m.ps}
\end{figure}

\begin{figure}
\figurenum{2n}
\epsscale{1.0}
\plotone{f2n.ps}
\end{figure}

\begin{figure}
\figurenum{2o}
\epsscale{1.0}
\plotone{f2o.ps}
\end{figure}

\begin{figure}
\figurenum{2p}
\epsscale{1.0}
\plotone{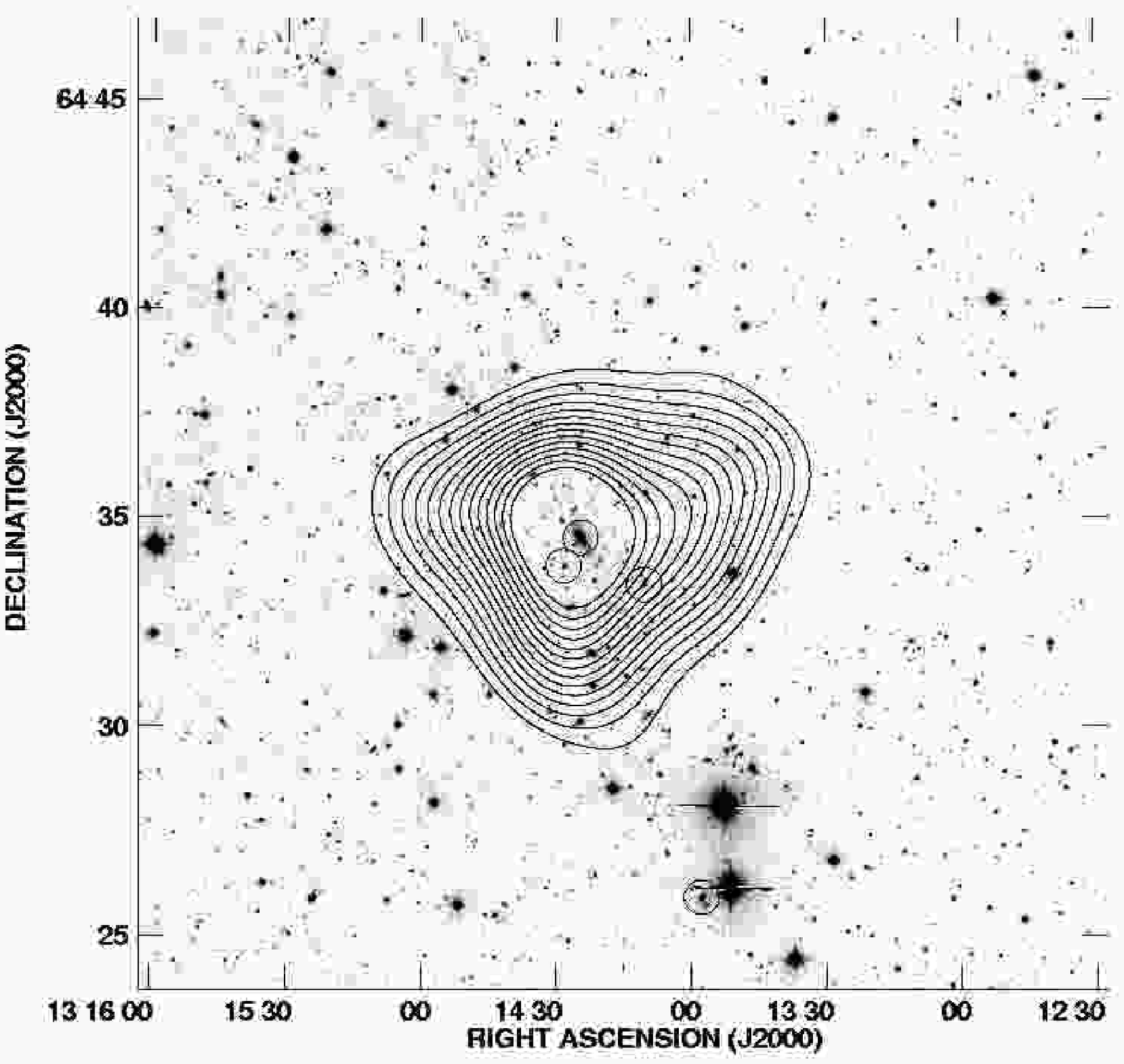}
\end{figure}
\clearpage

\begin{figure}
\figurenum{2q}
\epsscale{1.0}
\plotone{f2q.ps}
\end{figure}

\begin{figure}
\figurenum{2r}
\epsscale{1.0}
\plotone{f2r.ps}
\end{figure}

\begin{figure}
\figurenum{2s}
\epsscale{1.0}
\plotone{f2s.ps}
\end{figure}

\clearpage

\begin{figure}
\figurenum{3a}
\epsscale{1.0}
\plotone{f3a.ps}
\end{figure}

\begin{figure}
\figurenum{3b}
\epsscale{1.0}
\plotone{f3b.ps}
\end{figure}

\begin{figure}
\figurenum{3c}
\epsscale{1.0}
\plotone{f3c.ps}
\end{figure}

\begin{figure}
\figurenum{3d}
\epsscale{1.0}
\plotone{f3d.ps}
\end{figure}
\clearpage

\begin{figure}
\figurenum{3e}
\epsscale{1.0}
\plotone{f3e.ps}
\end{figure}

\begin{figure}
\figurenum{3f}
\epsscale{1.0}
\plotone{f3f.ps}
\end{figure}

\begin{figure}
\figurenum{3g}
\epsscale{1.0}
\plotone{f3g.ps}
\end{figure}

\begin{figure}
\figurenum{3h}
\epsscale{1.0}
\plotone{f3h.ps}
\end{figure}
\clearpage

\begin{figure}
\figurenum{3i}
\epsscale{1.0}
\plotone{f3i.ps}
\end{figure}

\begin{figure}
\figurenum{3j}
\epsscale{1.0}
\plotone{f3j.ps}
\end{figure}

\begin{figure}
\figurenum{3k}
\epsscale{1.0}
\plotone{f3k.ps}
\end{figure}

\begin{figure}
\figurenum{3l}
\epsscale{1.0}
\plotone{f3l.ps}
\end{figure}
\clearpage

\begin{figure}
\figurenum{3m}
\epsscale{1.0}
\plotone{f3m.ps}
\end{figure}

\begin{figure}
\figurenum{3n}
\epsscale{1.0}
\plotone{f3n.ps}
\end{figure}

\begin{figure}
\figurenum{3o}
\epsscale{1.0}
\plotone{f3o.ps}
\end{figure}

\begin{figure}
\figurenum{3p}
\epsscale{1.0}
\plotone{f3p.ps}
\end{figure}
\clearpage

\begin{figure}
\figurenum{3q}
\epsscale{1.0}
\plotone{f3q.ps}
\end{figure}

\begin{figure}
\figurenum{3r}
\epsscale{1.0}
\plotone{f3r.ps}
\end{figure}

\begin{figure}
\figurenum{3s}
\epsscale{1.0}
\plotone{f3s.ps}
\end{figure}

\end{document}